\documentclass[12pt,twoside, epsf]{article}

\usepackage{color,graphicx,times}
\pdfoutput=1

\makeatletter
\long\def\M#1{\leavevmode\setbox\@tempboxa\hbox{#1}\@tempdima\fboxrule
    \advance\@tempdima \fboxsep \advance\@tempdima \dp\@tempboxa
   \hbox{\lower \@tempdima\hbox
  {\vbox{\hrule \@height \fboxrule
          \hbox{  \hskip\fboxsep
          \vbox{\vskip\fboxsep \box\@tempboxa\vskip\fboxsep}\hskip
                 \fboxsep\vrule \@width \fboxrule}%
                  }}}}

\makeatother

\let \ttorg \tt \def \tt{\ttorg \obeyspaces}

\begin{document}

\title{\Large\bf Recursive Distinctioning }
\author{Joel Isaacson\\
and\\
Louis H. Kauffman\\ Department of Mathematics, Statistics \\ and Computer Science (m/c
249)    \\ 851 South Morgan Street   \\ University of Illinois at Chicago\\
Chicago, Illinois 60607-7045\\ $<$kauffman@uic.edu$>$\\}

\maketitle

\thispagestyle{empty}

\subsection*{\centering Abstract}

{\em In this paper we explore recursive distinctioning.}
\bigbreak

\noindent {\bf Keywords.} recursive distinctioning, logic, algebra, topology, biology, replication, celluar automaton, quantum, DNA, container, extainer

 
\section{Introduction to Recursive Distinctioning} 
{\it Recursive Distinctioning (RD)} is a name coined by Joel Isaacson in his original patent document \cite{JI} describing how fundamental patterns of process arise from the systematic 
application of operations of distinction and description upon themselves. See also his other papers referenced herein \cite{JI0, JI1, JI2, JI3, JI4}. Background papers by Louis H. Kauffman on recursion, knotlogic and biologic can be found at \cite{LK1, LK2, LK3, LK4, LK5, LK6, LK7, LK8, LK9, LK10, LK11, LK12, SRF, BL1, BL2, KL, KP}\\

\begin{figure}
     \begin{center}
     \begin{tabular}{c}
     \includegraphics[width=8cm]{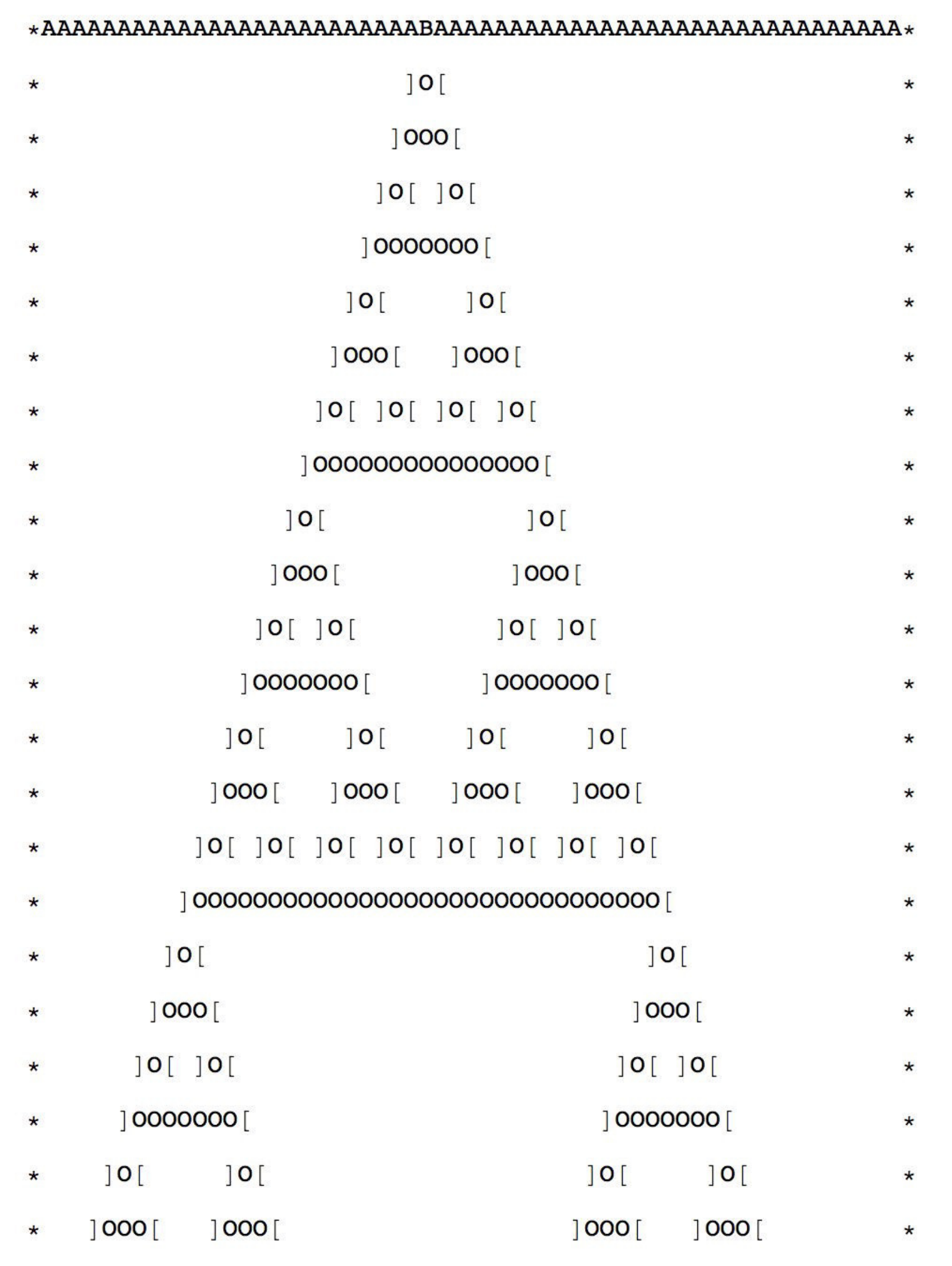}
     \end{tabular}
     \caption{\bf RD Replication}
     \label{rep}
\end{center}
\end{figure}

 Recursive Distinctioning means just what it says. A pattern of distinctions is given in a space based on a graphical structure (such as a line of print or a planar lattice or given graph). Each node of the graph is occupied by a letter from some arbitrary alphabet. A specialized alphabet is given that can indicate distinctions about neighbors of a given node. The neighbors of a node are all nodes that are connected to the given node by edges in the graph. The letters in the specialized alphabet (call it SA) are used to describe the states of the letters in the given graph and at each stage in the recursion, letters in SA are written at all nodes in the graph, describing its previous state. The recursive structure that results from the iteration of descriptions is called Recursive Distinctioning. Here is an example. We use a line graph and represent it just as a finite row of letters. The Special Alphabet is $SA = \{ =, [, ], O \}$ where $``="$ means that the letters to the left and to the right are equal to the letter in the middle. Thus if we had $AAA$ in the line then the middle $A$ would be replaced by $=.$ The symbol $``["$ means that the letter to the {\it left} is different. Thus in $ABB$ the middle letter would be replaced by $[.$ The symbol $``]"$ means that the letter to the right is different. And finally the symbol $``O"$ means that the letters both to the left and to the right are different. $SA$ is a tiny language of elementary letter-distinctions. Here is an example of this RD in operation where we use the proverbial three dots to indicate a long string of letters in the same pattern. For example,

$$... AAAAAAAAAABAAAAAAAAAA ...$$
 is replaced by
$$\cdots =========]O[========= \cdots$$is replaced by
$$\cdots ========]OOO[======== \cdots$$is replaced by 
$$\cdots=======]O[=]O[======= \cdots .$$
Note that the element $]O[$ appears and it has replicated itself in a kind of mitosis. See Figure~\ref{rep} for a more detailed example of this evolution.
In Figure~\ref{green} we show the evolution of the RD, starting from a more arbitrary string.
Elementary RD patterns are fundamental and will be found in many structures at all levels. To see a cellular automaton example of this phenomenon of patterns crossing levels of structure, we later will look at a replicator in ``HighLife" a modification of John Horton Conway's automaton ``Life". The Highlife Replicator follows the same pattern as our RD Replicator. However the entity in HighLife that is self-replicating requires twelve steps to do the replication. The resultant patterns of replication can be seen form Figure~\ref{hl1} to Figure~\ref{hl8}. In the successive figures twelve steps are hidden and we see the same basic pattern shown in Figure~\ref{rep}.  We can understand directly how the RD Replicator works. This gives a foundation for understanding how the more complex HighLife Replicator behaves in its context. We take this phenomenon of the simple and the complex to be generic for many systems. By finding a point of simplicity, we make possible the evolution of understandings that are otherwise impossible to obtain.\\
\\

\begin{figure}
     \begin{center}
     \begin{tabular}{c}
     \includegraphics[width=5cm]{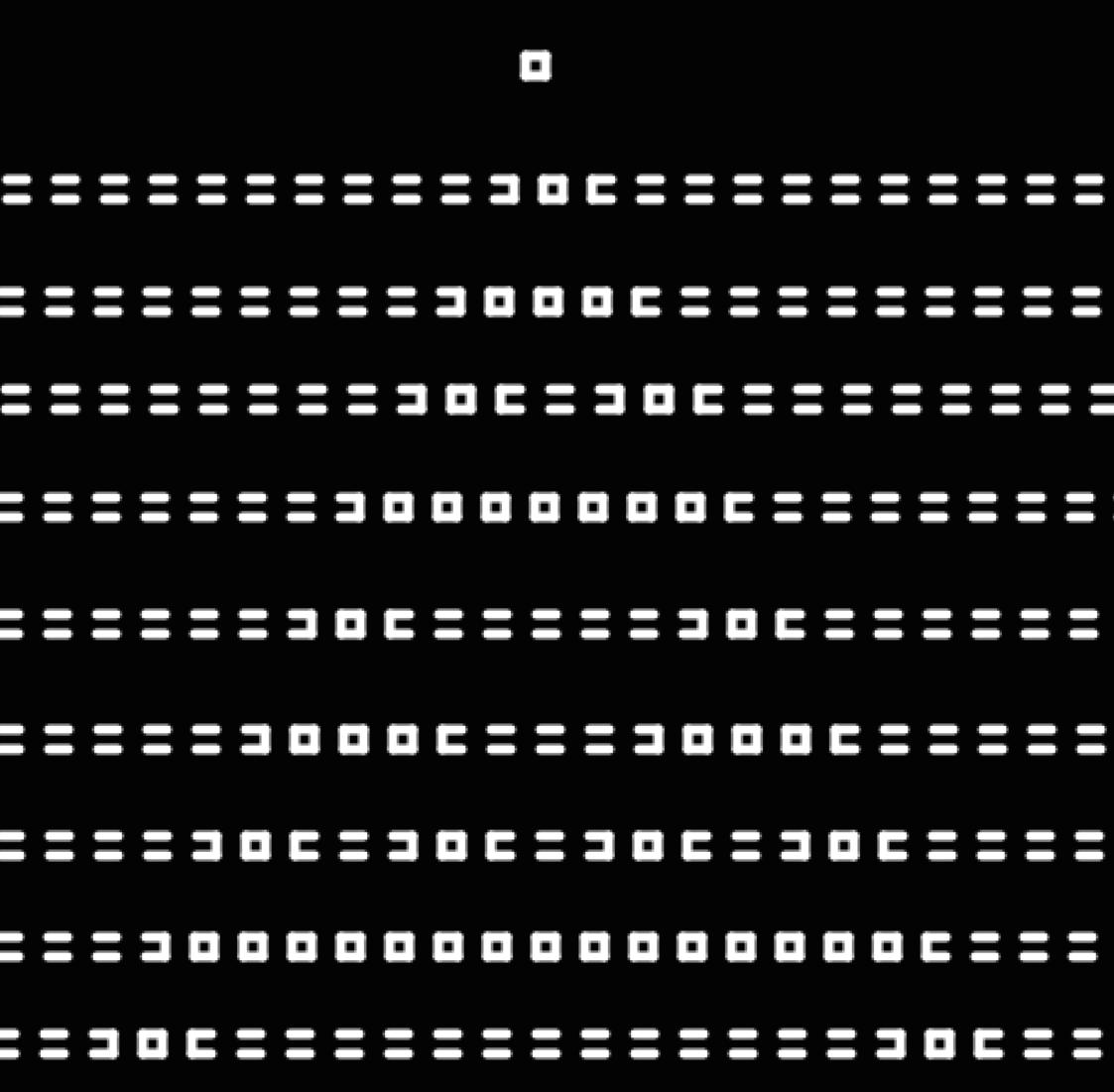}
     \end{tabular}
     \caption{\bf Second Picture of RD Replication}
     \label{replicate}
\end{center}
\end{figure}

\begin{figure}
     \begin{center}
     \begin{tabular}{c}
     \includegraphics[width=5cm]{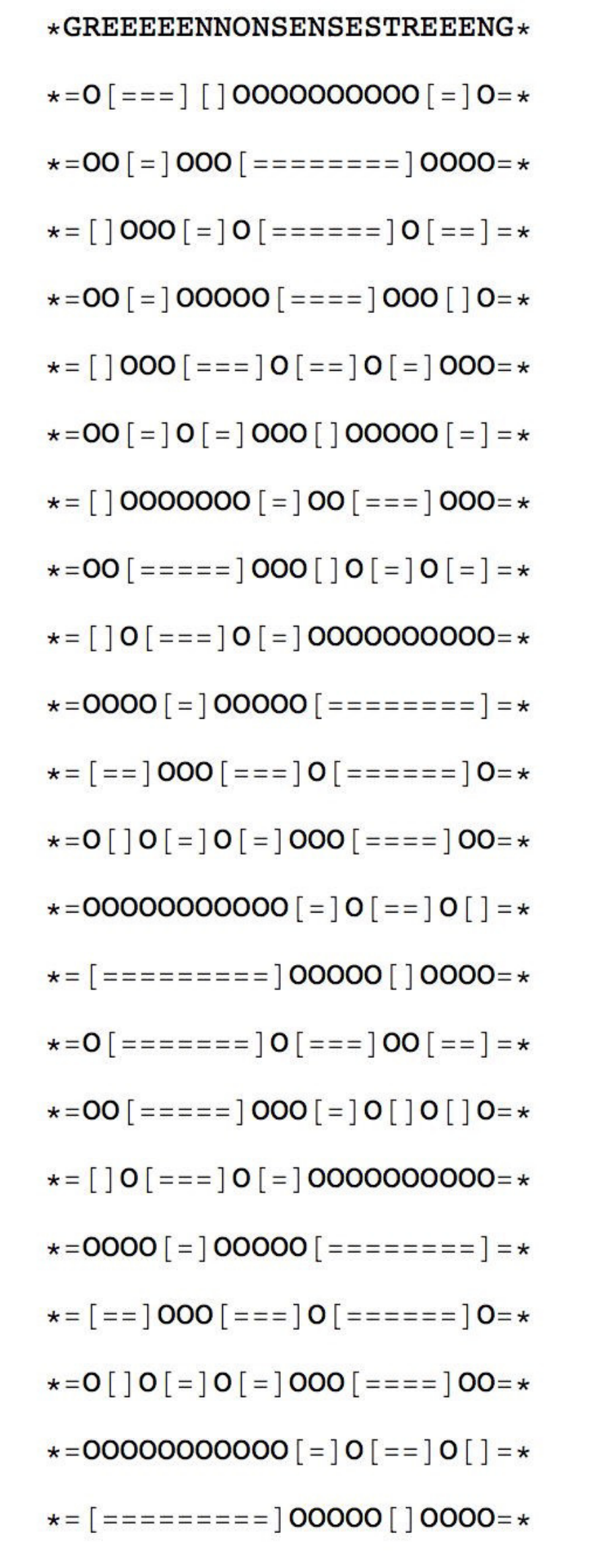}
     \end{tabular}
     \caption{\bf A String Evolution}
     \label{green}
\end{center}
\end{figure}

We can place the basic idea of recursive distinctioning with the context of cellular automata.
RD is distinct from other types of cellular automaton in that its basic recursion is based on direct distinctions made (locally) in relation to distinctions present in the given state of the automaton. In a typical cellular automaton  the next state is obtained on the basis of simple distinctions about the previous state. These distinctions are not necessarily at the letter-level. For example, in a Wolfram line automaton we have eight possible local neighborhoods consisting in triples of zeros and ones. Any distinction made among these eight, separating them into two classes, is acceptable as a rule for the Wolfram automaton. The operation of distinction is shifted to 
a higher level than the question of sameness or difference for nearby iconic elements of the state. This is the distinction between our ``orthodox" RD models and other recursive models. 
We are interested in rules that involve direct matters of sameness or difference. Such RD rules are very primitive rules. Nevertheless, we regard the orthodox RD models as part of the larger class of recursive cellular automata. We wish to explore the relationship between our primordial structures and the closely related structure of all cellular automata as they are understood at this time.\\

Everyone who works in science, mathematics, or computer science is familiar with the fundamental role of the concept of distinction and the making of distinctions in both theory and practice. For example, Einstein's relativity depends on a new distinction between space and time relative to an observer and a new unification of space and time that is part and parcel of this distinction. Every moment of using a digital computer depends upon the myriad of distinctions that are handled automatically by the computer, enabling the production and recording of these words and the computation and transmission of information. Distinctions act on other distinctions. Once a new distinction is born, it becomes the object of further action. Thus grows all the physics that comes from relativity and thus grows all the industry of computation that grows from the idea and implementation of the Turing machine, the programmed computer. \\

And yet it is not usually recognized that it is through recursive distinctioning that all such progress is made. We will discuss recursive distinctioning both in its human and its automatic aspects. In the automatic aspect we will give examples of automata that are based on making very simple distinctions of equality, right/left, that then, upon allowing these distinctions to act on themselves, produce periodic and dialectical patterns that suggest what are usually regarded as higher-level phenomena. In this way, and with these examples, we can illustrate and speculate on the nature of intelligence, evolution, and many themes of fundamental science. \\

The remarkable feature of these examples of recursive distinctioning is their great simplicity coupled with the complexity of behaviors that can arise from them. Notice that each successive string in the recursion can be regarded as describing its predecessor. It is remarkable that there should be such intricate structure in the process of description. Description is another word for making a distinction. The description of a given string is a string of individual distinctions that have been made. Each individual distinction is one that recognizes whether a given character in a string is equal to a left neighbor, a right neighbor, both, or neither. This elementary distinction becomes instantiated as a character in the new description string. The description string can be subjected to the same scrutiny and so the recursive process continues. \\

Note that this recursive process depends, at its base, on the most elementary distinctions possible for character strings. No mathematical calculations are performed. We should mention that distinction-making without mathematical computation is ubiquitous in natural neuronal processing. Joel Isaacson's collaboration with Eshel Ben- Jacob includes attempts to demonstrate RD in live neuronal tissue \cite{JI0}. One can also point to the molecular interactions of DNA and RNA as natural RD automata. Finally, we can point to the notion of chemlambda computation of Buliga and Kauffman \cite{BuligaKauff} as an abstract chemical combination computing that includes aspects of lambda calculus, but is based on direct and local action related to distinctions inherent in the system.\\

The epistemology behind this automaton is based directly on distinctions that can be made automatic. Other cellular automata are also based on distinctions. For example the well-known Wolfram line automata \cite{Wolfram}  are based on character strings with only two characters and the recognition of the eight possible triples of characters that can occur, including characters to the left and to the right of a given character. The automaton rule then replaces the middle character according to the structure of this neighborhood. 
There is a crucial difference in epistemology between a Wolfram line automaton and our recursive distinction program. We do not replace according to an arbitrary rule. We place a character that describes the distinctive structure of the neighborhood of the predecessor character. Our automaton engages in a meta-dialogue about its own structure. This dialogue is then entered as a string for the automaton to examine and act upon once again. The patterns produced by this recursive distinction are part of a dialogue that the strings hold with themselves. 
One can ask many questions about recursive distinctioning as presented here. The automaton we have demonstrated illustrates a concept that can be instantiated in many ways. We hope, in a paper to come, to demonstrate Turing universality for automata of this type. But in fact we feel that the paradigm of recursive distinctioning goes beyond (or around) the paradigm of the Turing machine, and we will discuss that issue as well. \\

There is another level to our automaton and that is the level of examining with human eyes and minds the output of the automaton, seeing patterns in the whole collection of strings and engaging in further design on this basis. This is where the recursive automatic distinctions meet the aware distinguishing of the observers of the system, connecting the automatic with the aware process and design level that goes on in the larger network of science.\\

It is the case that in the design of computing machines human beings have for centuries confronted the issue of repeatability for the sake of computation or for the sake of the production of pattern (as in weaving) or 
the reliability of manufacture (as in timekeeping). This means that elementary distinctions must be reproducible and comparable as in mathematical notations, written language and the mechanics of clocks and computing devices. Thus we shall refer to automatic distinctions when we speak of highly repeatable physical situations that can be regarded as reproductions of distinctions that are available to an observer. In some cases such 
distinctions are designed by someone who engineers them into the device. In other cases, we recognize computational and reproducible patterns in a natural situation. The earth goes around the sun periodically, the moon goes around the earth. Natural clocks arise from these periodicities and regularities observed in our world. Thus, in this essay, we do not restrict ourselves in the use of the word distinction to the meaning that a distinction 
is a distinction made by some human observer. We refer to distinctions that are ongoing in a device beyond our direct observation. Nevertheless, the buck stops at a human observer who recognizes the patterns of the device and who interprets the meaning of what has been produced. It is then possible to discuss the role of creativity in relation to deterministic and automatic actions.\\

\noindent {\bf Acknowledgement.} We thank Bernd Schmeikal for conversations and for sharing his own research in relation to our work. We thank Dan Sandin for a continuing collaboration with Lou Kauffman and particularly for sharing the computer program for 2DRD that has been evolved by the two of them. The graphical illustrations of 2DRD in this paper are all produced by that program.\\

\section{The Logic of Distinction and the Distinction of Logic}

\begin{figure}
     \begin{center}
     \begin{tabular}{c}
     \includegraphics[width=6cm]{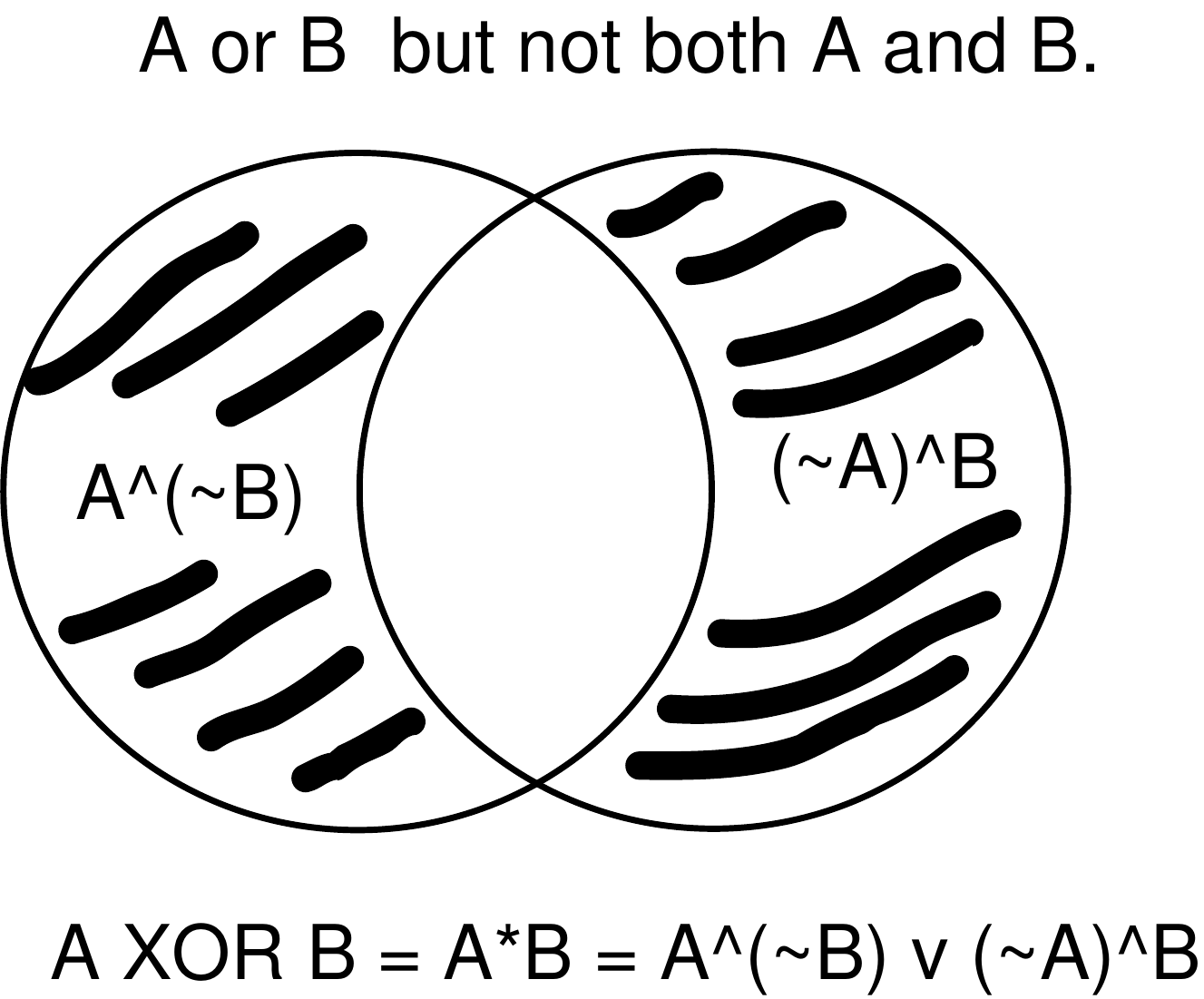}
     \end{tabular}
     \caption{\bf XOR in Venn Diagrams}
     \label{xor}
\end{center}
\end{figure}

\begin{figure}
     \begin{center}
     \begin{tabular}{c}
     \includegraphics[width=6cm]{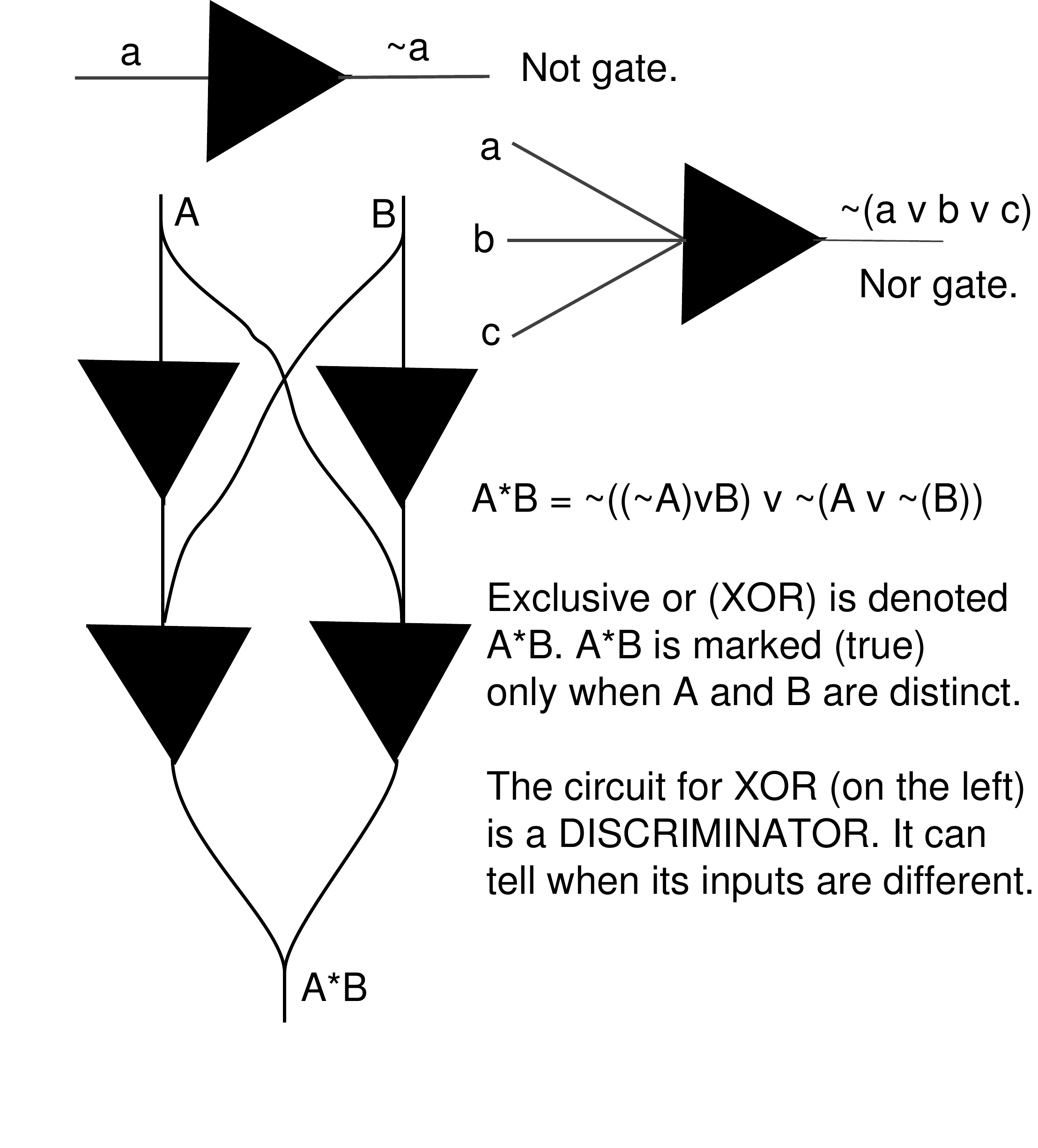}
     \end{tabular}
     \caption{\bf XOR Circuit}
     \label{xorcircuit}
\end{center}
\end{figure}

\begin{figure}
     \begin{center}
     \begin{tabular}{c}
     \includegraphics[width=6cm]{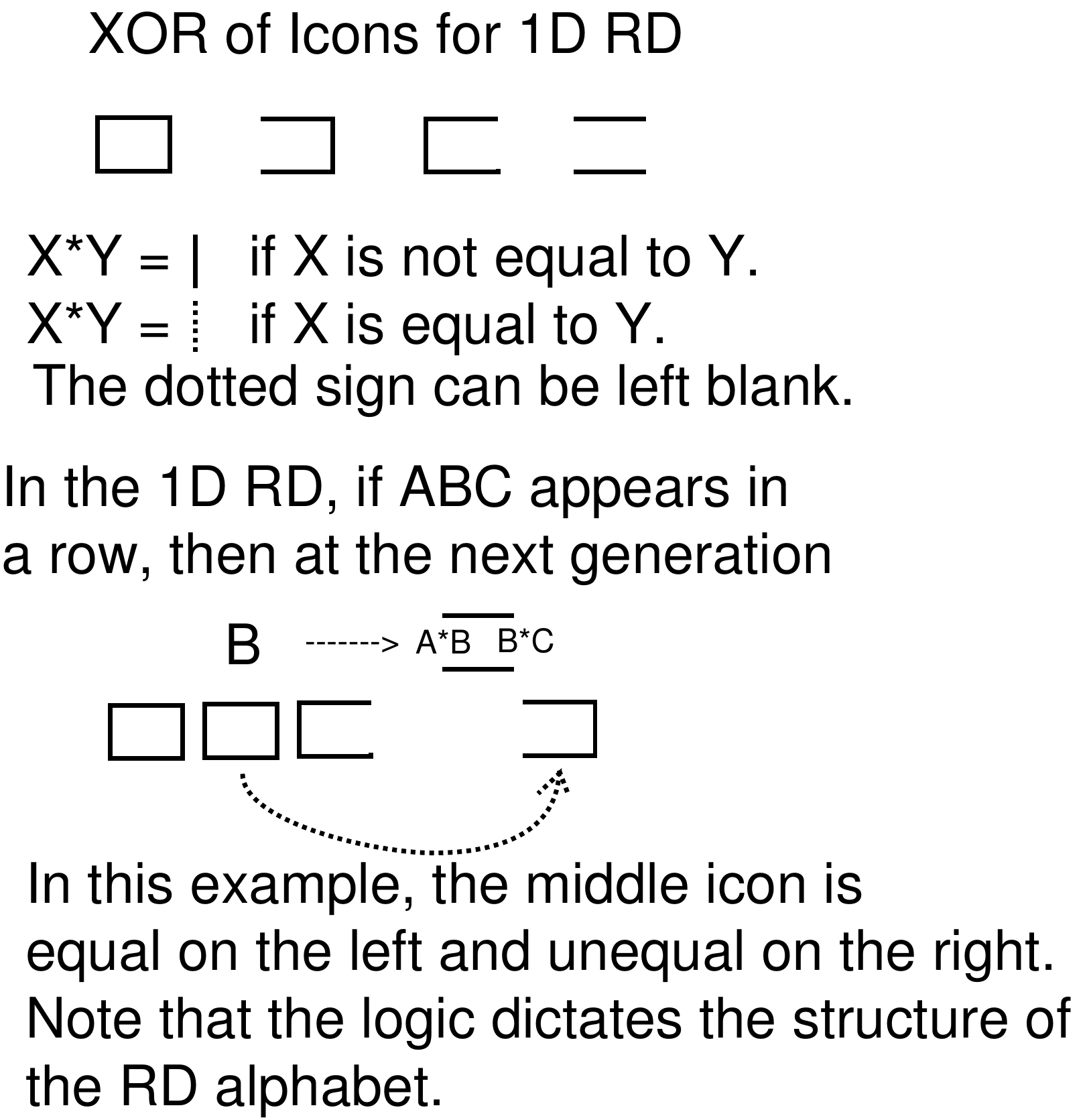}
     \end{tabular}
     \caption{\bf XOR of Icons}
     \label{xoricon}
\end{center}
\end{figure}

We have introduced the one dimensional $RD$ and its very simple alphabet based on the four iconic symbols shown in Figure~\ref{xoricon}.  In this figure we use a box rather than a circle for the icon that indicates difference to both the right and the left, and we use a box with a missing left vertical edge to denote sameness on the left, a box with a missing right vertical edge to denote sameness on the right. Sameness on both right and left is indicated by two parallel lines that remain when the two vertical edges of the box are removed. In this figure we give a logical justification for these icons in terms of the act of discrimination. That is, we give a 
logical construction for an icon that describes and embodies the discrimination itself. At a given point in the line of letters there is a given letter. This letter is either distinct or different from its neighbor to the left and/or its
neighbor to the right. We introduce a method to {\it manufacture} an icon that expresses these distinctions. In order to do this we insert a line segement in between the space for the given letter and the space  next to it
{\it if there is a difference between the given letter and its neighbor}. We take as given a line segment at the top of the space and a line segment at the bottom of the space. (This actually indicates the condition of the 
one dimensional RD where it is distinct from its context above and below the one dimension of operations.) As a result this process of discrimination constructs four possible icons that describe the condition of a given letter.
The icons are illustrated Figure~\ref{xoricon}, and the reader can see that they are: an equals sign when there is no distinction to the left or to the right, left and right brackets when there is a distinction to the left or the right but not both, and a rectangular box when there is a distinction to both the right and the left. In the next few paragraphs we describe this process further in terms of logical operations.\\

The fundamental underlying operation is ``exclusive or'', often denotes by $XOR.$ When we say ``A XOR B" we mean the statement
``A or B but not both A and B." This special version of OR has the property that it is true only when A and B have different truth values. Logically,
``A or B but not both A and B." is equivalent to ``A and not B, or B and not A.''. In this form we write the formula

$$A*B  = (A \wedge (\sim B)) \vee ((\sim A) \wedge B). $$ 
Here $A*B$ denotes ``A XOR B", $A \vee B$ denotes ``A or B'' and $A \wedge B$ denotes `` A and B".\\

When working with sets we can interpret $A*B$ as the intersection of $A$ with the complement of $B$  taken in union with the intersection of the complement of $A$ with $B.$ This is illustrated in 
Figure~\ref{xor}. In using the Venn diagrams, we have a very intuitive interpretation of XOR. A set is denoted by a shaded circle and when we XOR two sets, the part where they overlap vanishes.
Thus two identical sets will yield and empty diagram under this operation.  In this sense a set is its own negation! We will return to this point of view in Section 10 when we discuss the relationship of RD with Spencer-Brown's Laws of Form. In letting one shaded region operate upon another the parts that remain black after the XOR operation indicate the difference between the two sets. In this way, XOR is a logical exemplar of the operation of discrimination and it can be understood to underly all the RD operations that we describe. One can imagine that discrimination (as practiced by thinking beings) is more complex that XOR, but XOR is a backbone or skeletal aspect of all instances of discrimination.\\

Now view Figure~\ref{xoricon} once more. Here we show explicitly how the XOR operation acts on the icons for the 1D RD to produce the icons at the next iteration. We use a vertical slash $|$ and an unmarked vertical slash for the two
states of discrimination. We call these the {\em marked} and {\em unmarked} states, respectively. Given two such states we define $A*B$ to be marked if one of $A$ and $B$ are different. If both $A$ equals $B$, then $A*B$ is unmarked. This construction is then applied to the local interactions of the icons in the RD. If we have a row with $ABC$ in that row, then for the new $B$, we form $A*B$ and $B*C$. These vertical slashes or unmarked slashes become the left and right ends of the new icon that represents the new $B$ in the next row of the RD, one full time-step later. Thus the new icon is formed by the discriminations to its left and to its right in regard to those neighbor icons. The figure shows explicitly how we leave the horizontal lines of the icon unchanged while we change the vertical slashes. As mentioned at the beginnining of this section, this means that the logic of left and right naturally creates the four icons that are used in the 1D RD. The alphabet arises in the act of discrimination. The act of discrimination is quite general for the RD. Any letters or icons can be given to it at the start.
The XOR applies to make the discrimination and to produce a standard icon that indicates the left-right discrimination that was made.\\

Now view Figure~\ref{xorcircuit} where we indicate how the XOR process can be accomplished by digital circuitry. The figure  should be self-explanatory. There is a basic inverting element that will take states to their opposites and, with a multiplicity of inputs, this inverted is regarded as a NOR gate. That is, one starts with a collection of variables $\{a,b,c,d \}$ and the NOR gate returns $\sim (a \vee b \vee c \vee d).$ The circuit then implements the formula for the XOR operation that we have given above. This means that we could have an RD automaton that sampled signals inside a larger digital environment. It also means that we can look at the RD
as connected inside an information-processing environment that uses logical operations in great generality. In particular one could think of a sensing device that can detect differences in signals with which it otherwise has no direct access. Such external but not directly detectable signals have been called {\em fantomarks} by Isaacson \cite{JI1}. The information about their differences can become the initial data for an RD system that then amplifies and modifies these patterns allowing the possibility for communication (by letting another system find differences in the signals generated by this RD) between systems that have internal states that are fantomarked for the other system. Isaacson has speculated that this could be the basis for communicating with extraterrestrials. Here we point out that it can be regarded as a partial description of the situation of human-to-human communication with its mix of local-to-global discrimination based on the detection and articulation of differences.\\   

We regard this description of the process of discrimination to be fundamental. A ground that is subject to discrimination is given at the beginning. The XOR operations probe this ground and write naturally via marked and unmarked states in the geometry and alphabet of special icons that can be further discriminated by the same process. The icons record a neighborhood of discriminations. In the case of 1D RD this neighborhood is described in terms of  left and right. The process of discrimination alternates between the local indications of marked and unmarked states (the vertical slash and its absence) and the global examination of icons for their identity or difference. It is this crossing of levels that makes the structure of the RD process repeatable and unique.\\ 

In general an RD structure has alphabetic elements at specific loci. A process of discrimination generates an icon for that location that describes the distinctions between that letter and its neighbors. These icons of distinction become the letters of a special alphabet that is coherent with the geometry of the RD structure. The recursion replacing present icons or alphabetic elements with these icons of distinction is the process of 
recursive distinctioning. The process arises directly from the idea of description and the fundamental distinction of the given geometry. In the next section we show how this works for two dimensional RD.\\

\section {Two dimensional RD, a 16 letter alphabet, Quaternions and Spactime} 
\begin{figure}
     \begin{center}
     \begin{tabular}{c}
     \includegraphics[width=8cm]{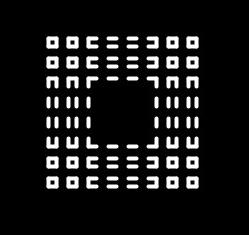}
     \end{tabular}
     \caption{\bf A Snapshot of a 2DRD}
     \label{2drd}
\end{center}
\end{figure}

\begin{figure}
     \begin{center}
     \begin{tabular}{c}
     \includegraphics[width=6cm]{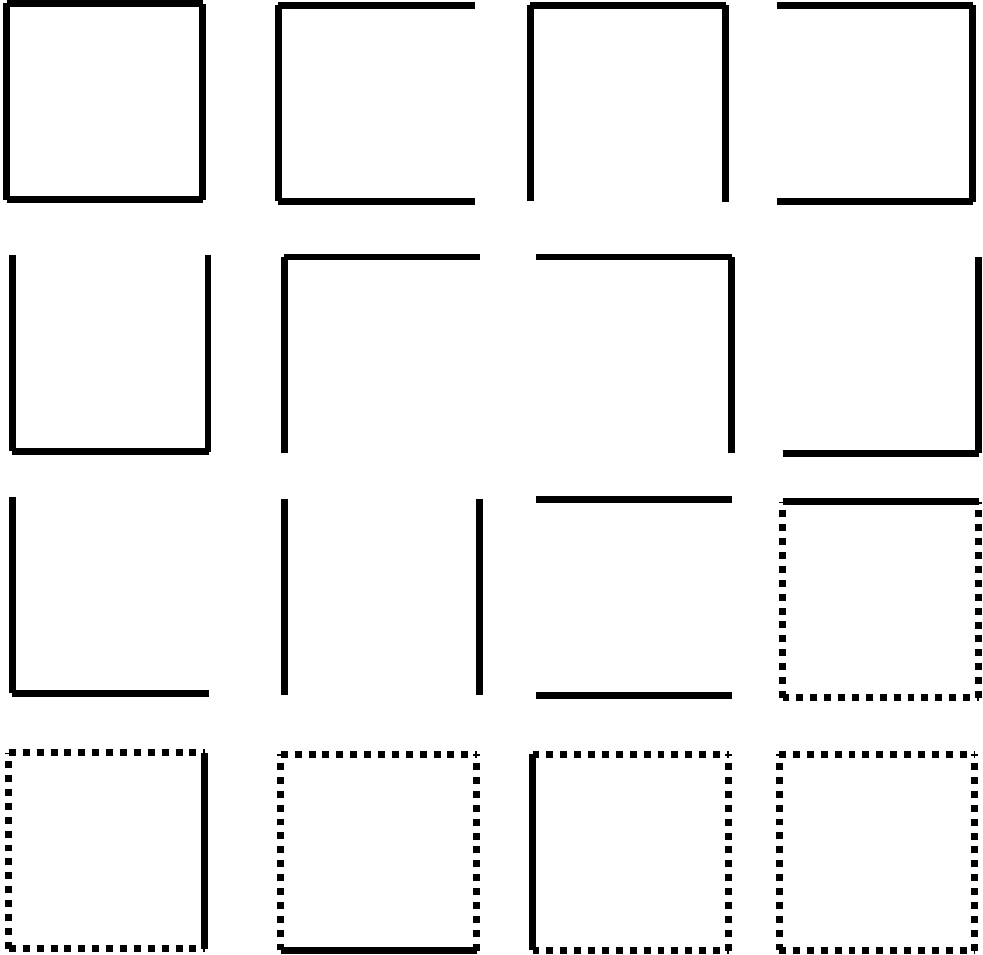}
     \end{tabular}
     \caption{\bf The 2D Alphabet1}
     \label{boxalpahbet}
\end{center}
\end{figure}

\begin{figure}
     \begin{center}
     \begin{tabular}{c}
     \includegraphics[width=8cm]{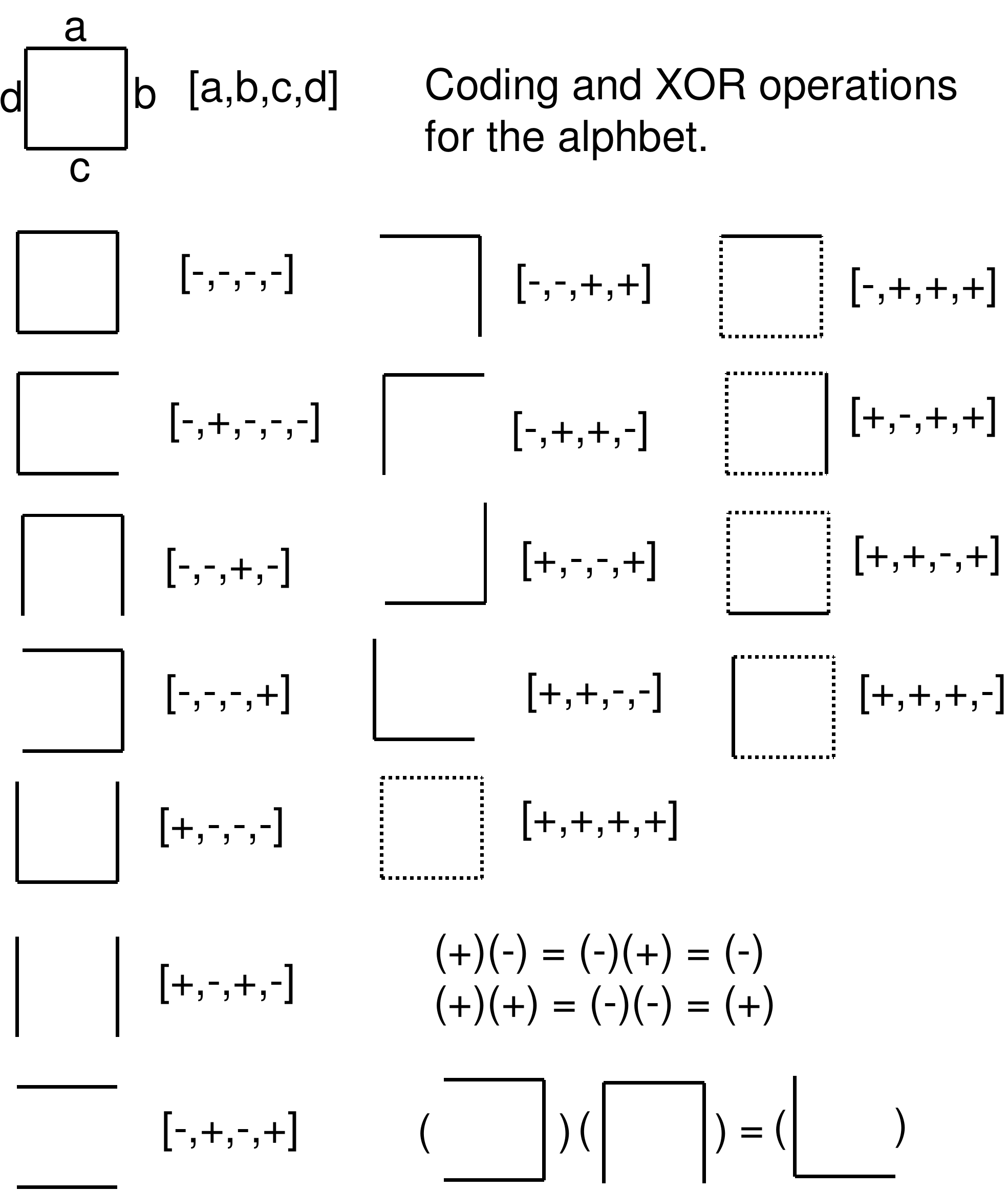}
     \end{tabular}
     \caption{\bf The 2D Alphabet2}
     \label{boxalphabetcodes}
\end{center}
\end{figure}

We now consider a natural generalization of the one dimensional RD to two dimensions. The geometry of the 2D RD will be a rectangular lattice with square cells. Each cell is regarded as having four neighbors, one to the 
north, one to the south, one to the east and one to the west, each sharing a one-dimensional interval of common boundary. The simplest occupant of such a cell will correspond to openings or closings of the four parts of the 
boundary. Thus one can block all of the boundary, or all but one edge of the boundary or all but two edges of the boundary and continue until one has the unique empty icon with no edges from the boundary. This makes a 16 letter alphabet, as illustrated in Figure~\ref{2drd} and Figure~\ref{boxalpahbet}. In Figure~\ref{boxalphabetcodes} we indicate how to code the letters as ordered sequences of four elements, each element a plus or a minus sign. In this figure
we also indicate how to make XOR combinations of these edges of the icons. The rule is simply that the superposition of two edges cancels them. With this, we can combine the letters to form other letters by superimposing
them. When two letters are identical, then there superpostion is the empty letter. Otherwise it is not empty and is a new resultant letter. Thus we see that this superposition of letters serves to distinguish one letter from another. Two letters are distinct if and only if their superposition is empty.\\

In the sequence from Figure~\ref{seq0} to Figure~\ref{seq8} we show eight steps from the first figure and returning to that figure. The first figure is an empty box with a fixed boundary condition that declares that its outer squares are different from the adjacent squares outside the box. Each successive figure is the result of one redescription by the RD process. In this case and with this initial condition the process has period eight.\\

In the sequence from Figure~\ref{seq0seed} to Figure~\ref{seq12seed} we show the same box with a different initial condition (some marked spaces inside). Now the evolution is more complex as is illustrated in the figures.
Remarkably, in this case the result is eventually periodic of period two. Figure~\ref{tdum} and Figure~\ref{tdee} illustrate two consecutive frames from this automaton after it has entered period two. The reader can compare these two frames and see that each describes the other. Focus on the pair of 2D patterns Tweedledum and Tweedledee in these two figures. 
What is remarkable about these two patterns is that they
mutually describe each other in such a way that they complement
each other, just like a positive and a negative in photography.
If separated, each would construct its complement, and the patterns would replicate indefinitely.
So these are ``antithetical'' and their superposition will yield
a ``synthesis.'' (A synthesis here would be the big square filled
completely with only little squares).  Note that they are typical in many  2DRD runs and are not exceptions.\\

The two strands of DNA are also complementary,
which allows their replication.  The reader will recognize how much more
complex this 2D complementarity is compared to the
1D complementarity of DNA.   Obviously, no one can dream
or design such intricate mutual description of patterns and
yet it is a by-product of an automatic RD automaton. 
One might speculate that the DNA molecule with its complementary Watson and Crick strands evolved through recursive chemical interactions.
 \\

\subsection{Quaternions and Iterants}
In this subsection we concentrate on the Figure~\ref{boxalphbetquaternions} and show how the 16 letter alphabet is related to the algebra of the quaternions and  concommitantly to the algebra of spacetime.
Before we do this however, it will be helpful to explain a way to think about such matters that is developed in the paper  \cite{LK10} by Kauffman (and references therein). In that paper one finds a temporal interpretation of 
the square root of minus one. The idea is that one starts with a simple oscillation such as $$\cdots +-+-+-+-+- \cdots.$$ In starting in this way, we can connect with RD simply by observing that some of the simplest 1DRD
with tight boundary conditions will oscillate in period two. Once recursion is on the scene the simplest oscillations are inevitably present. That said, let us make two abbreviations that correspond to two ways to distinguish a period two oscillation: $$[+,-]=[+1,-1]$$ and  $$[-,+]=[-1,+1].$$ These two ordered pairs correspond to distinguishing the oscillation as proceediing from plus to minus or as proceeding from minus to plus.\\

 Call an ordered pair such as $[a,b]$ an {\it iterant}. We can combine iterants by adding their coordinates or by multiplying their coordinates.
 $$[a,b]+[c,d] = [a+c,b+d],$$
 $$[a,b][c,d]= [ac,bd].$$
 We add to this structure an operator $\eta$ that participates in the {\it time shift} that relates one iterant to the other.
 $$\eta^{2} = 1,$$
 $$[a,b]\eta = \eta[b,a].$$
 Formally, $\eta$ acts as a permutation of order two, exchanging $[a,b]$ to $[b,a]$ when it is commuted with an iterant.
 We regard an element of the form $[a,b]\eta$ as a {\it temporally sensitive iterant}. Note what happens when we multiply 
 $$i = [+1,-1]\eta$$ by itself.
 $$i^2 = ii = [+1,-1]\eta[+1,-1]\eta = [+1,-1][-1,+1]\eta\eta = [(+1)(-1), (-1)(+1)]1 = [-1,-1] = -1.$$
 Thus $$i^2 = -1.$$
 We have produced a square root on minus one as a temporally sensitive iterant associated with an elementary oscillation.\\
 
 In fact, we have produced an algebra containing $$\{ \eta, 1=[1,1], -1=[-1,-1], \alpha=[1,-1], -\alpha = [-1,1]\}.$$ Note that $$\eta^2 = \alpha^2 = 1$$ and that $$\alpha\eta + \eta \alpha = 0.$$
 This is a first example of a {\it Clifford algebra}, an algebra generated by elements of square one that anticommute with one another. We have
 $i = \alpha\eta$ and  $$i^2 = \alpha\eta \alpha\eta = \alpha(-\alpha)\eta^2 = -\alpha^2 = -1.$$ Thus we can also see our temporal interpretation of the square root of minus one as a Clifford algebra phenomenon.\\
 
 Clifford algebras are deeply connected with physics. To see a hint of this we consider a fundamental formula from special relativity theory (we use the convention that the speed of light is $c=1.$).
 Let $E$ denote  energy, $p$ momentum and $m$ the mass of a particle. Now let $$E = \alpha p + \eta m.$$ assume that $p$ and $m$ commute with $\alpha$ and $\beta.$ You can easily prove by multiplying it out that 
 $$E^2 = ( \alpha p + \eta m) ( \alpha p + \eta m) = \alpha^2 p^2 + \eta^2 m^2 + (\alpha \eta + \eta \alpha) pm = p^2 + m^2 + 0 pm = p^2 + m^2.$$
 This formula $E^2 = p^2 + m^2$ is fundamental to special relativity and we have shown that it follows from a Clifford algebra representation of the energy. This way of writing the energy is due to the great physicist Dirac and is the beginning of the deep relationship of Clifford algebra with physics. Our point is that by looking at this through the lens of iterants, we can draw the connection of fundamental recursion with quantum and relativistic
 physics. The reader will find more about this in \cite{LK10} aand in \cite{Bernd}.\\
 
Now we turn to the {\it quaternions.} Sir William Rowan Hamilton discovered the quaternion algebra in 1843, after 15 years of trying to find a three dimensional analog for the complex numbers. When he realized the key was a four dimensional space, the pattern fell into place. Recall that the quaternions are generated by $\{1,-1,I,J,K\}$ so that $I^2 = J^2 = K^2 = IJK = -1$ from which it follows that $IJ=K, JK=I, KI=J$ and $IJ=-JI, JK=-KJ, KI=-IK.$
There is a natural iterant structure for the quaternions. See Figure~\ref{iterantquat}. In this figure we show the order four iterant sequences that correspond to each of $I$, $J$ and $K$ and the analogy of the simple time shifter $\eta$ that is associated with each one. These analogs are diagrammed as permuations and they act when one composes the iterants by attaching their braided forms together. The new temporal shift operators generate the so-called Klein Four Group, the symmetries of a square. See \cite{LK10} for more details. We now show how this iterant version of the quaternions is related to our 16 letter alphabet and how the symmetries of the square come into play directly.\\

Now we turn to Figure~\ref{boxalphbetquaternions} where we show how there is a natural quaternion structure associated with the sixteen letter alphabet.
What you see is a subset of the $16$ letter alphabet and the operations $A,B,C$ (and $1$) of the Klein Four Group.
We define $I, J, K$ each of the form $I = aA, J = bB, K = cC$ where $a,b,c$ are certain  elements of the $16$ letter alphabet.
We then define  e.g. $xA = A x^{A}$ where $x^{A}$ is the operation of the symmetry element $A$ on the letter $x.$
We define $xy$ (on letters) via $XOR$ of the corresponding letters in the alphabet.
We find that $I,J,K$ give the quaternions. 
Thus the quaternions are a combination of $XOR$ operations and symmetry operations in the alphabet.
Note that $xy = XOR(x,y) =$ result of superimposing $x$ and $y$ as letters and canceling common occurrences.
Once we have the quaternions we have an entry into spacetime algebra as follows.
We have $II = JJ = KK = IJK = -1.$
Let $E= (x,y,z,t) = xI + yJ + zK+ t1$ where $x,y,z,t$ are real numbers.
Then think of $E$ as a point in spacetime.
We have $$E^{2} = =(xI + yJ + zK+ t1)(xI + yJ + zK+ t1)=-x^{2} -y^{2} - z^{2} + t^{2}$$ is the Minkowski metric (it is often written as the negative of this expression) for spacetime.
Electromagnetism and much other physics can be written in quaternionic language.
One can start with a Clifford algebra with generators $e_{1}, e_{2}, e_{3},e_{4}$ with $(e_{i})^{2} = 1$ and distinct elements anti-commuting and construct spacetime algebra, the quaternions and more.
The iterant structure that we have hinted at here is part of a reformulation of the mathematics of matrix algebra that puts it into a temporal framework and a framwork that respects the ubiquitous appearance of the 
symmetries of permutation groups. It is likely that in another generation of the RD concept we shall include more about the role of symmetry.
In this way we have the beginnings of relationship of RD structure and fundamental frameworks for physical theory. For more about these relationships, see \cite{Bernd} and \cite{LK10}.\\

All this said, we have made only a superficial connection between the spacetime algebra of the quaternions and the actions or operations of the 2DRD.\\

The iterant process is in back of the quaternion multiplication where the symmetry group acts on the alphabetical letters. This  could become part of an extension of RD operations.
Then the RD would not just compare and describe. It would also interact with its own descriptions and change them by certain symmetry operations. This is one possibility for adding rules,
but we do not yet have a clear picture of what extra structure can be added  naturally to the very simple base at which we have started.\\

\begin{figure}
     \begin{center}
     \begin{tabular}{c}
     \includegraphics[width=8cm]{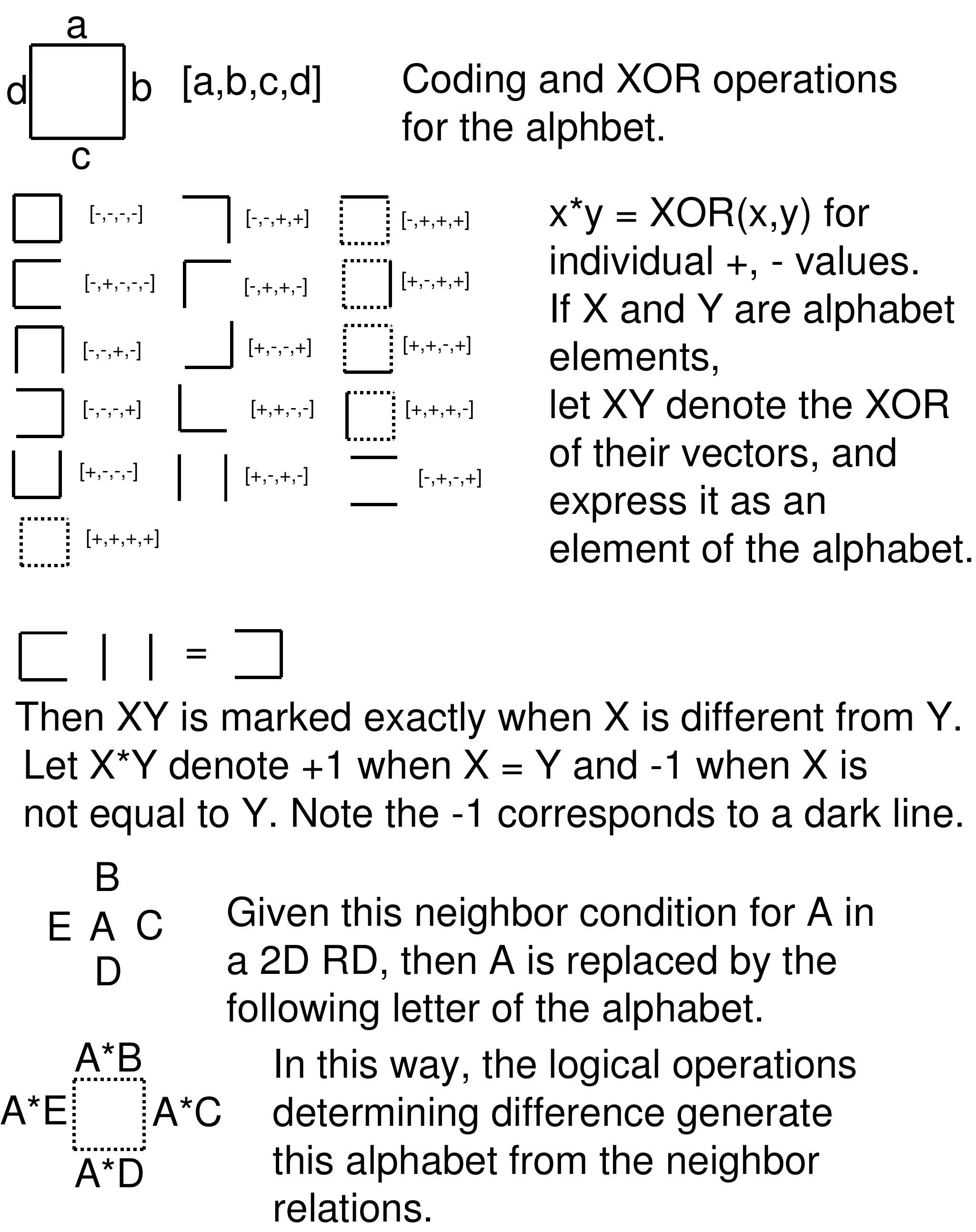}
     \end{tabular}
     \caption{\bf The 2D Alphabet3}
     \label{boxalphbetneighbors}
\end{center}
\end{figure}

\begin{figure}
     \begin{center}
     \begin{tabular}{c}
     \includegraphics[width=8cm]{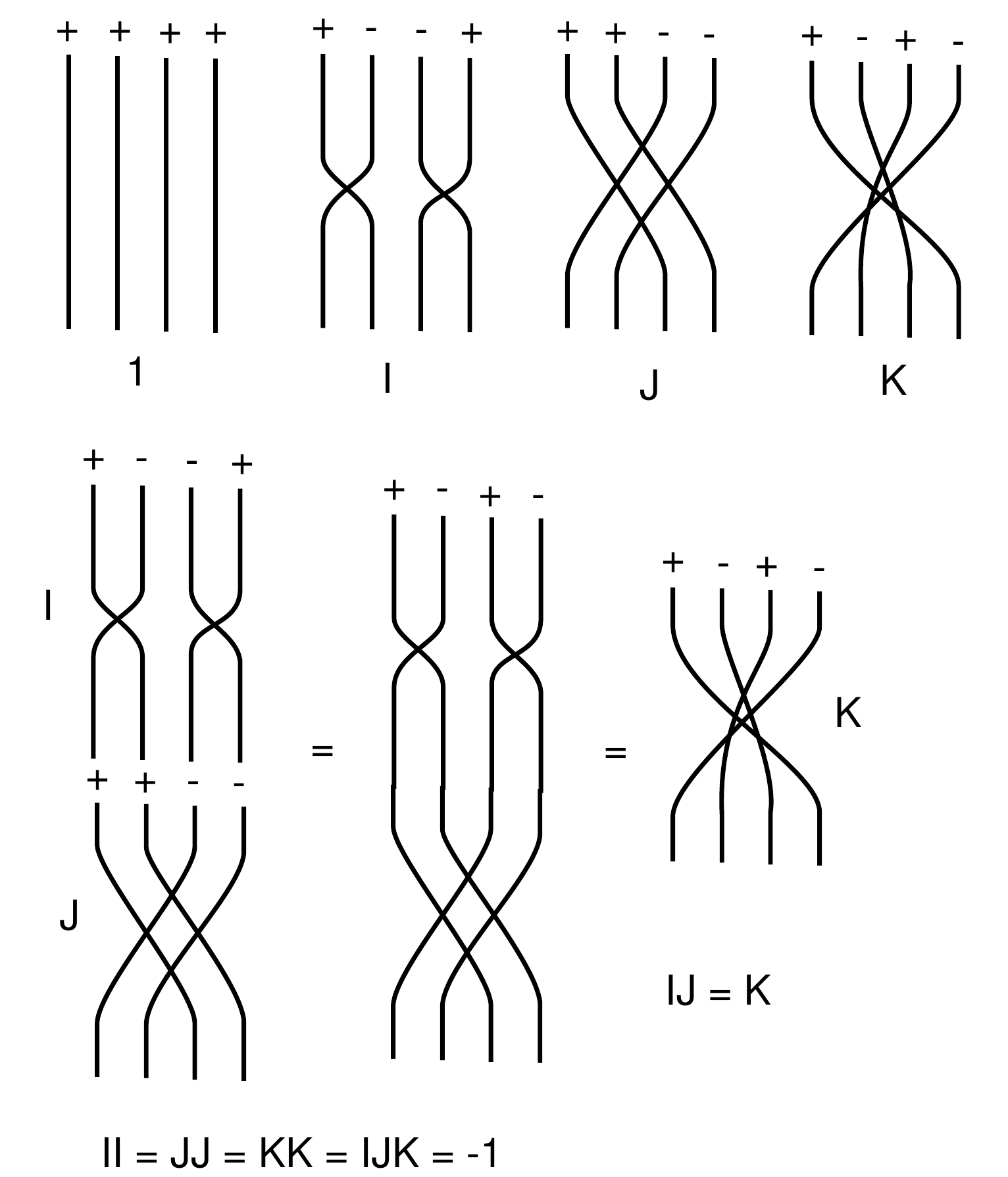}
     \end{tabular}
     \caption{\bf Iterant Representation of the Quaternions}
     \label{iterantquat}
\end{center}
\end{figure}

\begin{figure}
     \begin{center}
     \begin{tabular}{c}
     \includegraphics[width=8cm]{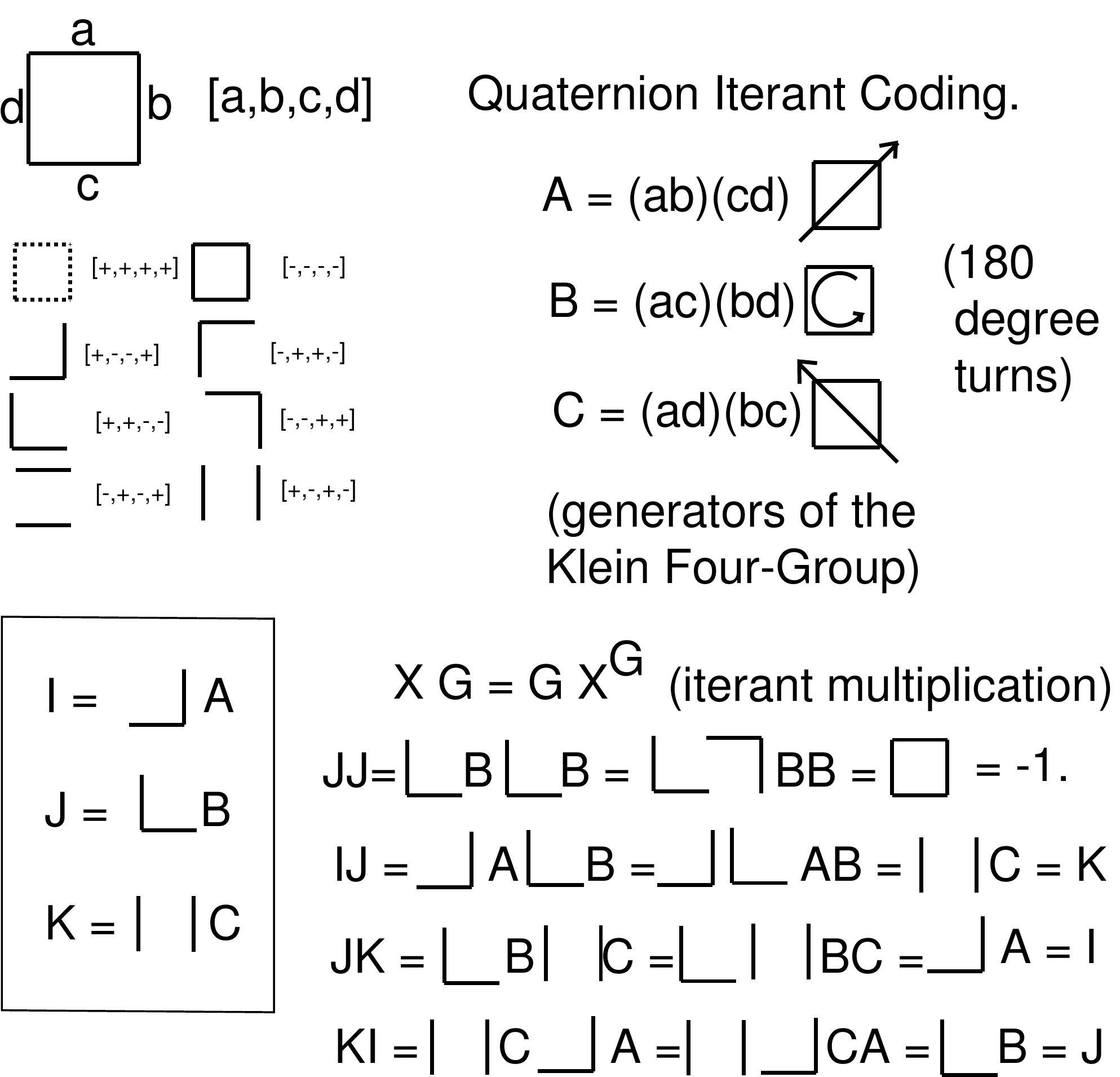}
     \end{tabular}
     \caption{\bf The 2D Alphabet4}
     \label{boxalphbetquaternions}
\end{center}
\end{figure}

\begin{figure}
     \begin{center}
     \begin{tabular}{c}
     \includegraphics[width=8cm]{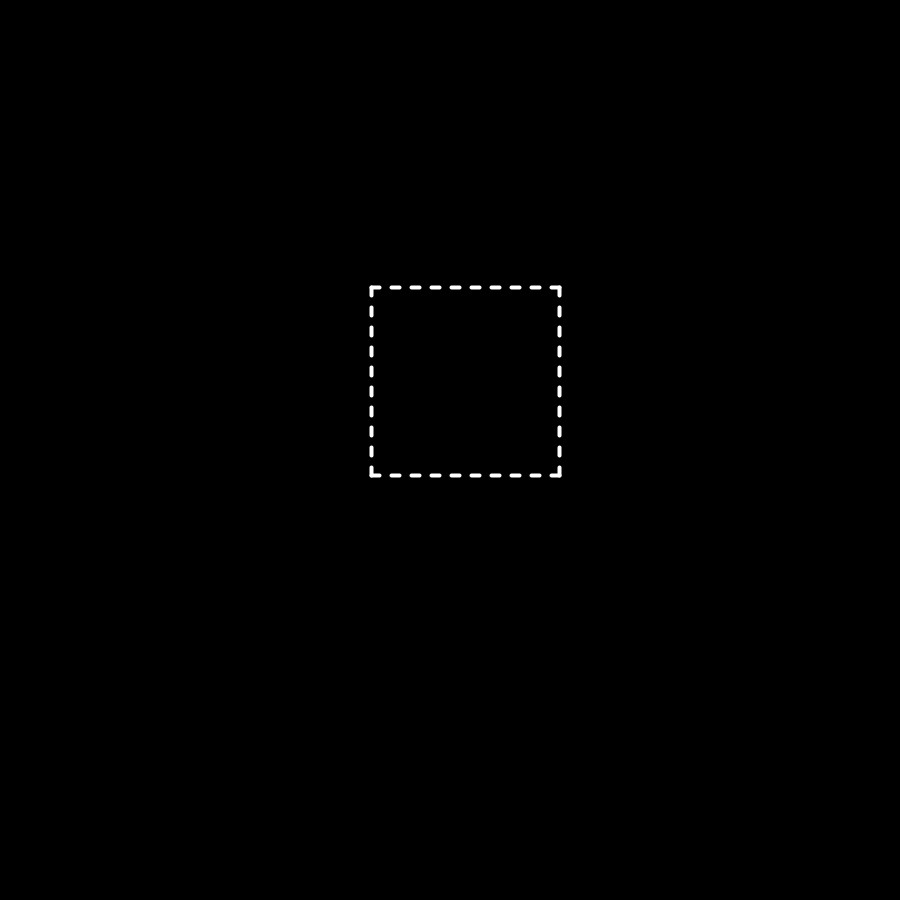}
     \end{tabular}
     \caption{\bf 2DRD  Box, No Seed}
     \label{seq0}
\end{center}
\end{figure}

\begin{figure}
     \begin{center}
     \begin{tabular}{c}
     \includegraphics[width=8cm]{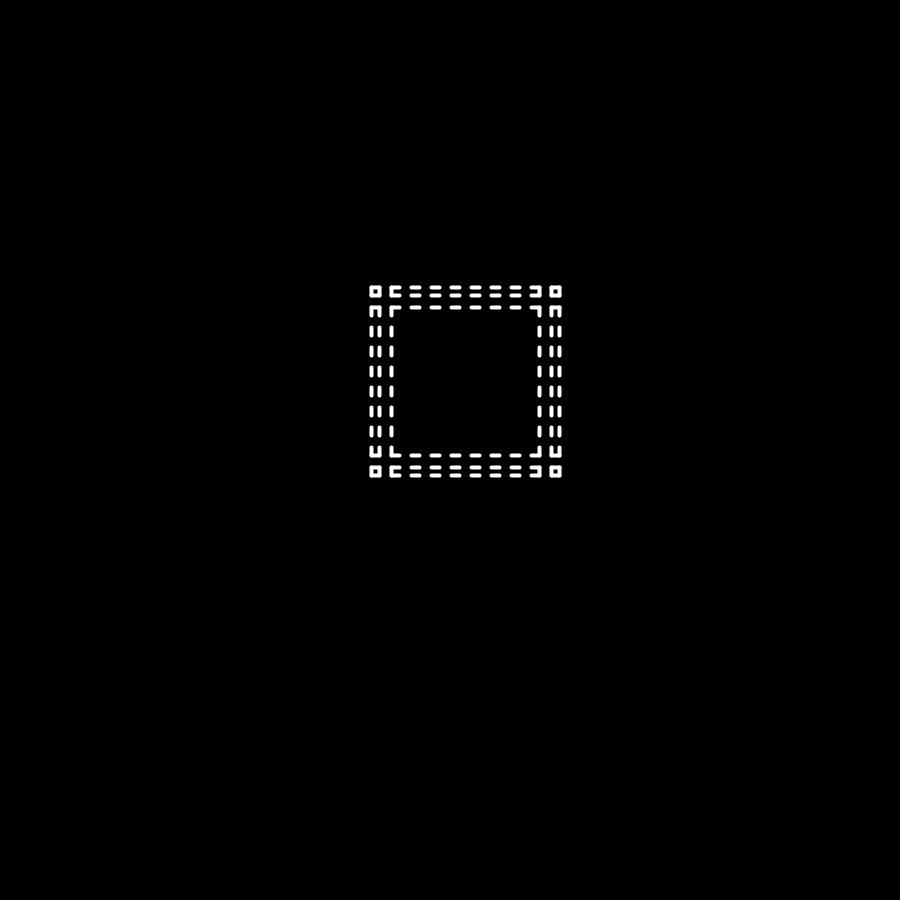}
     \end{tabular}
     \caption{\bf 2DRD  Box, No Seed}
     \label{seq1}
\end{center}
\end{figure}

\begin{figure}
     \begin{center}
     \begin{tabular}{c}
     \includegraphics[width=8cm]{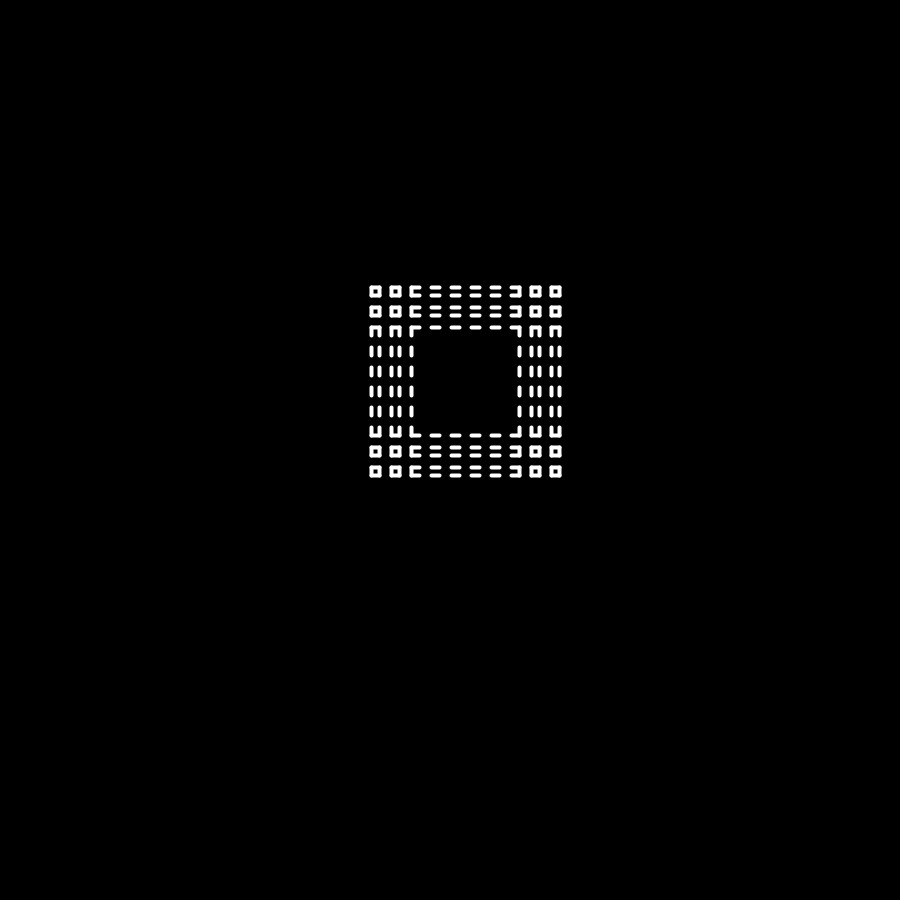}
     \end{tabular}
     \caption{\bf 2DRD  Box, No Seed}
     \label{seq2}
\end{center}
\end{figure}

\begin{figure}
     \begin{center}
     \begin{tabular}{c}
     \includegraphics[width=8cm]{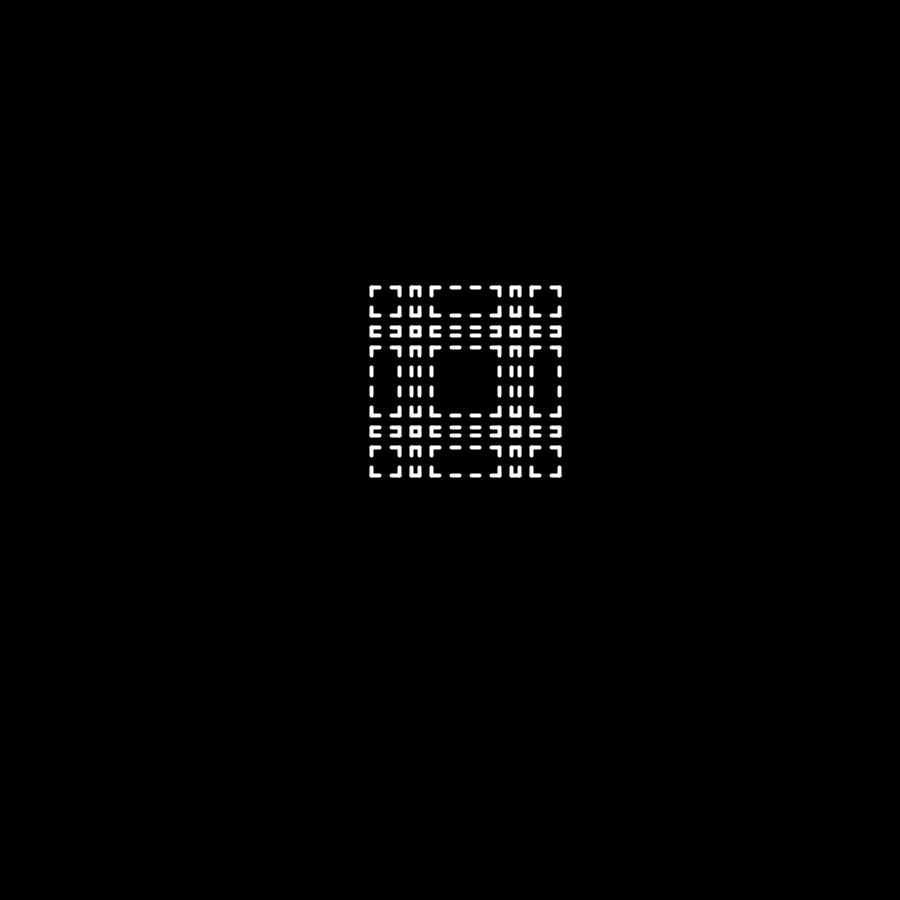}
     \end{tabular}
     \caption{\bf 2DRD  Box, No Seed}
     \label{seq3}
\end{center}
\end{figure}

\begin{figure}
     \begin{center}
     \begin{tabular}{c}
     \includegraphics[width=8cm]{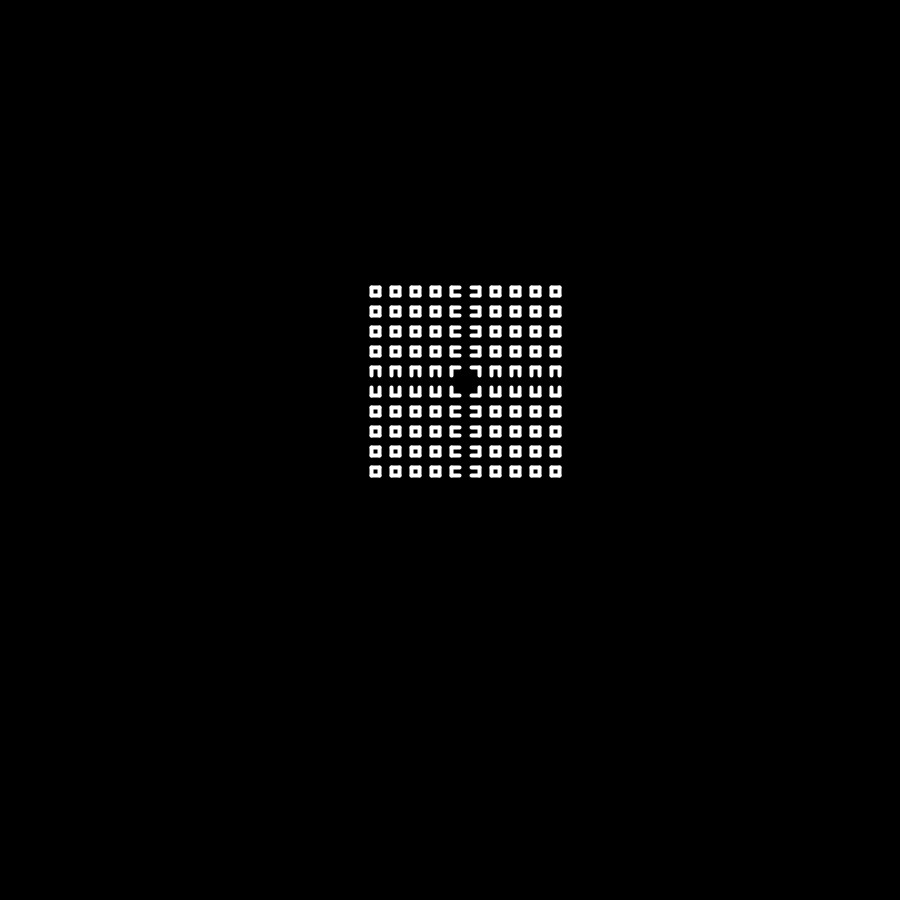}
     \end{tabular}
     \caption{\bf 2DRD  Box, No Seed}
     \label{seq4}
\end{center}
\end{figure}

\begin{figure}
     \begin{center}
     \begin{tabular}{c}
     \includegraphics[width=8cm]{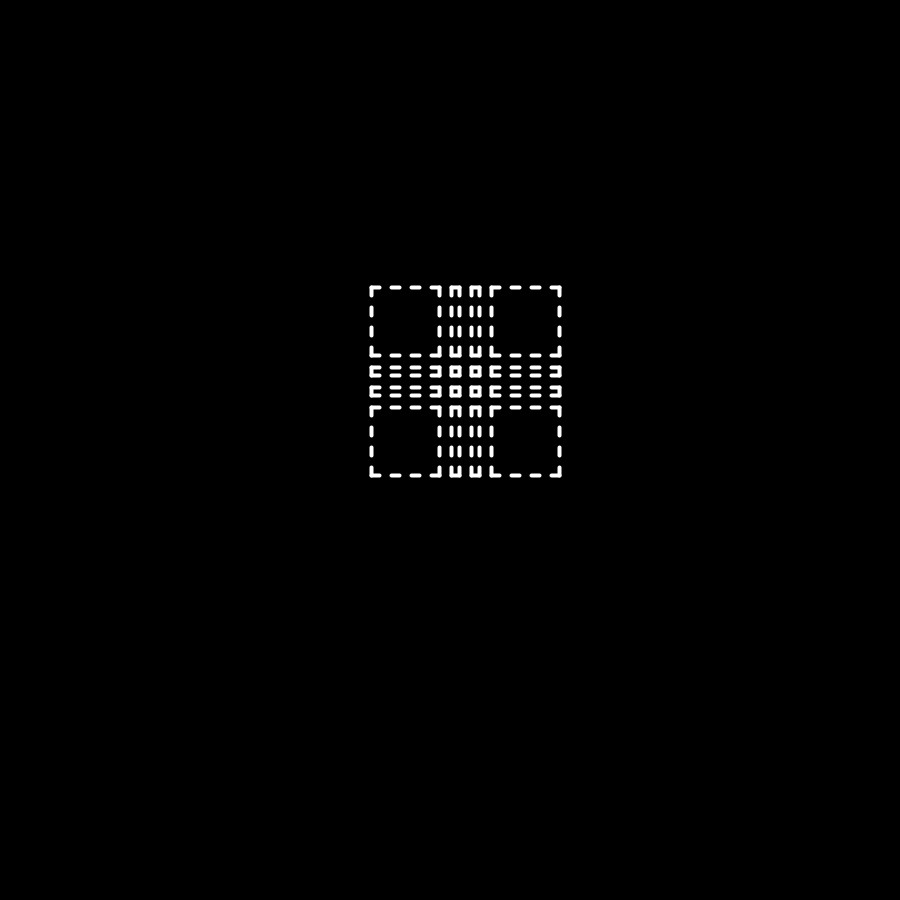}
     \end{tabular}
     \caption{\bf 2DRD  Box, No Seed}
     \label{seq5}
\end{center}
\end{figure}

\begin{figure}
     \begin{center}
     \begin{tabular}{c}
     \includegraphics[width=8cm]{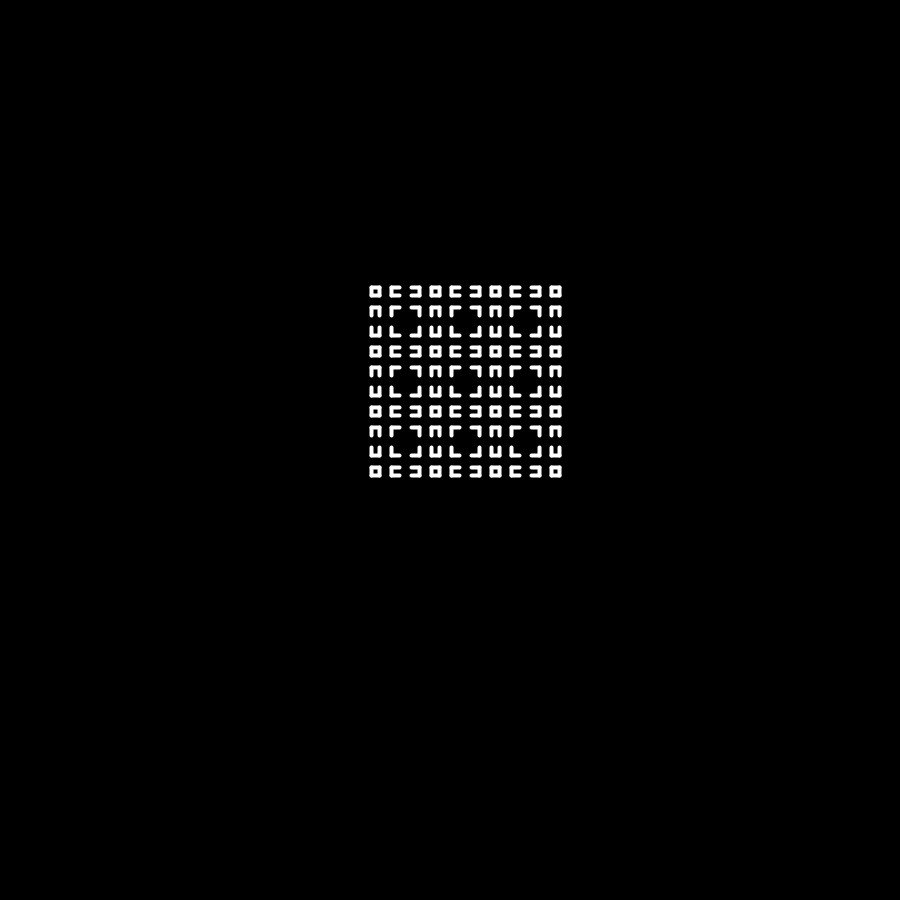}
     \end{tabular}
     \caption{\bf 2DRD  Box, No Seed}
     \label{seq6}
\end{center}
\end{figure}

\begin{figure}
     \begin{center}
     \begin{tabular}{c}
     \includegraphics[width=8cm]{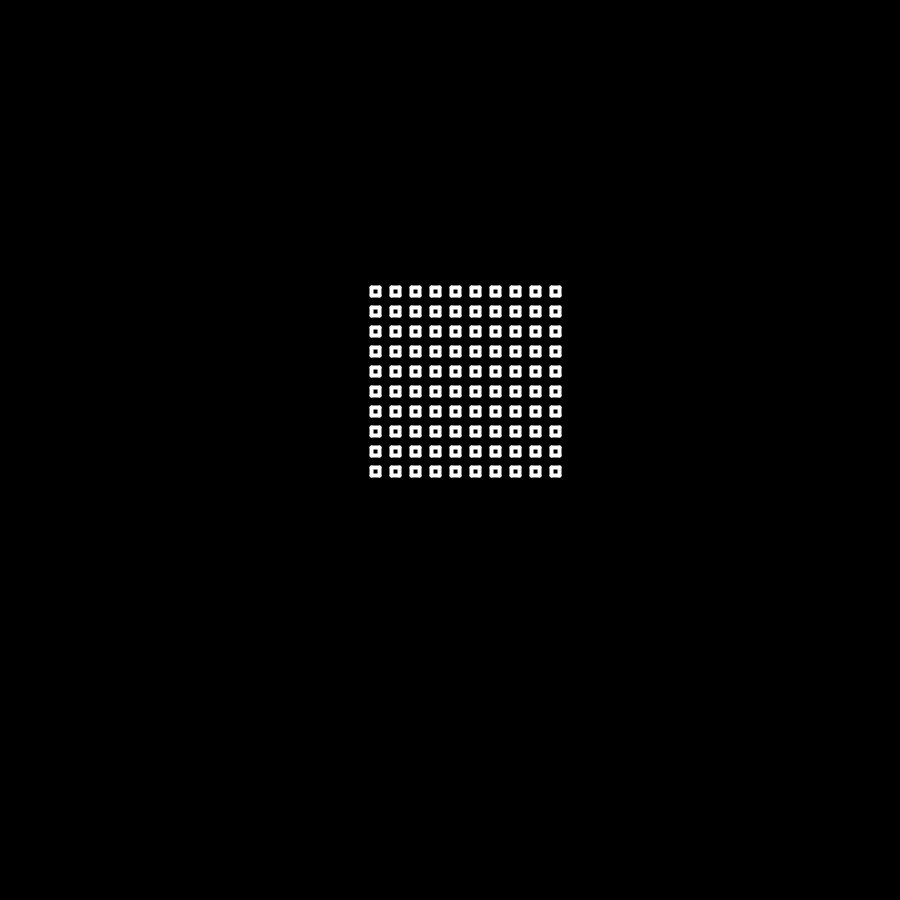}
     \end{tabular}
     \caption{\bf 2DRD  Box, No Seed}
     \label{seq7}
\end{center}
\end{figure}

\begin{figure}
     \begin{center}
     \begin{tabular}{c}
     \includegraphics[width=8cm]{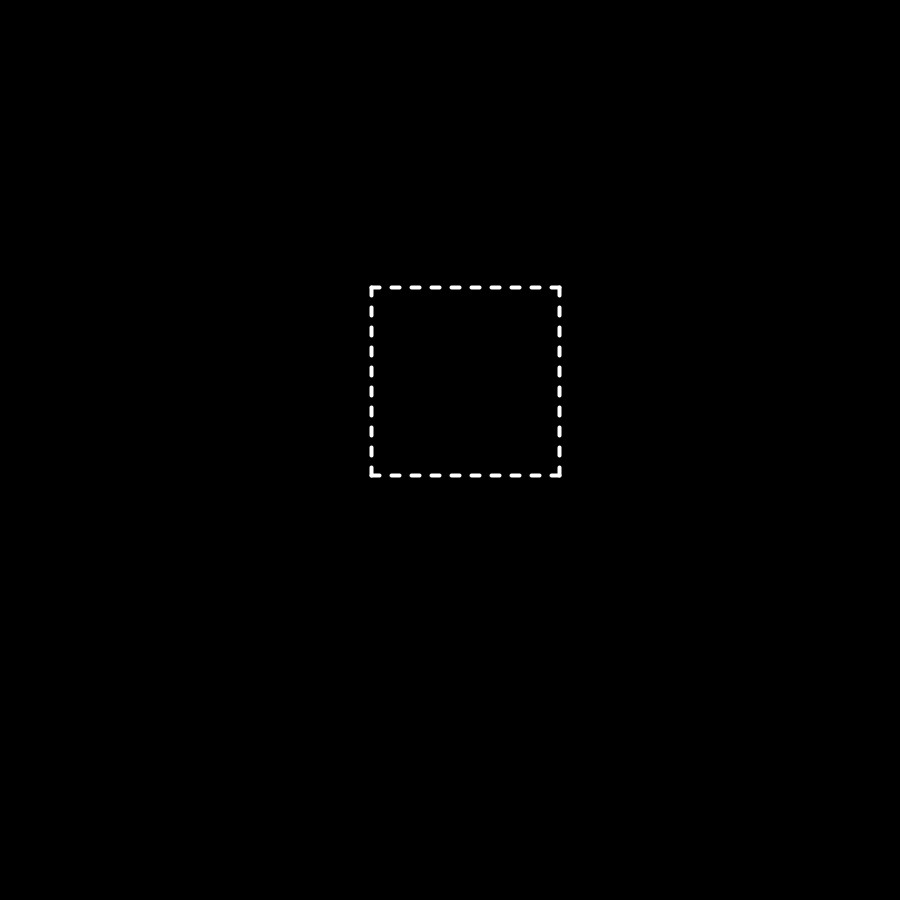}
     \end{tabular}
     \caption{\bf 2DRD  Box, No Seed}
     \label{seq8}
\end{center}
\end{figure}

\clearpage

\begin{figure}
     \begin{center}
     \begin{tabular}{c}
     \includegraphics[width=8cm]{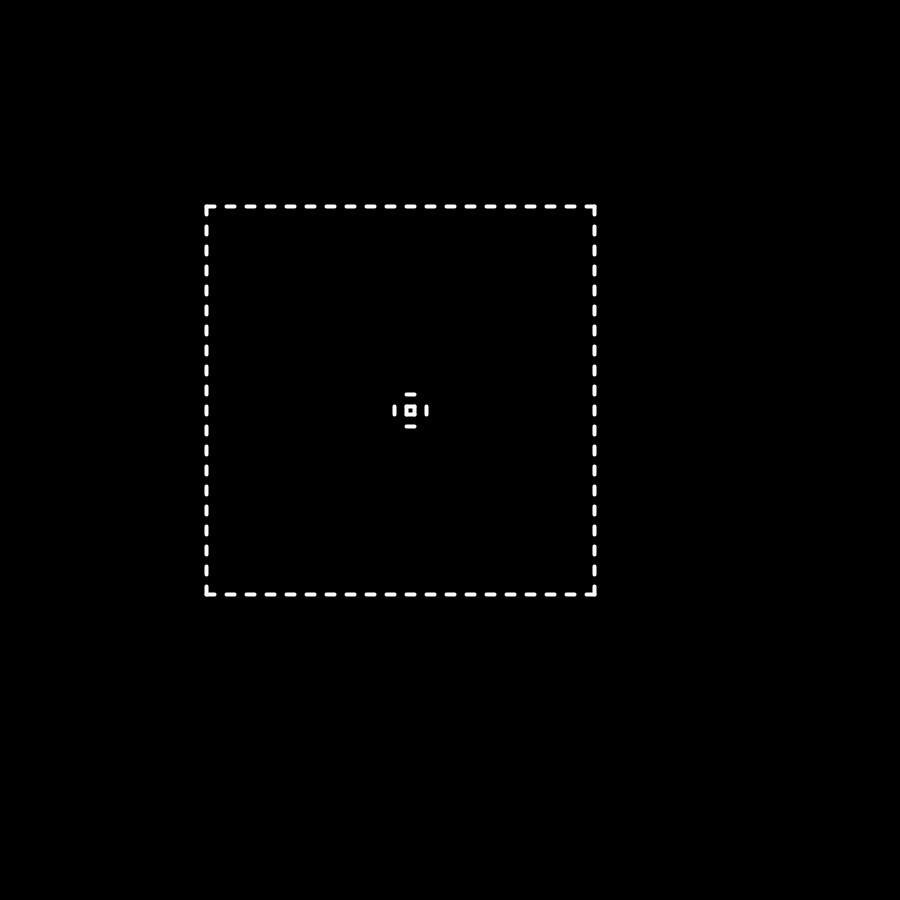}
     \end{tabular}
     \caption{\bf 2DRD  Box With Seed}
     \label{seq0seed}
\end{center}
\end{figure}

\begin{figure}
     \begin{center}
     \begin{tabular}{c}
     \includegraphics[width=8cm]{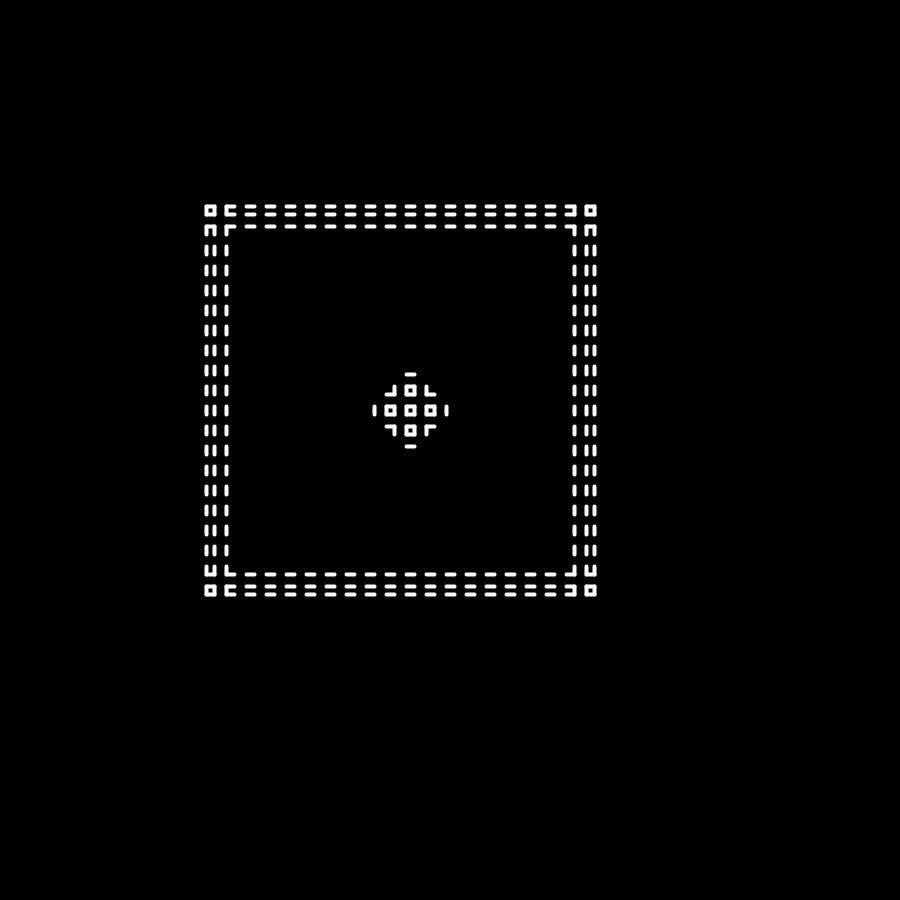}
     \end{tabular}
     \caption{\bf 2DRD  Box With Seed}
     \label{seq1seed}
\end{center}
\end{figure}

\begin{figure}
     \begin{center}
     \begin{tabular}{c}
     \includegraphics[width=8cm]{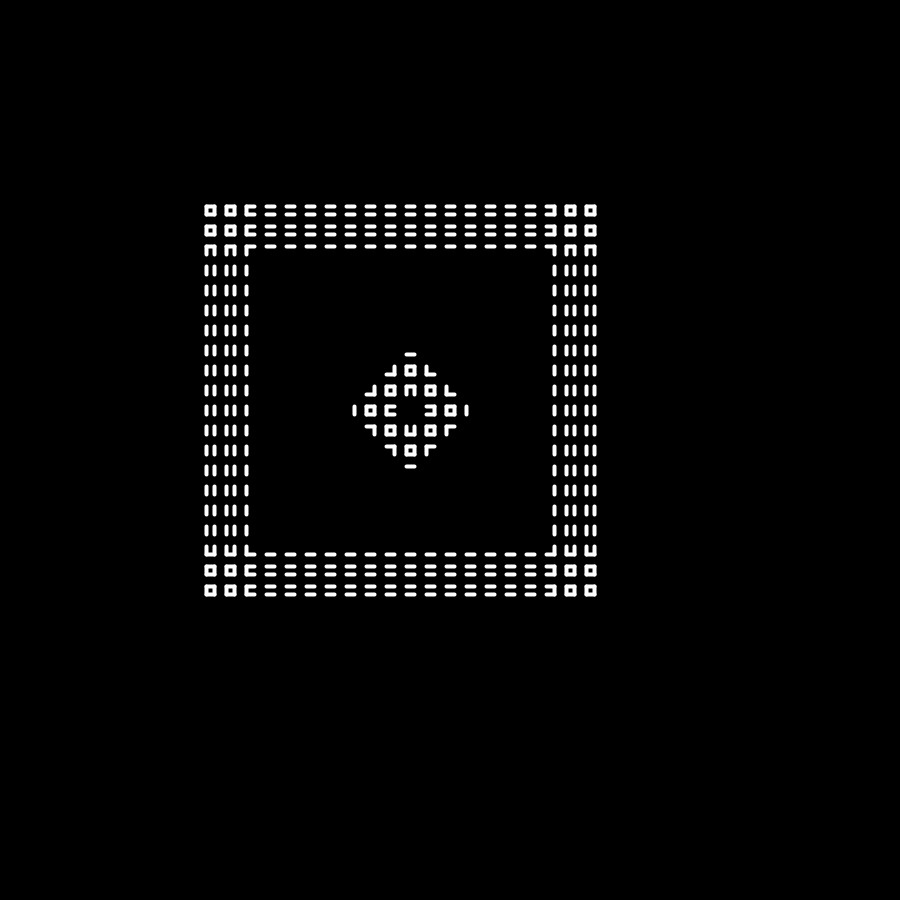}
     \end{tabular}
     \caption{\bf 2DRD  Box With Seed}
     \label{seq2seed}
\end{center}
\end{figure}

\begin{figure}
     \begin{center}
     \begin{tabular}{c}
     \includegraphics[width=8cm]{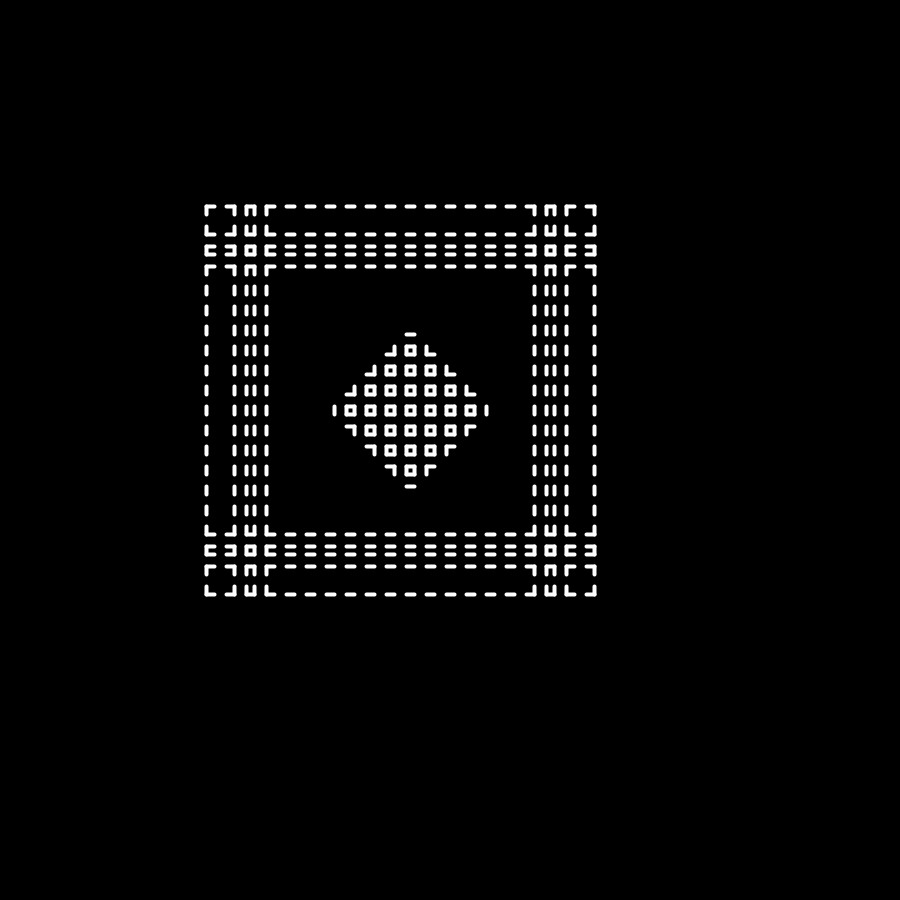}
     \end{tabular}
     \caption{\bf 2DRD  Box With Seed}
     \label{seq3seed}
\end{center}
\end{figure}

\begin{figure}
     \begin{center}
     \begin{tabular}{c}
     \includegraphics[width=8cm]{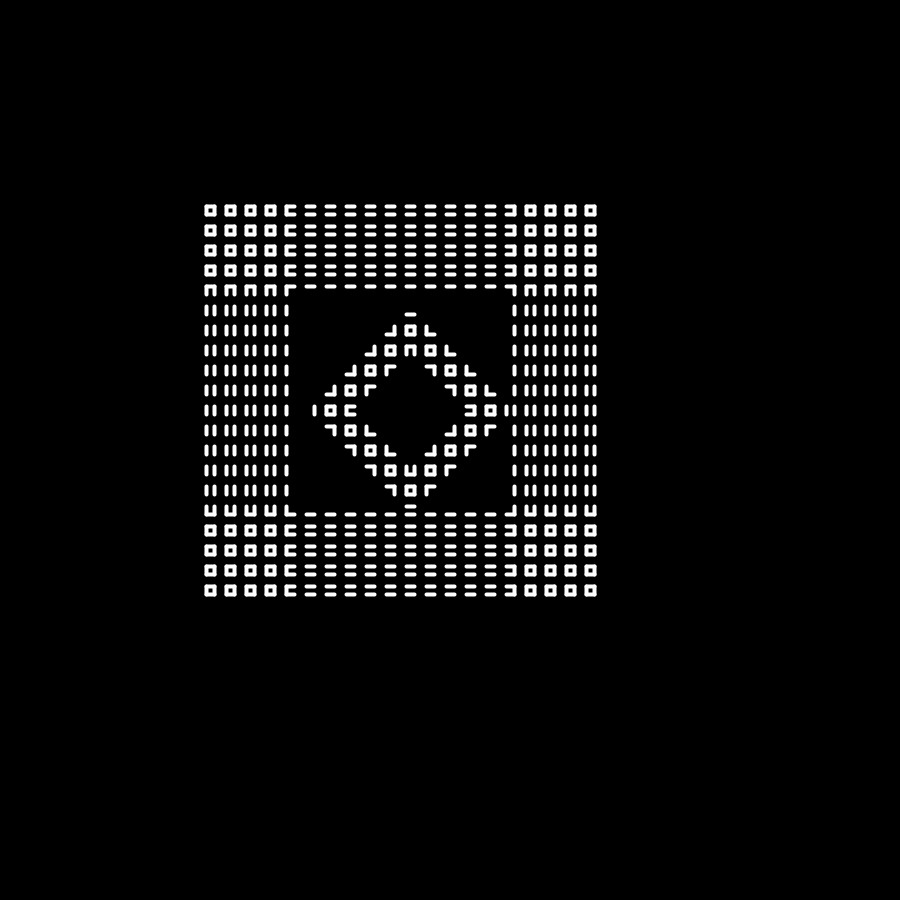}
     \end{tabular}
     \caption{\bf 2DRD  Box With Seed}
     \label{seq4seed}
\end{center}
\end{figure}

\begin{figure}
     \begin{center}
     \begin{tabular}{c}
     \includegraphics[width=8cm]{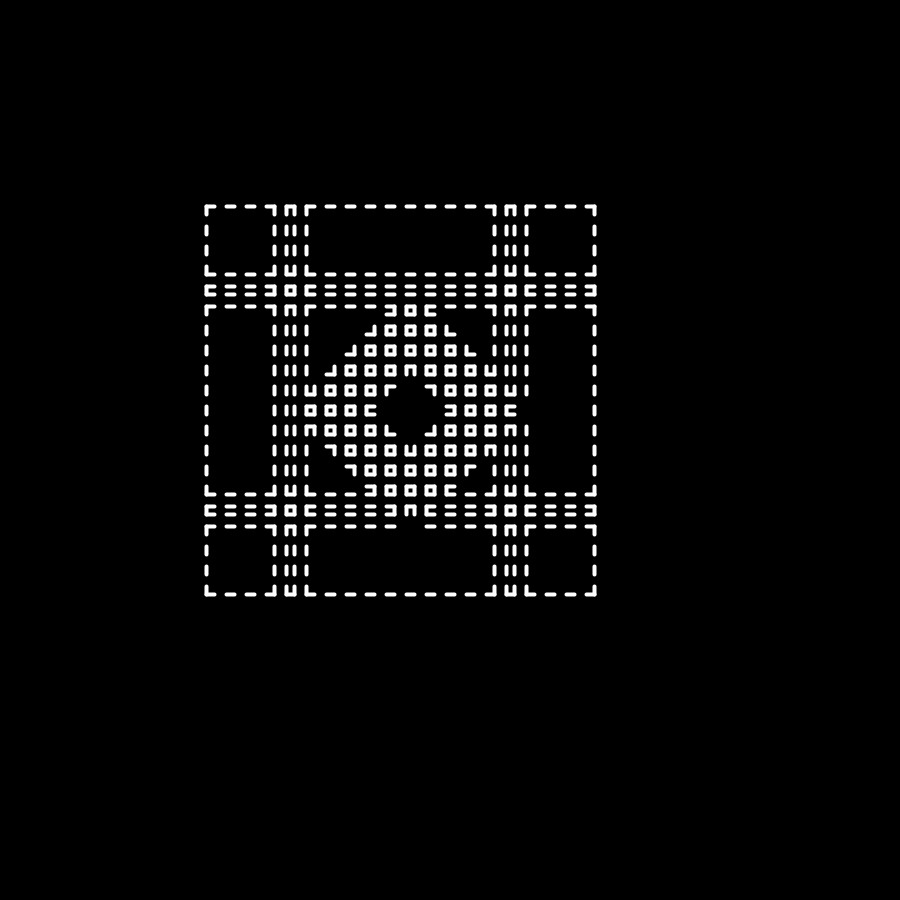}
     \end{tabular}
     \caption{\bf 2DRD  Box With Seed}
     \label{seq5seed}
\end{center}
\end{figure}

\begin{figure}
     \begin{center}
     \begin{tabular}{c}
     \includegraphics[width=8cm]{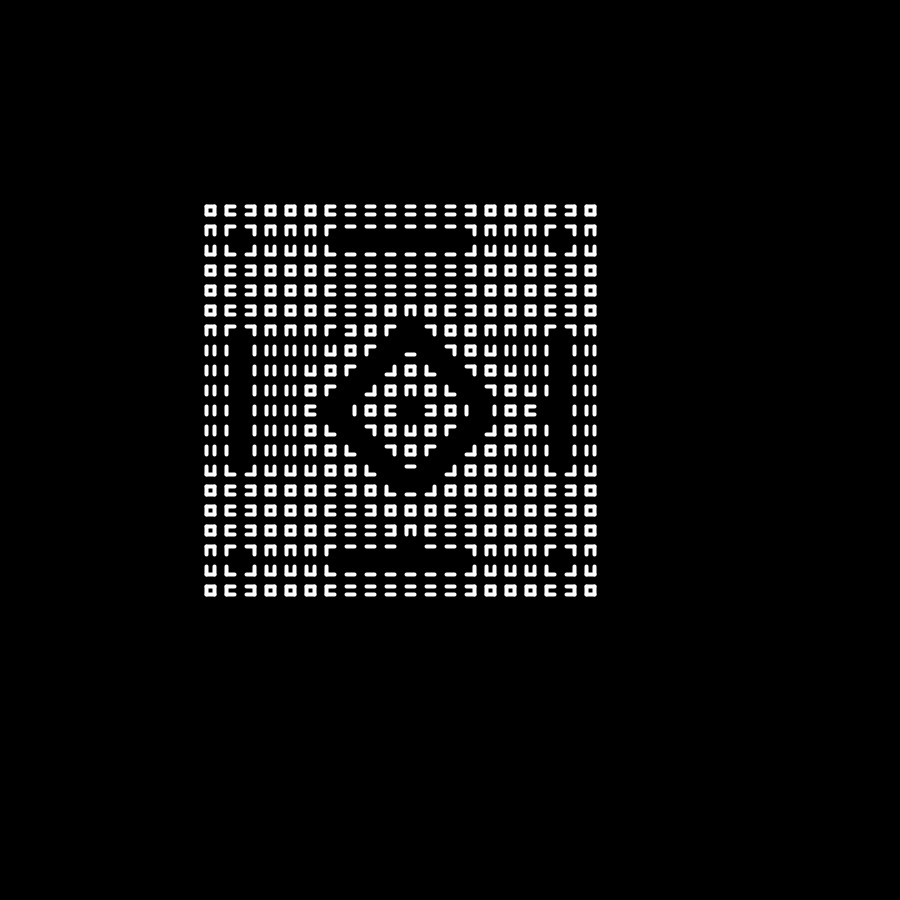}
     \end{tabular}
     \caption{\bf 2DRD  Box With Seed}
     \label{seq6seed}
\end{center}
\end{figure}

\begin{figure}
     \begin{center}
     \begin{tabular}{c}
     \includegraphics[width=8cm]{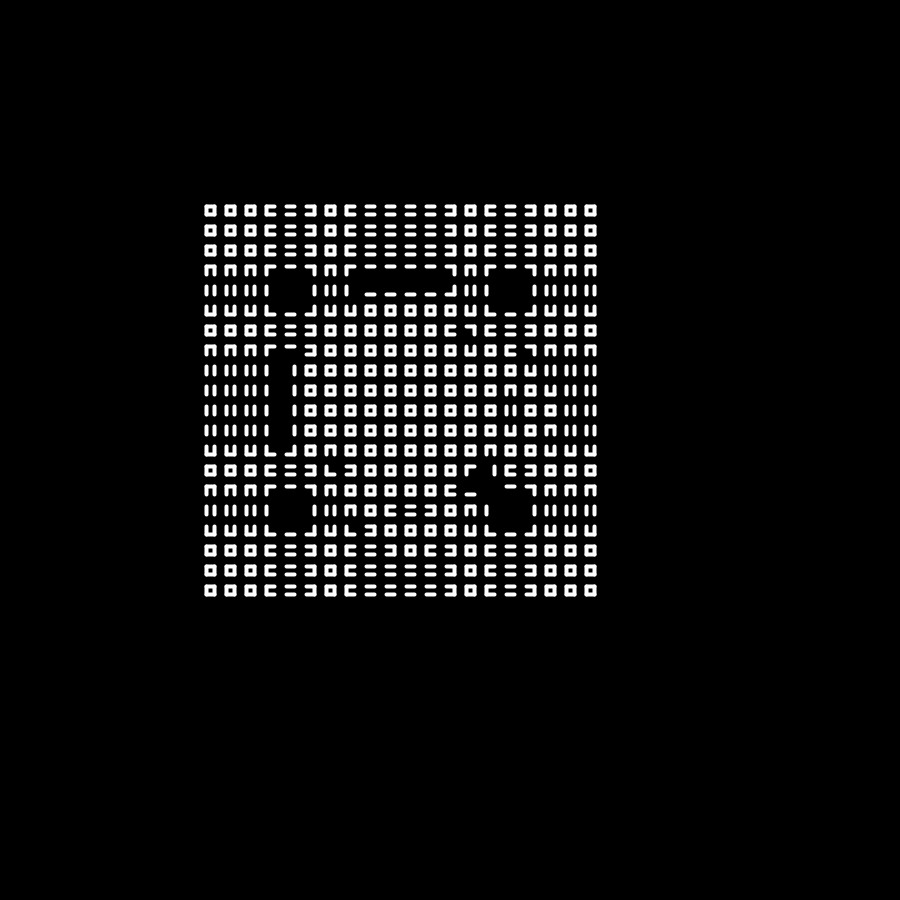}
     \end{tabular}
     \caption{\bf 2DRD  Box With Seed}
     \label{seq7seed}
\end{center}
\end{figure}

\begin{figure}
     \begin{center}
     \begin{tabular}{c}
     \includegraphics[width=8cm]{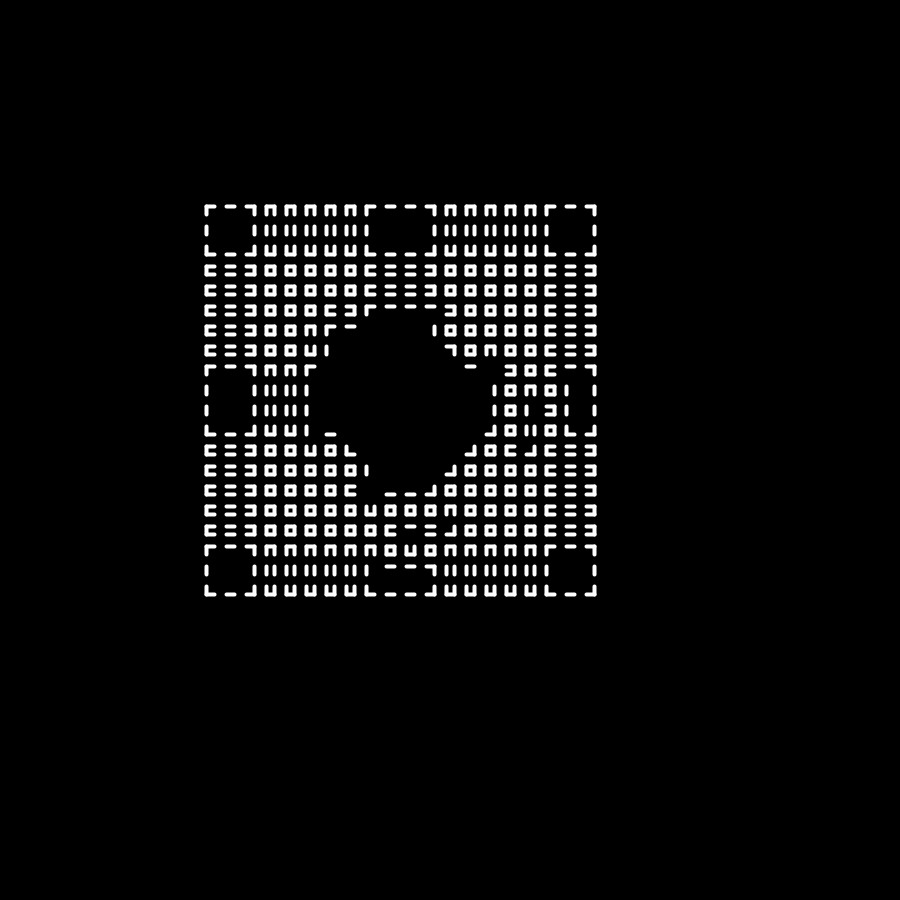}
     \end{tabular}
     \caption{\bf 2DRD  Box With Seed}
     \label{seq8seed}
\end{center}
\end{figure}

\begin{figure}
     \begin{center}
     \begin{tabular}{c}
     \includegraphics[width=8cm]{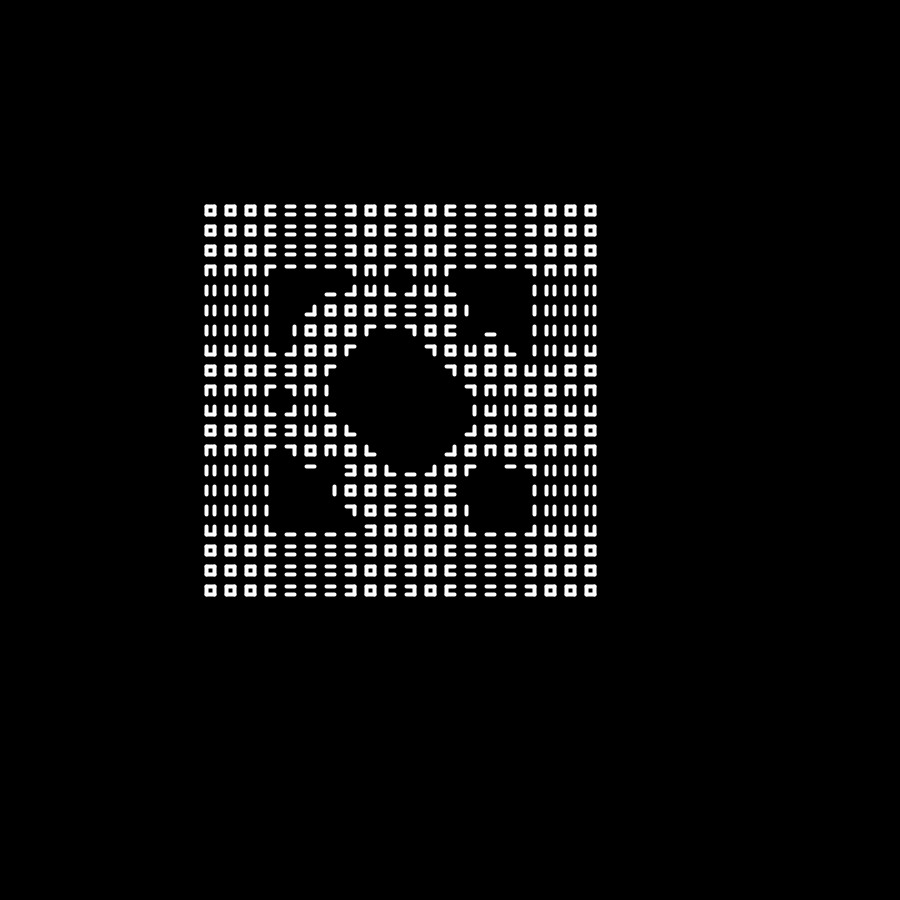}
     \end{tabular}
     \caption{\bf 2DRD  Box With Seed}
     \label{seq9seed}
\end{center}
\end{figure}

\begin{figure}
     \begin{center}
     \begin{tabular}{c}
     \includegraphics[width=8cm]{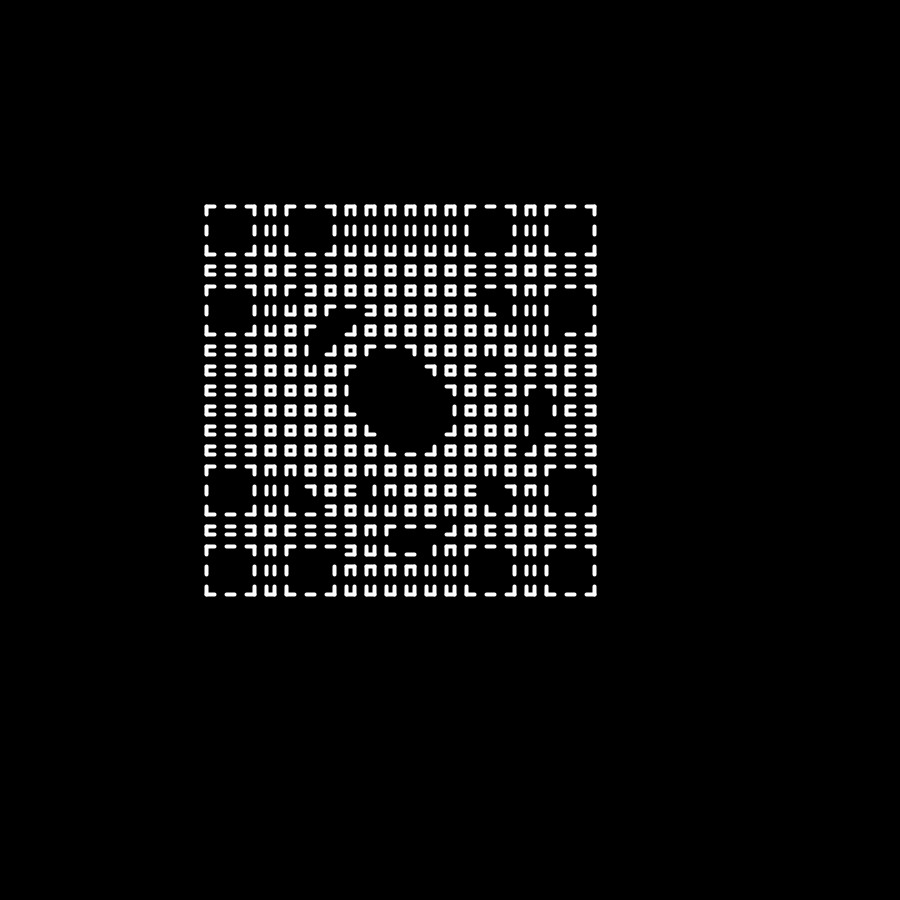}
     \end{tabular}
     \caption{\bf 2DRD  Box With Seed}
     \label{seq10seed}
\end{center}
\end{figure}

\begin{figure}
     \begin{center}
     \begin{tabular}{c}
     \includegraphics[width=8cm]{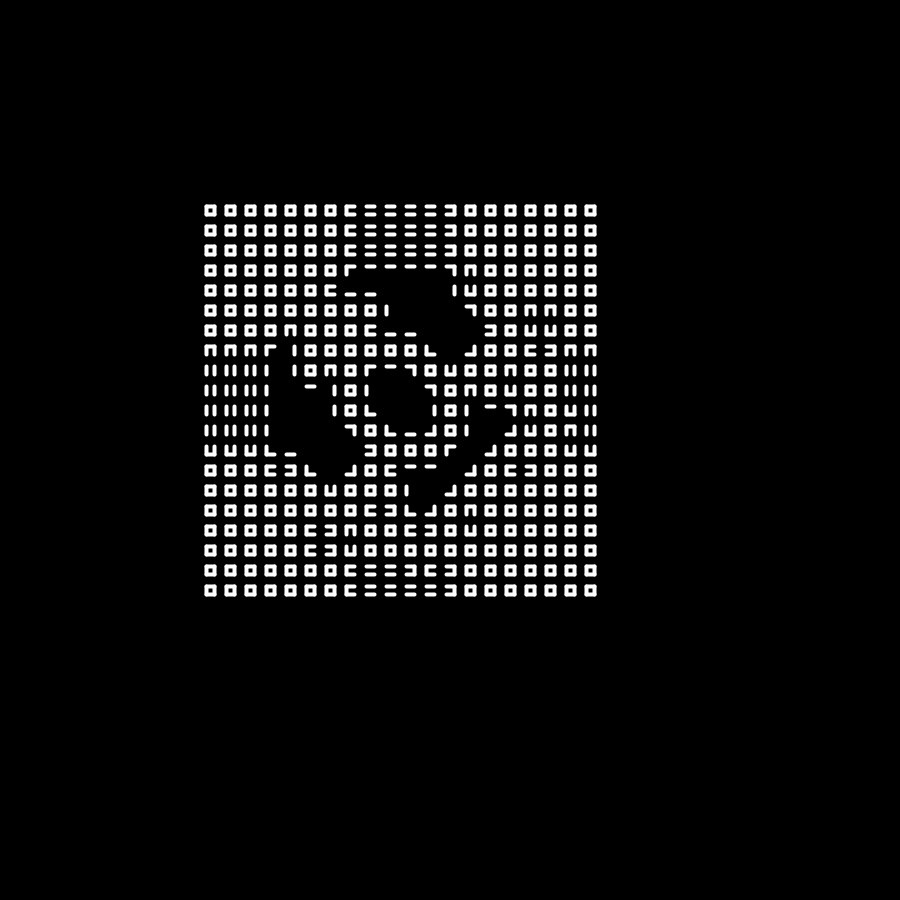}
     \end{tabular}
     \caption{\bf 2DRD  Box With Seed}
     \label{seq11seed}
\end{center}
\end{figure}

\begin{figure}
     \begin{center}
     \begin{tabular}{c}
     \includegraphics[width=8cm]{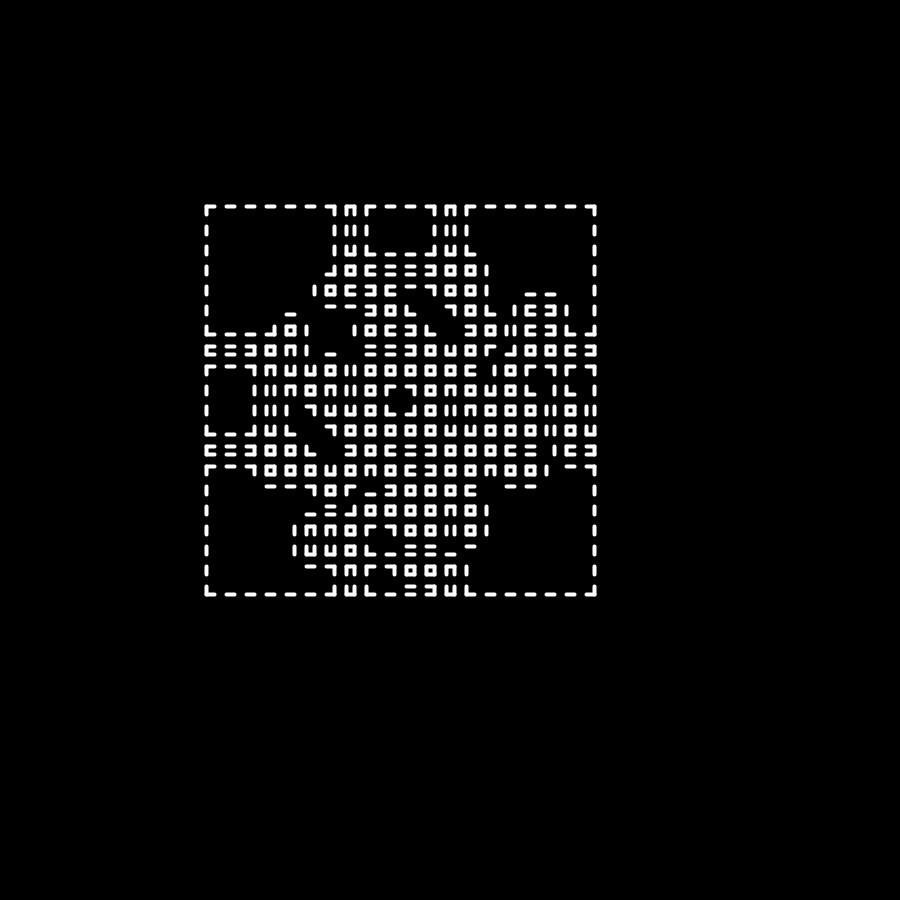}
     \end{tabular}
     \caption{\bf 2DRD  Box With Seed}
     \label{seq12seed}
\end{center}
\end{figure}

\begin{figure}
     \begin{center}
     \begin{tabular}{c}
     \includegraphics[width=6cm]{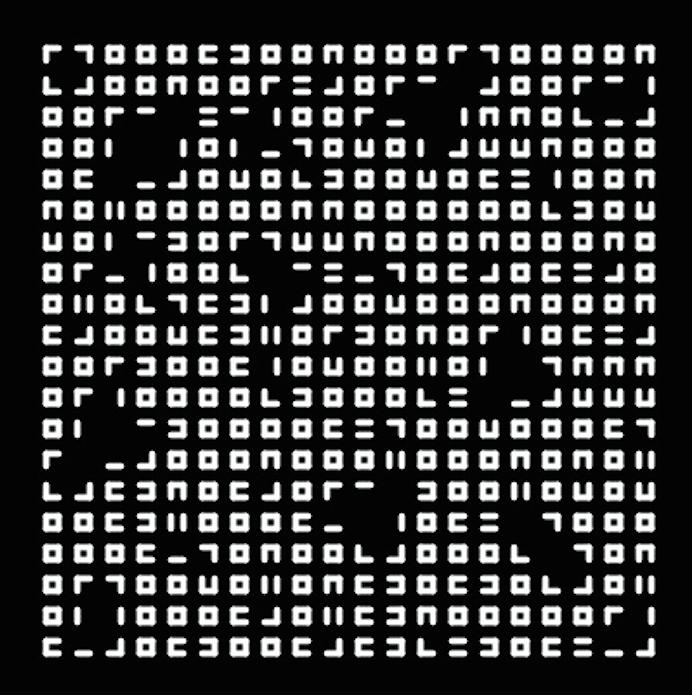}
     \end{tabular}
     \caption{\bf Tweedledum}
     \label{tdum}
\end{center}
\end{figure}

\begin{figure}
     \begin{center}
     \begin{tabular}{c}
     \includegraphics[width=6cm]{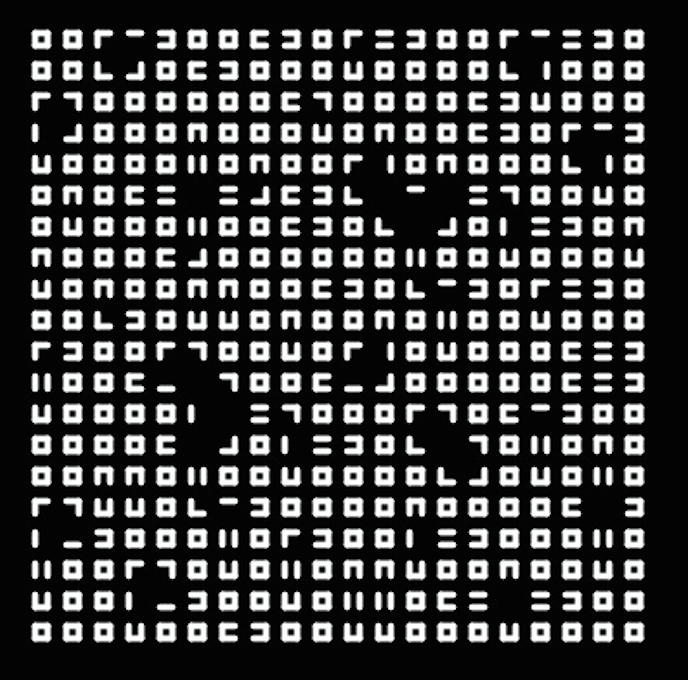}
     \end{tabular}
     \caption{\bf Tweedledee}
     \label{tdee}
\end{center}
\end{figure}

\section{Distinctions, Distinctioning and Wolfram Automata}
In this section we make a comparison with the general structure of Wolfram line automata \cite{Wolfram}. The Wolfram automata use a very simple alphabet consisting of two letters (black and white, or $0$ and $1$).
At every stage in the process a distinction is applied to the $8$ possible states consisting of a square and its neighbors to the left and to the right. The distinction assigns $0$ or $1$ to each of these states and the fate of the middle square in the next row is decided by that distinction. We see that these line automata are certainly RD automata, but that they are not strictly orthodox in our sense, in that the alphabet is not descriptive of all the local 
distinctions under consideration. The alphabet is simple but the distinctions that can be made are complex. The result of this choice leads to a large and interesting body of phenomena.\\

\begin{figure}
     \begin{center}
     \begin{tabular}{c}
     \includegraphics[width=8cm]{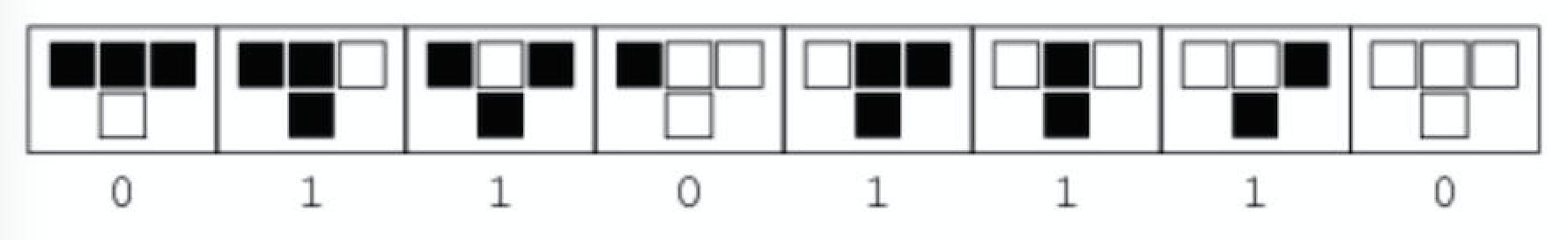}
     \end{tabular}
     \caption{\bf Rule 110}
     \label{prule110}
\end{center}
\end{figure}

\begin{figure}
     \begin{center}
     \begin{tabular}{c}
     \includegraphics[width=8cm]{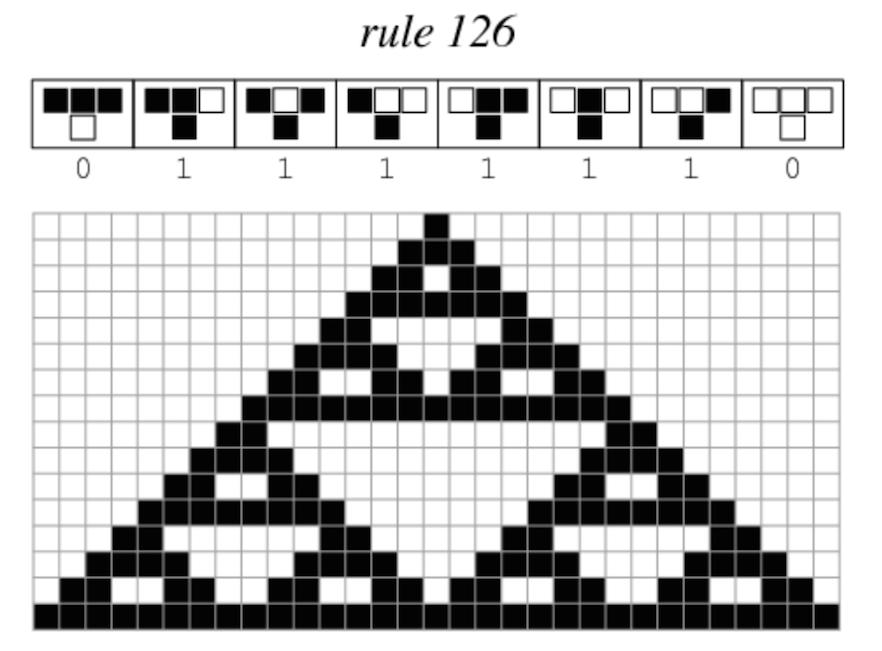}
     \end{tabular}
     \caption{\bf Wolfram Rule 126}
     \label{wrule126}
\end{center}
\end{figure}

\begin{figure}
     \begin{center}
     \begin{tabular}{c}
     \includegraphics[width=8cm]{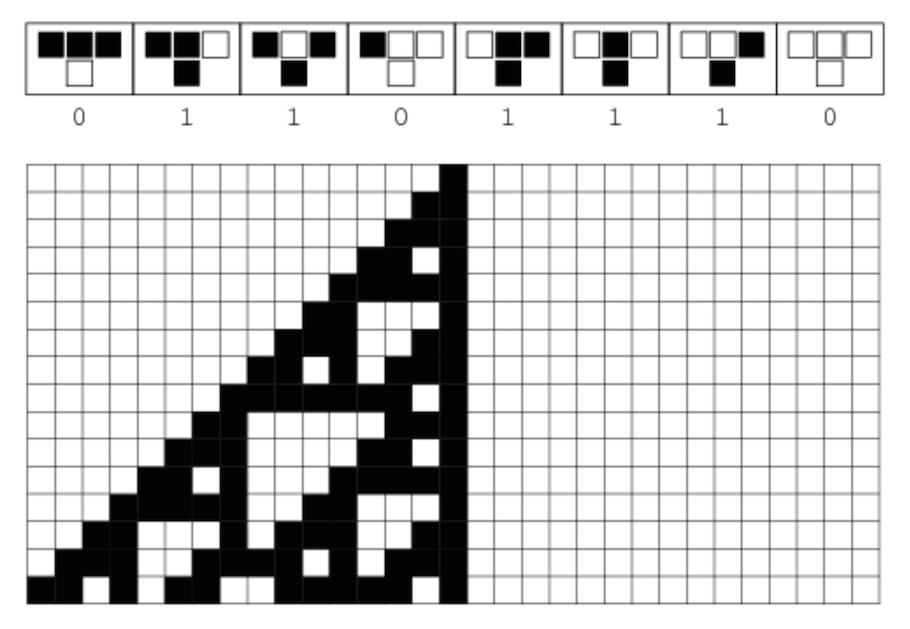}
     \end{tabular}
     \caption{\bf Rule 110}
     \label{rule110}
\end{center}
\end{figure}

\begin{figure}
     \begin{center}
     \begin{tabular}{c}
     \includegraphics[width=8cm]{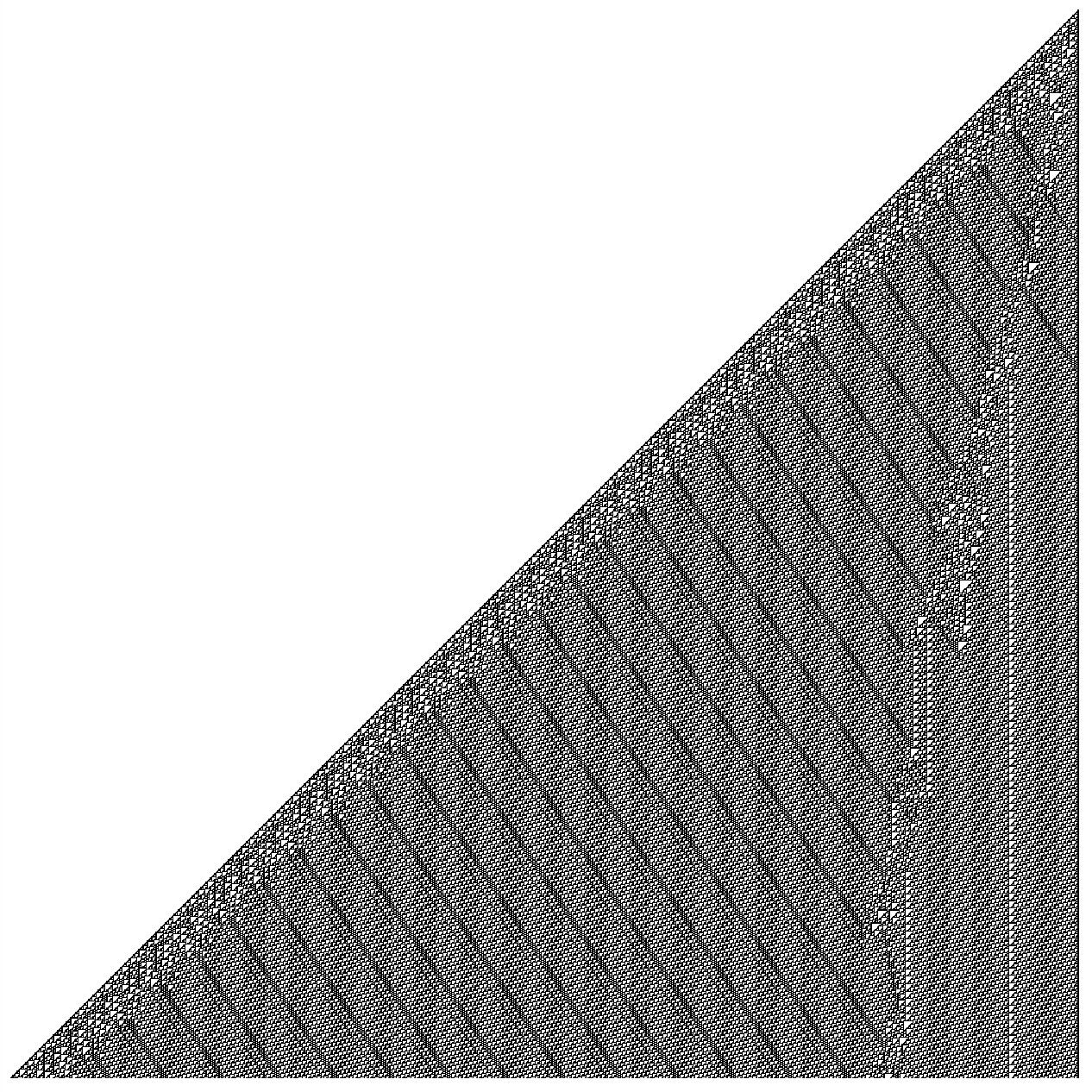}
     \end{tabular}
     \caption{\bf Wolfram Rule 110}
     \label{wrule110}
\end{center}
\end{figure}

In Figure~\ref{wrule126}we see a depiction of the results of applying Wolfram Rule 126. As the reader can see, by comparison with Figure~\ref{rep} and Figure~\ref{replicate}, the overall pattern resulting from Rule 126 is essentially the same as that obtained from our 1DRD. The underlying structure of alphabet and distinction is different. This is a first example indicating the need for more detailed comparison between orthodox RD rules and cellular automata. We will leave such analysis for further work. Note that we show in Figure~\ref{prule110} the Rule 110. It differs from Rule 126 in only one place. This de-symmetrization of Rule 126 results in
very complex behaviour. In Figure~\ref{rule110} and Figure~\ref{wrule110} we illustrate Rule 110 and show how its iteration looks. Here we are farther from the simple 1DRD. The Rule 110 takes full advantage of the very simple alphabet of  zero and one, and it uses an asymetrical distinction on the set of eight triples of zeros and ones. The result is a very complex pattern of evolution and an automaton that has been proved to be Turing universal. One can certainly regard Rule 110 as a highly successful application of non-orthodox RD. We will return to this Rule in a subsequent paper and examine it further in the light of RD structure.\\

\clearpage

\section{The HighLife Replicator}
\begin{figure}
     \begin{center}
     \begin{tabular}{c}
     \includegraphics[width=2cm]{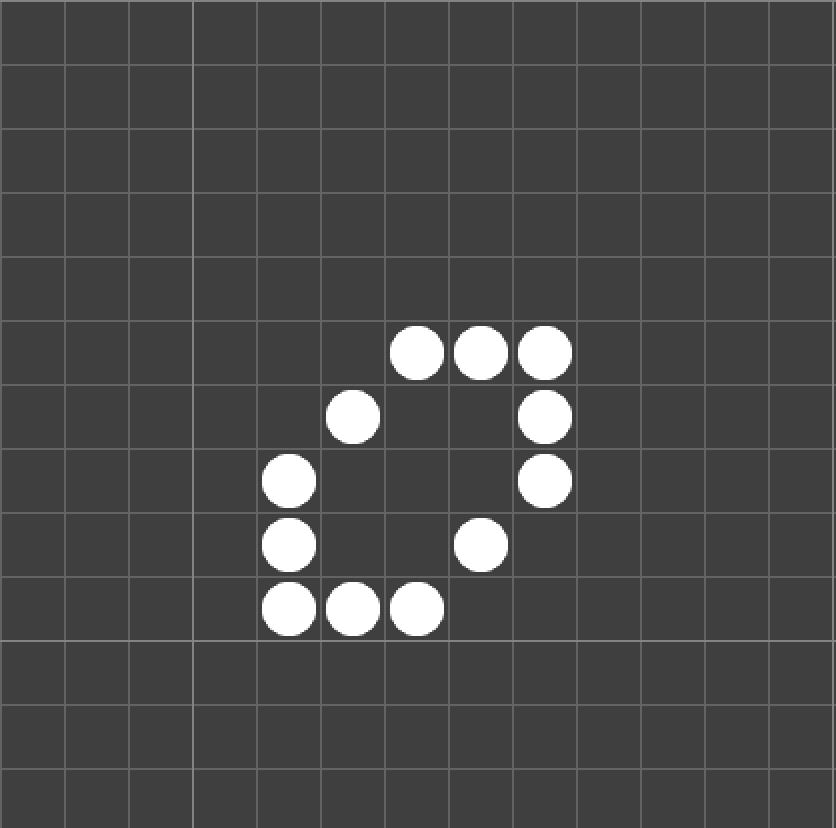}
     \end{tabular}
     \caption{\bf The High Life Replicator}
     \label{hlr1}
\end{center}
\end{figure}

This section is a comparison of patterns of the self-replicating element in the 1DRD and a very similar pattern in the much more complicated enviroment of the two-dimensional cellular automaton called HighLife,
a variant of John Horton Conway's Game of Life. In Highlife the environment is a rectangular lattice and each square is regarded as having eight neighbors. We could analyze an orthodox RD with an alphabet that 
generalizes the $16$ letter alphabet to a $256 = 2^8 $ letter alphabet for this geometry. This analysis is a future project for us. HighLife uses a simple binary rule. Each square in the lattice is either occupied (by a marker) or it is unoccupied (unmarkd). We say that a square has $n$ neighbors (where $n$ is between $0$ and $8$) if $n$ of its neighboring squares are occupied. The rule for HighLife is that an occupied square will survive (remain occupied) only if it has $2$ or $3$ neighbors. Otherwise it will beome unmarked (``die"). An unoccupied square will become occupied (be ``born") if it has $3$ or $6$ neighbors. In HighLife there is a remarkable small comfiguration that can reproduce itself. It takes $12$ steps for this replication process to take place. See Figure~\ref{hlr1}. And quite remarkably, the pattern that these replicators follow is essentially the same as the pattern that is followed by the self-replicating element in the 1DRD. See Figure~\ref{hl1}.

\begin{figure}
     \begin{center}
     \begin{tabular}{c}
     \includegraphics[width=2cm]{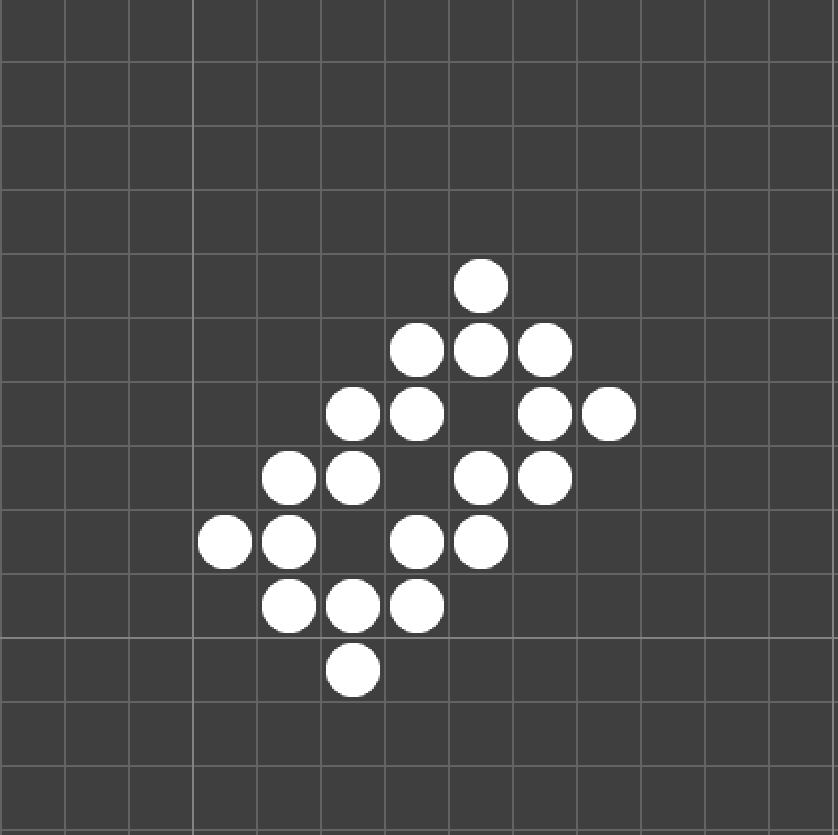}
     \end{tabular}
     \caption{\bf The High Life Replicator}
     \label{hlr2}
\end{center}
\end{figure}

\begin{figure}
     \begin{center}
     \begin{tabular}{c}
     \includegraphics[width=2cm]{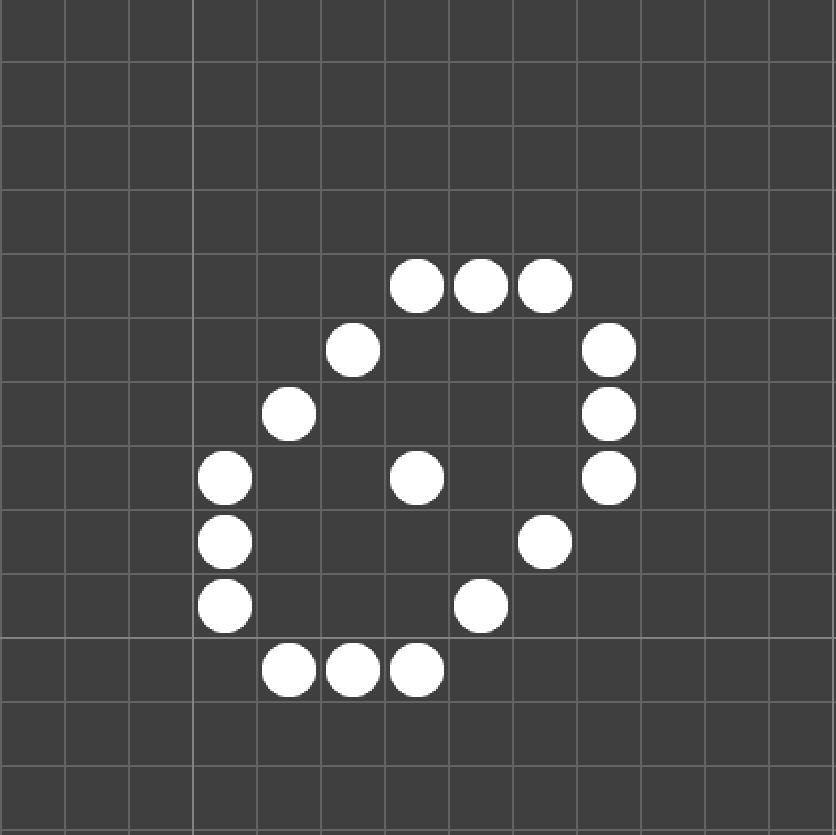}
     \end{tabular}
     \caption{\bf The High Life Replicator}
     \label{hlr3}
\end{center}
\end{figure}

\begin{figure}
     \begin{center}
     \begin{tabular}{c}
     \includegraphics[width=2cm]{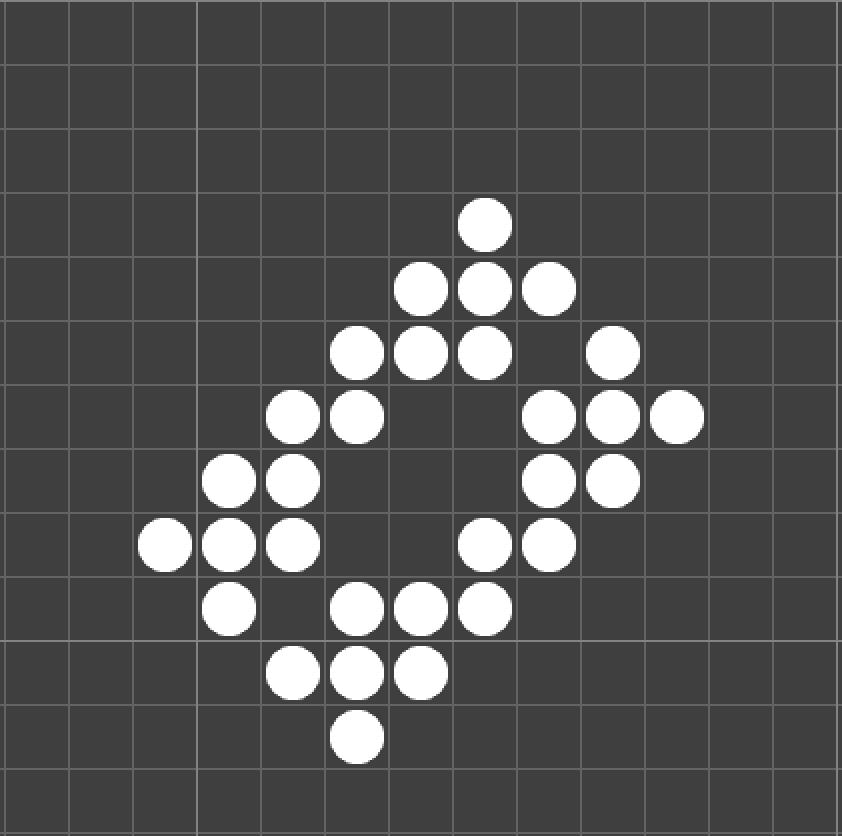}
     \end{tabular}
     \caption{\bf The High Life Replicator}
     \label{hlr4}
\end{center}
\end{figure}

\begin{figure}
     \begin{center}
     \begin{tabular}{c}
     \includegraphics[width=2cm]{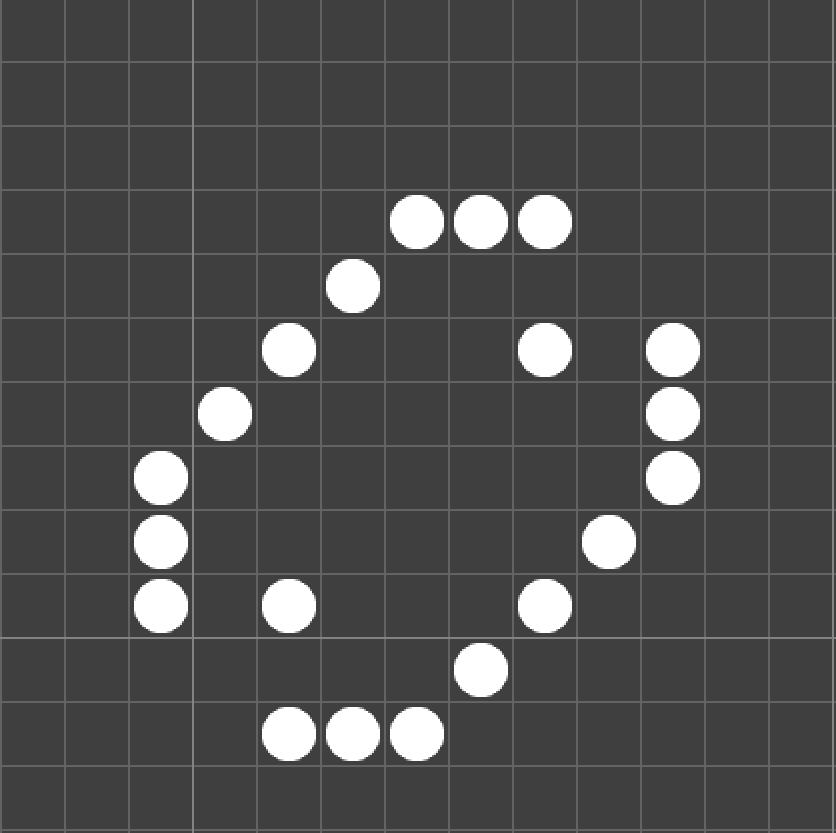}
     \end{tabular}
     \caption{\bf The High Life Replicator}
     \label{hlr5}
\end{center}
\end{figure}

\begin{figure}
     \begin{center}
     \begin{tabular}{c}
     \includegraphics[width=2cm]{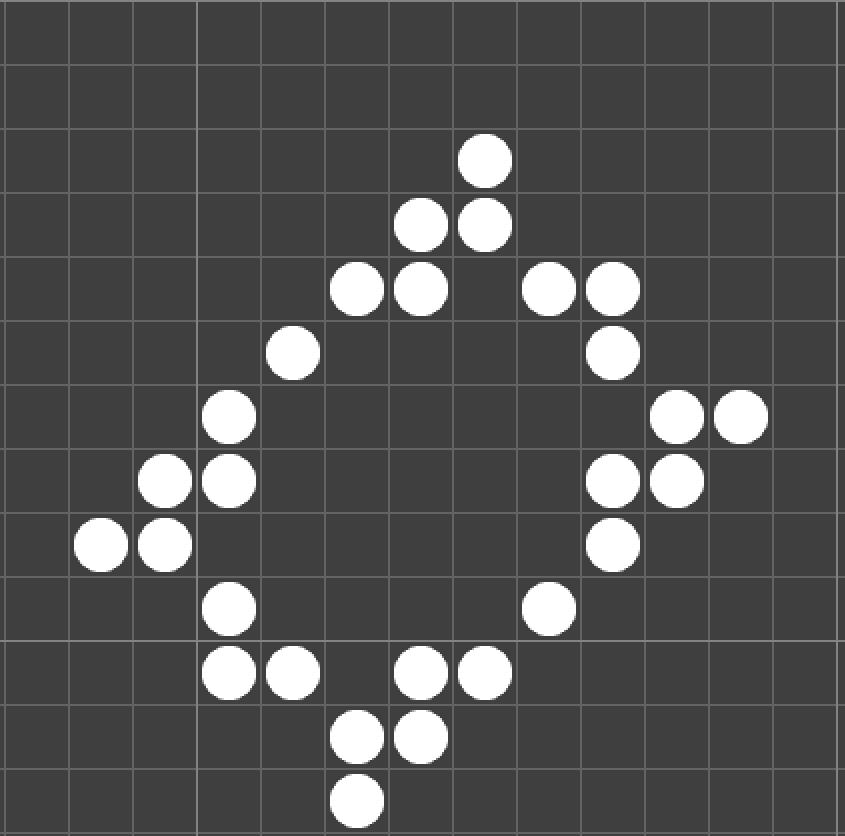}
     \end{tabular}
     \caption{\bf The High Life Replicator}
     \label{hlr6}
\end{center}
\end{figure}

\begin{figure}
     \begin{center}
     \begin{tabular}{c}
     \includegraphics[width=2cm]{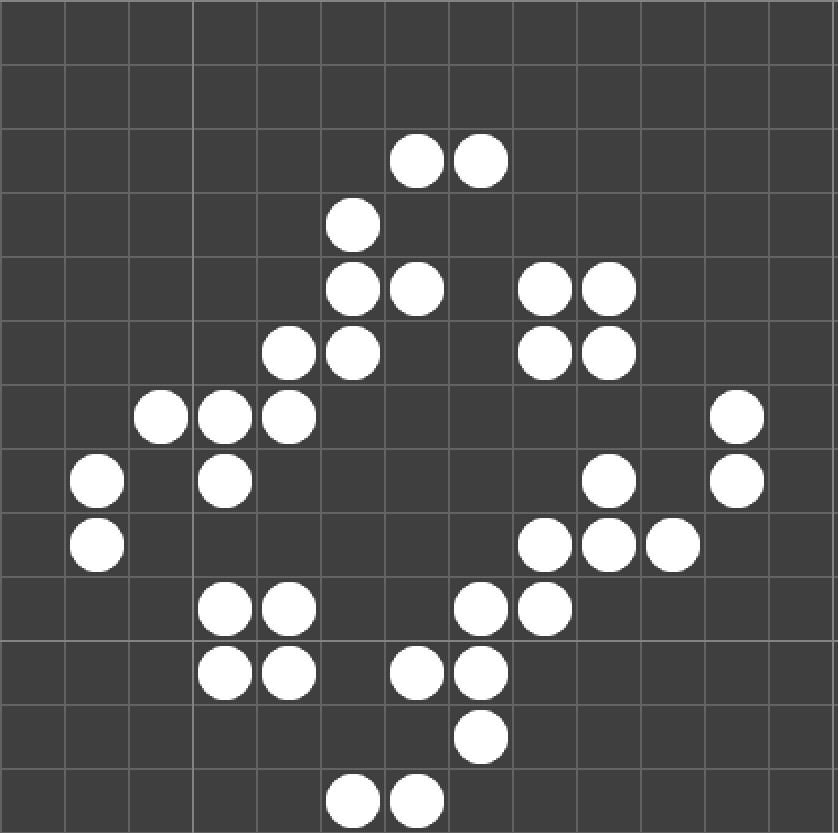}
     \end{tabular}
     \caption{\bf The High Life Replicator}
     \label{hlr7}
\end{center}
\end{figure}

\begin{figure}
     \begin{center}
     \begin{tabular}{c}
     \includegraphics[width=2cm]{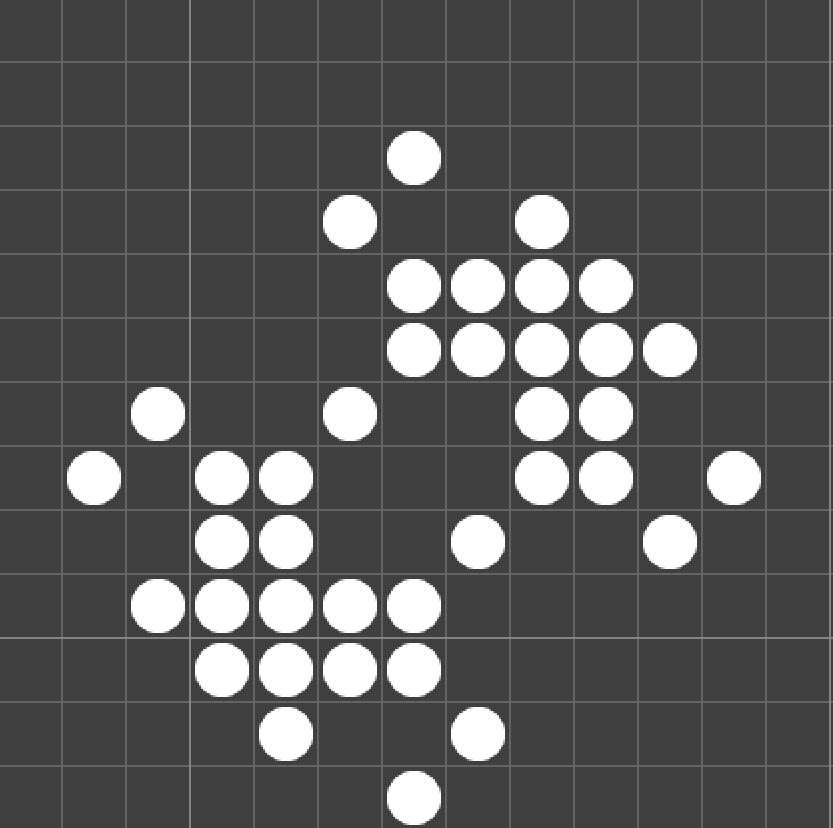}
     \end{tabular}
     \caption{\bf The High Life Replicator}
     \label{hlr8}
\end{center}
\end{figure}

\begin{figure}
     \begin{center}
     \begin{tabular}{c}
     \includegraphics[width=2cm]{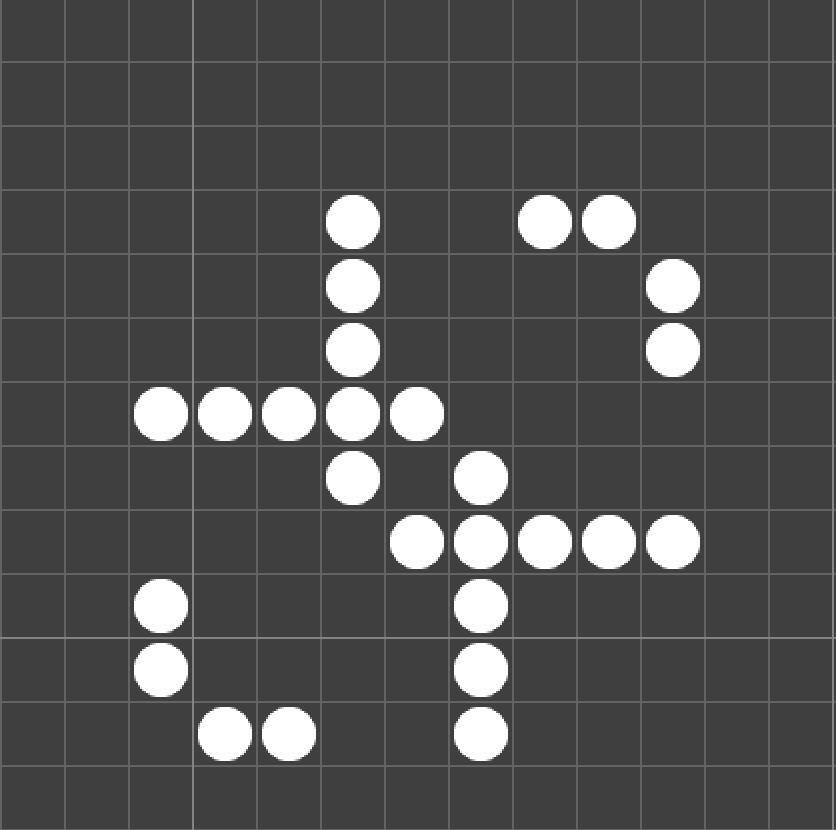}
     \end{tabular}
     \caption{\bf The High Life Replicator}
     \label{hlr9}
\end{center}
\end{figure}

\begin{figure}
     \begin{center}
     \begin{tabular}{c}
     \includegraphics[width=2cm]{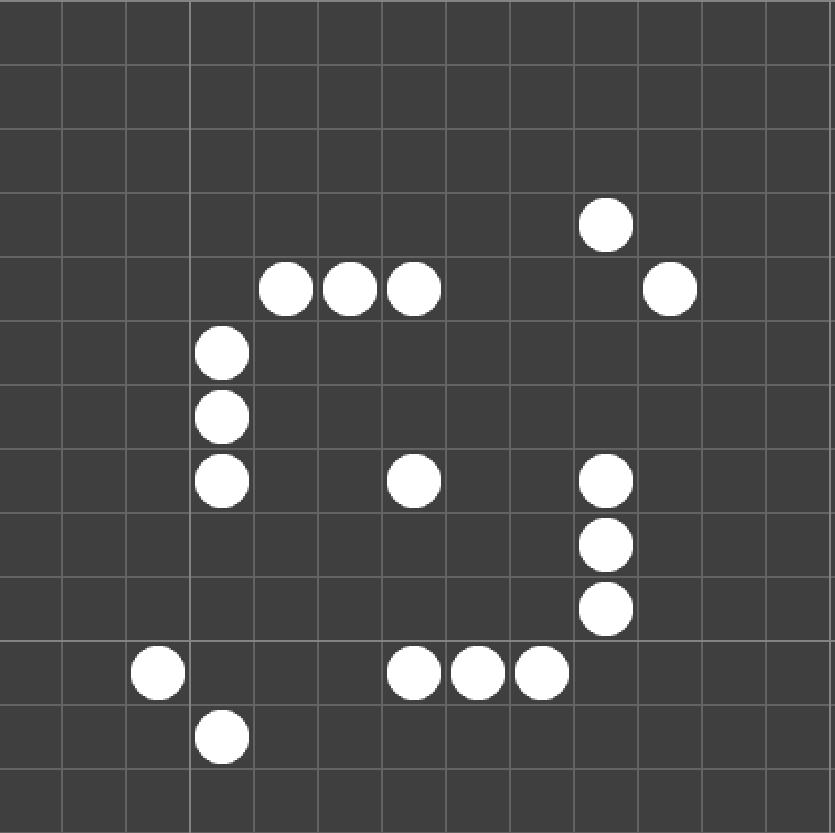}
     \end{tabular}
     \caption{\bf The High Life Replicator}
     \label{hlr10}
\end{center}
\end{figure}

\begin{figure}
     \begin{center}
     \begin{tabular}{c}
     \includegraphics[width=2cm]{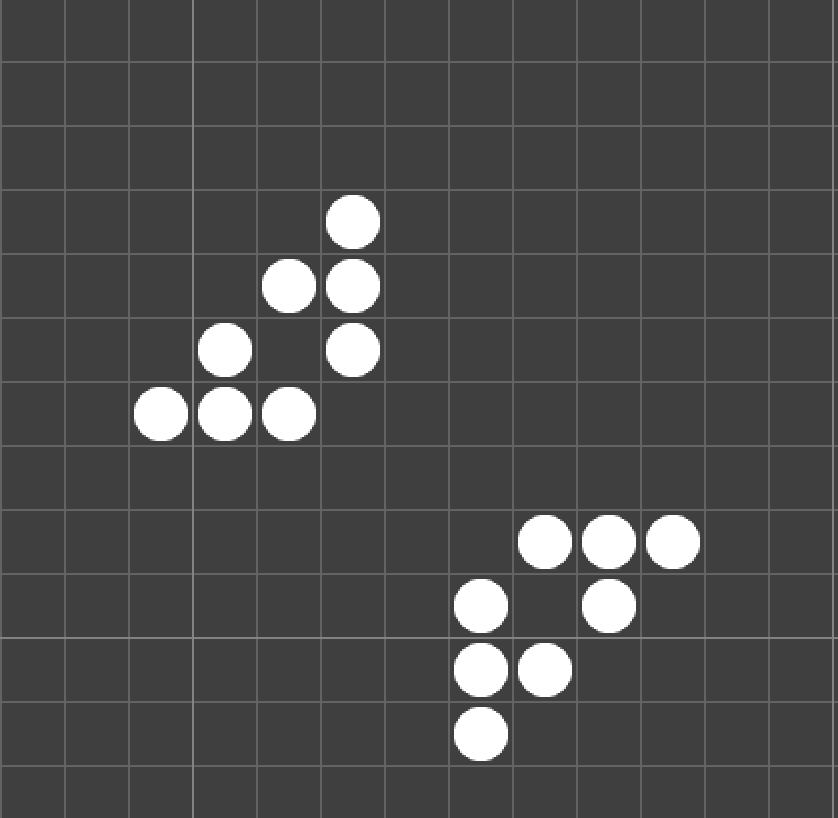}
     \end{tabular}
     \caption{\bf The High Life Replicator}
     \label{hlr11}
\end{center}
\end{figure}

\begin{figure}
     \begin{center}
     \begin{tabular}{c}
     \includegraphics[width=2cm]{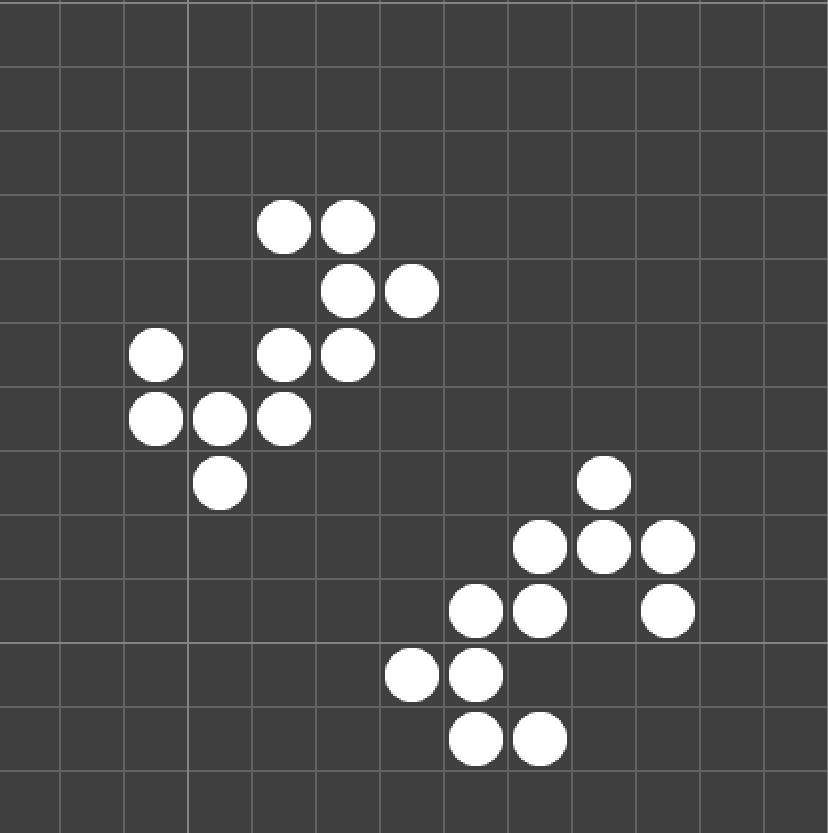}
     \end{tabular}
     \caption{\bf The High Life Replicator}
     \label{hlr12}
\end{center}
\end{figure}

\begin{figure}
     \begin{center}
     \begin{tabular}{c}
     \includegraphics[width=2cm]{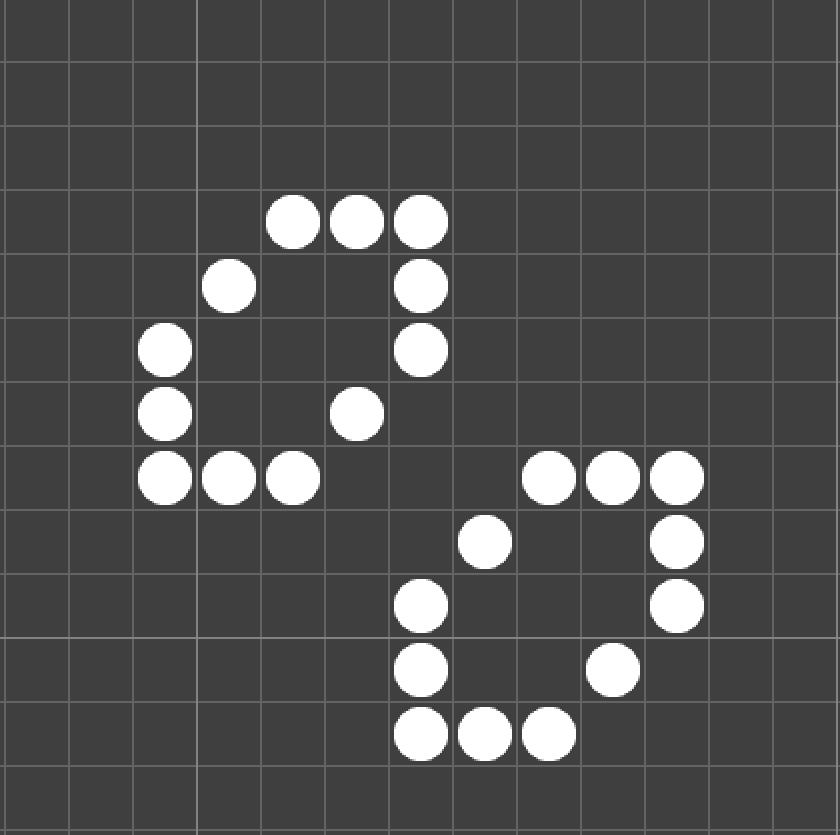}
     \end{tabular}
     \caption{\bf The High Life Replicator}
     \label{hlr13}
\end{center}
\end{figure}

\clearpage

  \begin{figure}
     \begin{center}
     \begin{tabular}{c}
     \includegraphics[width=2cm]{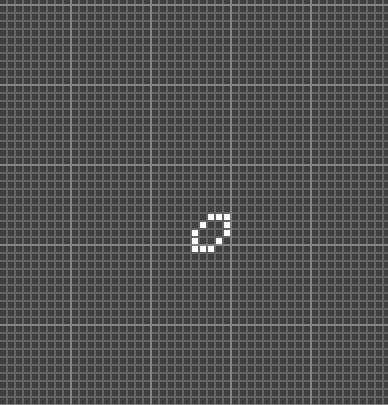}
     \end{tabular}
     \caption{\bf The High Life Replicator Twelve-Step}
     \label{hl1}
\end{center}
\end{figure}

  \begin{figure}
     \begin{center}
     \begin{tabular}{c}
     \includegraphics[width=2cm]{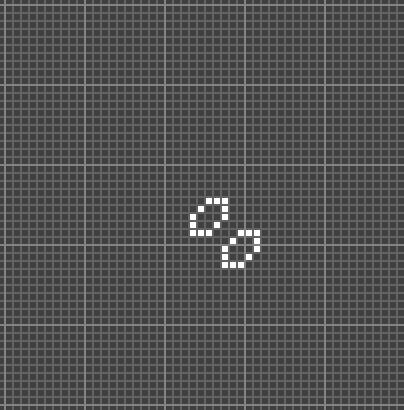}
     \end{tabular}
     \caption{\bf The High Life Replicator Twelve-Step}
     \label{hl2}
\end{center}
\end{figure}

  \begin{figure}
     \begin{center}
     \begin{tabular}{c}
     \includegraphics[width=2cm]{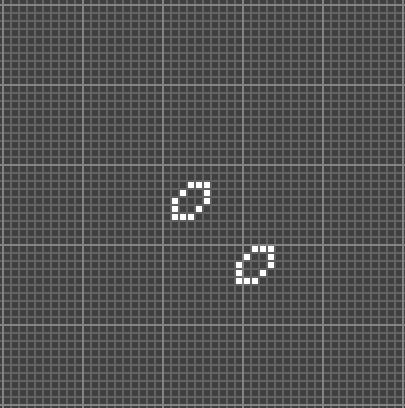}
     \end{tabular}
     \caption{\bf The High Life Replicator Twelve-Step}
     \label{hl3}
\end{center}
\end{figure}

  \begin{figure}
     \begin{center}
     \begin{tabular}{c}
     \includegraphics[width=2cm]{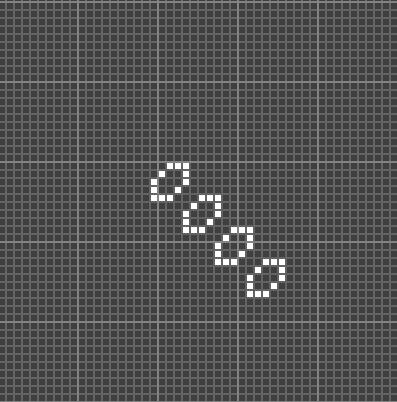}
     \end{tabular}
     \caption{\bf The High Life Replicator Twelve-Step}
     \label{hl4}
\end{center}
\end{figure}

  \begin{figure}
     \begin{center}
     \begin{tabular}{c}
     \includegraphics[width=2cm]{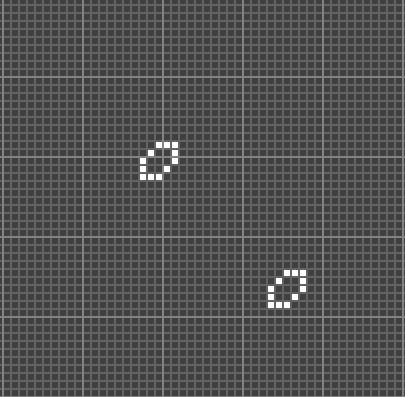}
     \end{tabular}
     \caption{\bf The High Life Replicator Twelve-Step}
     \label{hl5}
\end{center}
\end{figure}

  \begin{figure}
     \begin{center}
     \begin{tabular}{c}
     \includegraphics[width=2cm]{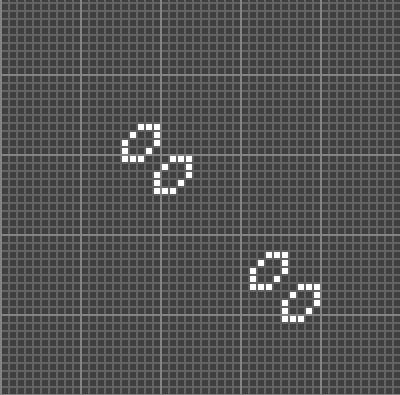}
     \end{tabular}
     \caption{\bf The High Life Replicator Twelve-Step}
     \label{hl6}
\end{center}
\end{figure}

  \begin{figure}
     \begin{center}
     \begin{tabular}{c}
     \includegraphics[width=2cm]{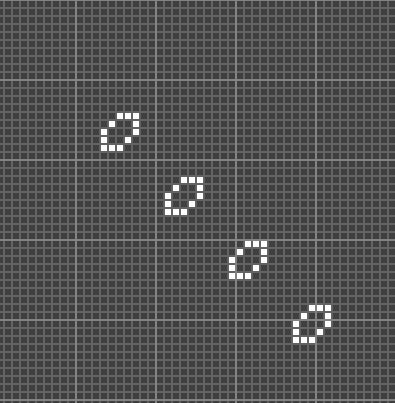}
     \end{tabular}
     \caption{\bf The High Life Replicator Twelve-Step}
     \label{hl7}
\end{center}
\end{figure}

  \begin{figure}
     \begin{center}
     \begin{tabular}{c}
     \includegraphics[width=2cm]{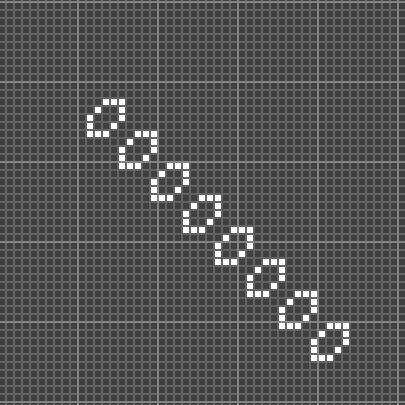}
     \end{tabular}
     \caption{\bf The High Life Replicator Twelve-Step}
     \label{hl8}
\end{center}
\end{figure}

\section{RD and DNA}
We begin this section with a review of material from the introduction to the paper.
In this section we will describe one version RD process, and we will show how it gives rise to a pattern self-replication that is recognizable as a case of replication that we have called {\it DNA Replication} \cite{LK11,LK12}.\\

The rules for the RD process are very simple. We begin with an arbitrary finite text string delimited by the character * at both ends. The RD process creates a new string from the given string by {\it describing the distinctions in the initial string.} Each character in the initial string is examined together with its
left and right neighbors. Let $LCR$ denote a character $C$ with neighbors $L$ and $R.$ Then we replace $C$ by a new character according to the following rules:
\begin{enumerate}
\item $C \longrightarrow  \, = \,$ if $L = C$ and $C = R$  (no distinction).
\item $C \longrightarrow   \, [ \,$ if $L \neq C$ but $C = R$ (distinction on the left).
\item $C \longrightarrow  \, ] \,$ if  $L = C$ but $C \neq R$ (distinction on the right).
\item $C \longrightarrow  \, O \,$ if  $L \neq C$ and $C \neq R$ (distinction on both the left and the right).
\item If $C$ is adjacent to $*$ change $C$ to $=$. (This is just a choice of boundary behavior.)
\end{enumerate} 

See Figure~\ref{green} for the result of applying the $RD$ process to a chosen text string.\\

In Figure~\ref{rep} we show the result of starting with a very simple text string. In this figure we do not print the character $=$, so that the resulting strings have empty space where this character would appear. As the reader  can see, the string $*======]O[======*$
has a long sequence of transformations under the RD process. 
the pattern $]O[$ is replicated by the sequence below.\\
\begin{enumerate}
\item  $É=======]O[=======É.$
\item  $É======]OOO[=======É$
\item $ É=====]O[=]O[======É$
\end{enumerate}

Remarkably, this self-replication has the same patten as an abstract description of DNA replication. We will explain this below and in a separate section.\\

\begin{figure}
     \begin{center}
     \begin{tabular}{c}
     \includegraphics[width=6cm]{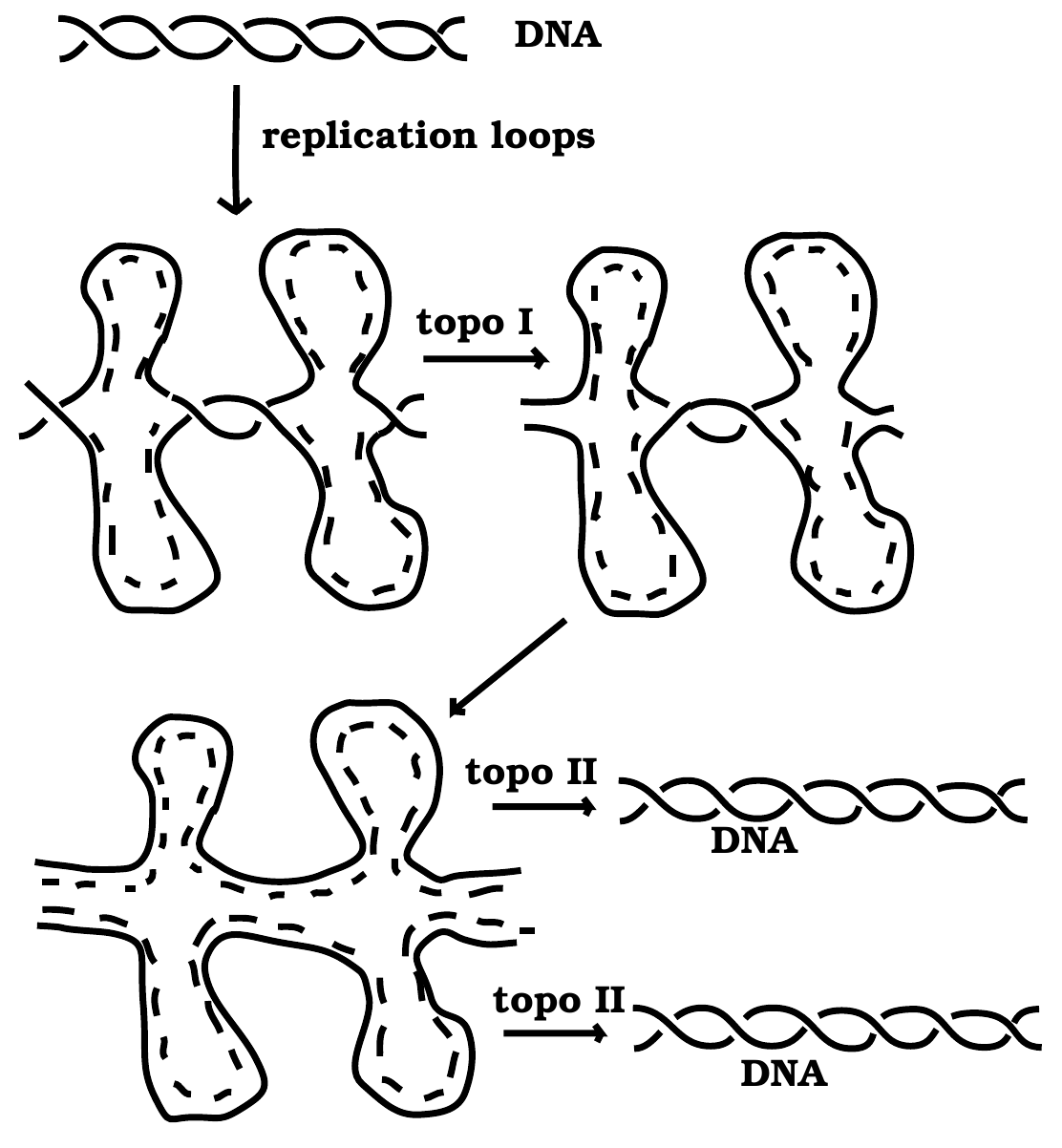}
     \end{tabular}
     \caption{\bf DNA Replication}
     \label{DNArep}
\end{center}
\end{figure}

\noindent {\bf A Quick Review of the Pattern of DNA Replication.} DNA consists in two strands of base-pairs wound helically around a phosphate backbone.  It is 
customary to call one of these strands the ``Watson" strand and the other the ``Crick" strand.
Abstractly we can write  $$DNA = <W|C>$$ \noindent to symbolize the binding of the two strands into the single
DNA duplex. Replication occurs via the separation of the two strands via polymerase enzyme.
This separation occurs locally and propagates. Local sectors of separation can amalgamate into
larger pieces of separation as well. Once the strands are separated, the environment of the cell 
can provide each with complementary bases to form the base pairs of new duplex DNA's. Each strand,
separated {\em in vivo}, finds its complement being built naturally in the environment. This picture ignores
the well-known topological difficulties present to the actual separation of the daughter strands. See Figure~\ref{DNArep}. In this figure we give some hints about the topological complexities that are not discussed here.
Biologists discovered enzymes that cut and reconnect strands of DNA, resulting in the release of topological linking that would otherwise obstruct the separation of the newly produced strands of DNA. All this is subject to 
another discussion of its relationship with RD concepts.
\bigbreak

The base pairs in the DNA sequence are $AT$ (Adenine and Thymine) and $GC$ (Guanine and Cytosine). Thus if
 
$$<W| = <...TTAGAATAGGTACGCG... |$$ \noindent then $$|C> = |...AATCTTATCCATGCGC...>.$$ 

\noindent Symbolically we can oversimplify the whole process as 

$$<W| + E \longrightarrow <W|C> = DNA$$

$$E + |C> \longrightarrow <W|C> = DNA$$

$$<W|C> \longrightarrow <W| + E + |C> = <W|C> <W|C>$$

\noindent Either half of the DNA can, with the help of the environment, become a full DNA.
We can let $E \, \longrightarrow \, |C><W|$ be a symbol for the process by which the environment supplies the 
complementary base pairs $AG$, $TC$ to the Watson and Crick strands. In this oversimplification
we have cartooned the environment as though it contained an already-waiting strand $|C>$ to 
pair with $<W|$ and an already-waiting strand $<W|$ to pair with $|C>.$ 
\smallbreak 

{\em In fact it is the opened
strands themselves that command the appearance of their mates. They conjure up their mates from 
the chemical soup of the environment.} 
\smallbreak 

The environment $E$ is an identity element in this algebra of 
cellular interaction. That is, $E$ is always in the background and can be allowed to appear spontaneously
in the cleft between Watson and Crick:

$$<W|C> \longrightarrow  <W| |C> \longrightarrow <W| E |C>$$ 

$$\longrightarrow <W| |C><W| |C> \longrightarrow <W|C><W|C>$$

\noindent This is the formalism of DNA replication.  
\bigbreak

We are now in a position to compare the formalism of the DNA replication with the RD replication.

\begin{enumerate}
\item  $É=======]O[=======É.$
\item  $É======]OOO[=======É$
\item $ É=====]O[=]O[======É$
\end{enumerate}

In the RD replication, we start with $]O[$ in its RD-environment. Matters of distinction of this entity from its surroundings leads to the production of $]OOO[$, and then we see that the identity of the internal $0$
with its neighbors leads to the splitting $]O[=]O[.$ There is no question that the basis of this replication is not the same as the DNA replication, but thematically, the two patterns are certainly related. The RD pattern 
is at a different level than the DNA pattern. In the RD replication, that environment for the symbol string is the larger symbol string. Thus it is only in the eyes of the observer of the RD that the ``entity" $]O[$ is 
distinguished and is seen as an actor against the background of ``declarations of identity" $...========...$. These declarations of identity are indeed equal to one another and so form an invariant background
or void from which patterns arise in the presence of any difference. This is, in fact how our entity came into being.
$$...AAAAAAAAAAAAABAAAAAAAAAAAAA...$$
$$...==============]0[===============...$$
Our entity $]0[$ is the first description of sameness on left, difference in middle, samness on the right. The left and right icons $]$ and $[$ form a carapace for the indicator of difference $O.$ Thus a bare difference 
of B from its equal neighbors A evolves by description, at once into a {\it proto-cell} with a carapace. It is this protocell that then undergoes mitosis in the next two rounds of description. The cell-division or mitosis is
enabled by the production of new carapace ($]OOO[ \longrightarrow ]O[=]O[$) from within the cell. It is important to note that this production does not come from an ``inner mechanism" of the ``cell", but rather from the 
global recursive/descriptive situation of these entities in the entire line of the RD structure. It is the ``influence" of the ``surrounding void" that makes all this happen in the course of recursive description and distinction.
It is a fortuitous accident of working 
in one dimension that the carapace is seen in a left portion paired with a right portion, analogous to the two strands of the DNA. {\it At this condensed creation scenario, we find that the patterns of DNA replication, cell formation and mitosis all appear at once in the first few steps away from a marking (B) in the void ( of repeated A's).} \\

For DNA replication, we can interpret the correspondence as:
\begin{enumerate}
\item $ ] = $ Watson, $[ \, =$ Crick, $O = $ backbone or binding.
\item  RD action results in the opening of the backbone so that binding $O$ is replaced by environment $OOO$.
\item  RD action relative to the environment results in placement of new Watson and new Crick. 
So we have the self-replication of $]O[$.
\end{enumerate}

Note that there is another level at which we can think about this! Regard $]$ and $[$ as ``cell-wallsÓ. Then we are witnessing not $DNA$ reproduction, but {\it mitosis} itself!
The little fellow $]O[$ is a cell and we are watching how he reproduces in the line environment $ É=============É$. of the ``void" where there are no distinctions.
The reader should now look again at Figure~\ref{green} and note the many appearances and interactions related to this elementary cell.\\

Of course the interpretations of ``backbone", ``strand", ``environment" , ``cell"  are different from what happens in the biology, but it is very interesting that the basic principles are similar.\\

Note how we get  $ \cdots ===]OOOOO \cdots$ goes to   $ \cdots ==]O[=== \cdots$
So actually the whole ``environment" flips here.
But it is contained in the above scenario.
Everything that happens in RD is non-local since a single event affects the whole string.\\

Perhaps it is clear to the reader that Recursive Distinctioning in the sense of this section is a potentially explosive topic that will grow to influence all the aspects of biology and computing.
We believe that this is the case. The principle of [distinction/description in recursive process] applies at all levels of biology, coginition, information science and computing. \\

\section{Maturana, Uribe and Varela and the Game of Life}
\begin{figure}
     \begin{center}
     \begin{tabular}{c}
     \includegraphics[width=6cm]{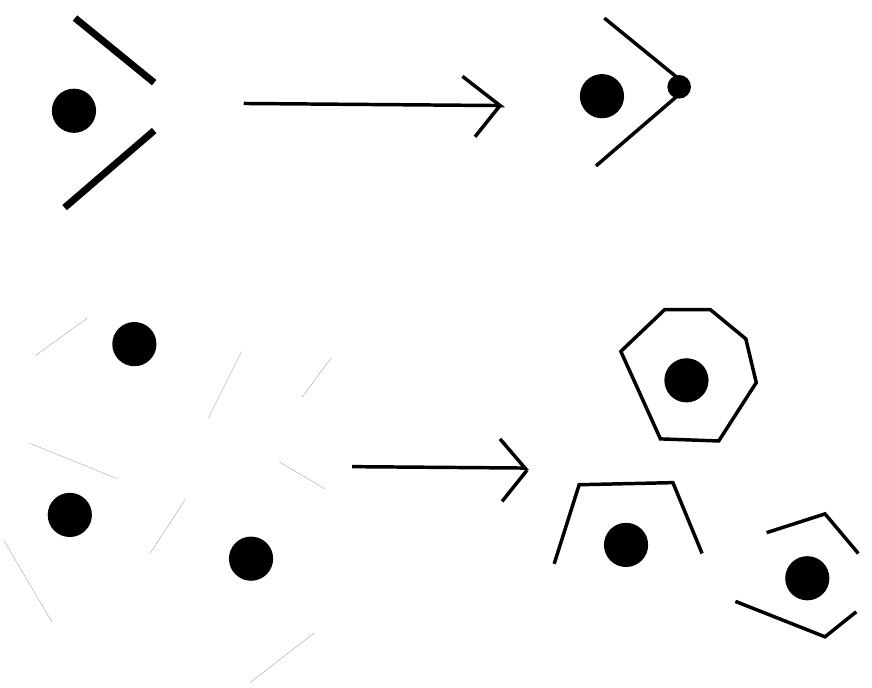}
     \end{tabular}
     \caption{\bf Proto-Cells of Maturana, Uribe and Varela}
     \label{Figure 5}
\end{center}
\end{figure}

Some examples from cellular automata clarify many of the issues about replication and the relationship of logic and biology. 
Here is an example due to 
Maturana, Uribe and Varela \cite{MUV}. See also \cite{FV} for a global treatment of related issues.
The ambient space is two dimensional and in it there are
``molecules" consisting in ``segments" and ``disks" (the catalysts) (See Figure~\ref{Figure 5}). There is a minimum distance among the 
segments and the disks (one can place them on a discrete lattice in the plane). And ``bonds" can form with
a probability of creation and a probability of decay between segment molecules with minimal
spacing. There are two types of molecules: ``substrate" (the segments) and ``catalysts" (the disks). The catalysts are not
susceptible to bonding, but their presence (within say three minimal step lengths) enhances the 
probability of bonding and decreases the probability of decay. Molecules that are not bonded
move about the lattice (one lattice link at a time) with a probability of motion.
In the beginning there is a randomly placed soup of molecules with a high percentage of substrate 
and a smaller percentage of catalysts. What will happen in the course of time?
\bigbreak

In the course of time the catalysts (basically separate from one another due to lack of bonding) 
become surrounded by circular forms of bonded or partially bonded substrate. A distinction 
(in the eyes of the observer) between inside (near the catalyst) and outside 
(far from a given catalyst) has spontaneously arisen through the ``chemical rules". Each catalyst
has become surrounded by a proto-cell. No higher organism has formed here, but there is a hint of the 
possibility of higher levels of organization arising from a simple set of rules of interaction.
{\em The system is not programmed to make the proto-cells.} They arise spontaneously in the evolution of 
the structure over time.
\bigbreak

\section{Conway Life}
One might imagine that organisms could be induced to arise as the evolutionary 
behavior of formal systems. There are difficulties, not the least of which is that there are 
nearly always structures in such systems 
whose probability of spontaneous emergence is vanishingly small. A good example is given by another 
automaton --  John H. Conway's ``Game of Life".  In ``Life" the cells  appear and disappear as
marked squares in a rectangular planar grid. A newly marked cell is said to be ``born". An unmarked cell
is ``dead". A cell dies when it goes from the marked to the unmarked state. A marked cell 
survives if it does not become unmarked in a given time step.
According to the rules of Life, an unmarked cell is born if and only if it has three neighbors.
A marked cell survives if it has either two or three neighbors. All cells in the lattice are updated in 
a single time step. The Life automaton is one of many automata of this type and indeed it is 
a fascinating exercise to vary the rules and watch a panoply of different behaviors. 
\bigbreak

\begin{figure}
     \begin{center}
     \begin{tabular}{c}
     \includegraphics[width=6cm]{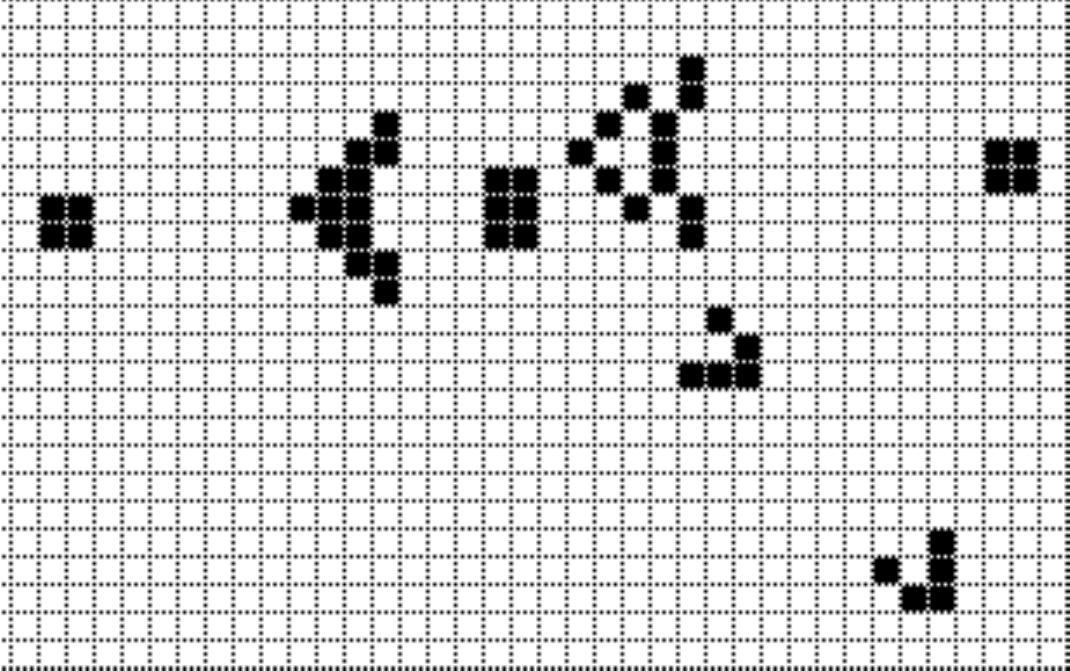}
     \end{tabular}
     \caption{\bf Glider Gun and Gliders}
     \label{Figure 6}
\end{center}
\end{figure}

For this 
discussion we concentrate on some particular features. There is a configuration in Life called a
``glider". See Figure~\ref{Figure 6}, which illustrates a series of gliders
going diagonally from left to right down the Life lattice, as well as a "glider
gun" (discussed below) that has produced them.
The glider consists in five cells in one of two basic configurations.
Each of these configurations produces the other (with a change in orientation). After four steps the
glider reproduces itself in form, but shifted in space. Gliders appear as moving entities in 
the temporality of the Life board. The glider is a complex entity that arises naturally from a 
small random selection of marked cells on the Life board. Thus the glider is a ``naturally 
occurring entity" just like the proto-cell in the Maturana-Uribe-Varela automaton. 
\bigbreak

But Life 
contains potentially much more complex phenomena. For example, there is the ``glider gun" (See
Figure~\ref{Figure 6}) which perpetually creates new gliders.  The ``gun" was invented by the Gosper Group, a group of researchers
at MIT in the 1970's. It is highly unlikely that a gun would appear 
spontaneously in the Life board. Of course there is a tiny probability of this, but we would guess
that the chances of the appearance of the glider gun by random selection or evolution from a 
random state is similar to the probability of all the air in the room collecting in one corner.
Nervertheless, the gun is a natural design based on forms and patterns that do appear spontaneously 
on small Life boards.  The glider gun emerged through the coupling of the power of human cognition and the automatic behavior of 
a mechanized formal system.
\bigbreak
  
Cognition is in fact an attribute of our biological system at an 
appropriately high level of organization. Cognition itself looks as improbable as the glider gun!  
Do patterns as complex as cognition or the glider gun arise 
spontaneously in an appropriate biological context? 
\bigbreak

There is a middle ground.  If one examines cellular automata of a given type and varies the
rule set randomly rather than varying the initial conditions for a given automaton, then a very wide
variety of phenomena will present themselves. In the case of molecular biology at the level of the 
DNA there is exactly this possibility of varying the rules, in the sense of varying the sequences in 
the genetic code. So it is possible at this level to produce a wide range of remarkable complex systems.
\bigbreak

\section{Other Forms of Replication}

Other forms of self-replication are quite revealing. For example, one might point out that a
stick can be made to reproduce by breaking it into two pieces. This may seem satisfactory on the 
first break, but the breaking cannot be continued indefinitely. In mathematics on the other hand, 
we can divide an interval into two intervals and continue this process ad infinitum. For a 
self-replication to have meaning in the physical or biological realm there must be a genuine 
repetition of structure from original to copy. At the very least the interval should grow to 
twice its size before it divides (or the parts should have the capacity to grow independently).
\bigbreak

\begin{figure}
     \begin{center}
     \begin{tabular}{c}
     \includegraphics[width=6cm]{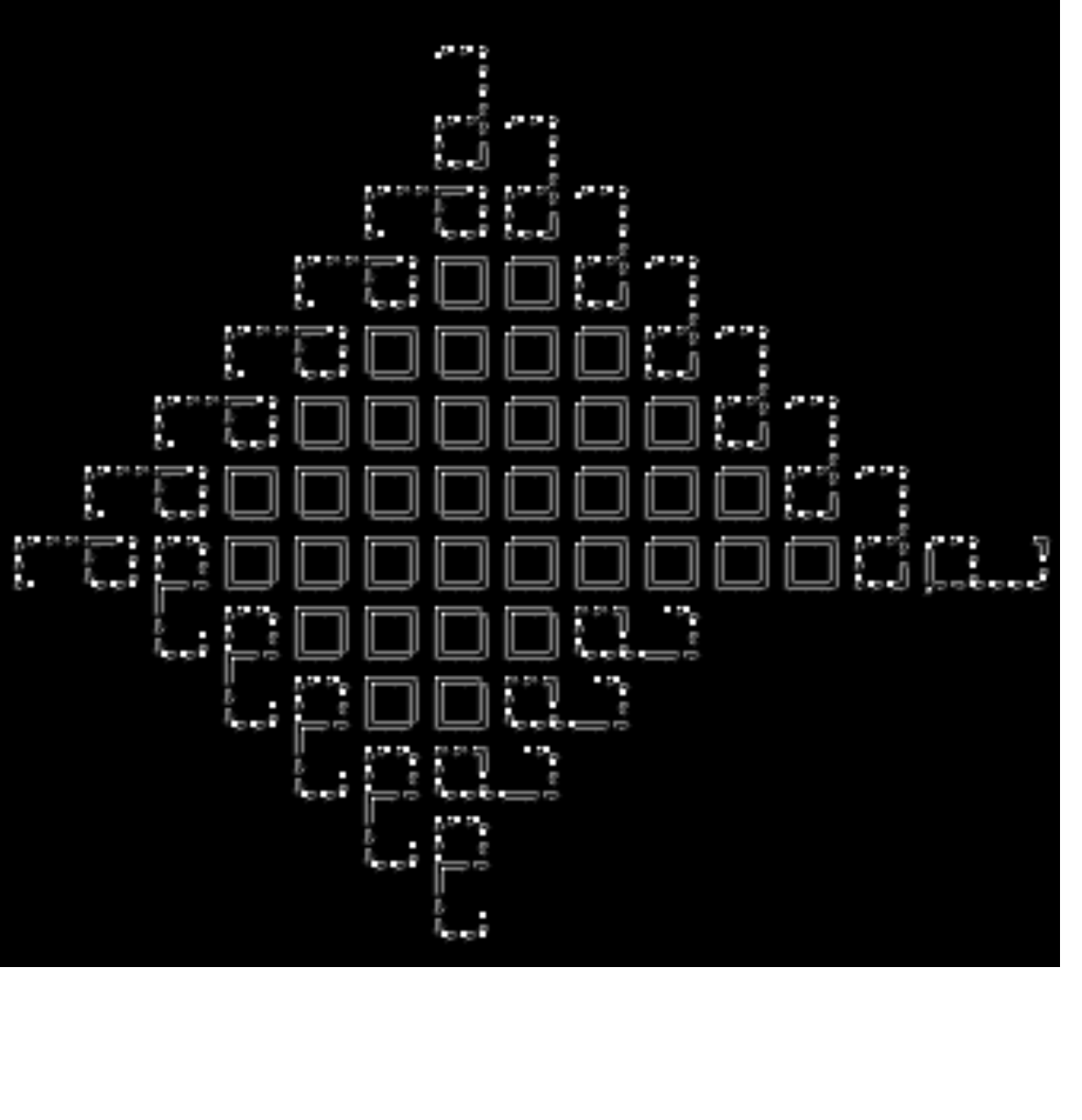}
     \end{tabular}
     \caption{\bf Langton's Automaton}
     \label{Figure 7}
\end{center}
\end{figure}

A clever automaton, due to Chris Langton, takes the initial form of a square in the plane.
The square extrudes an edge that grows to one edge length and a little more, 
turns by ninety degrees, grows one edge length, turns by ninety degrees grows one edge length,
turns by ninety degrees and when it grows enough to collide with the original extruded edge, 
cuts itself off to form a new adjacent square, thereby reproducing itself. This scenario is 
repeated as often as possible producing a growing cellular lattice. See Figure~\ref{Figure 7}. 
\bigbreak

\begin{figure}
     \begin{center}
     \begin{tabular}{c}
     \includegraphics[width=6cm]{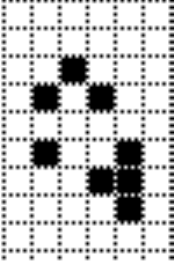}
     \end{tabular}
     \caption{\bf Condition D with geometric period $48$}
     \label{Figure 8}
\end{center}
\end{figure}

The replications that happen in automata such as Conway's Life are all really instances of periodicity
of a function under iteration. The glider is an example where the Life game function $L$ applied to an
initial condition $G$ yields $L^{5}(G) = t(G)$ where $t$ is a rigid motion of the plane. Other intriguing
examples of this phenomenon occur. For example the initial condition $D$ for Life shown in Figure~\ref{Figure 8} has the property that 
$L^{48}(D) = s(D) + B$ where $s$ is a rigid motion of the plane and $s(D)$ and the residue $B$ are disjoint
sets of marked squares in the lattice of the game. $D$ itself is a small configuration of eight marked
squares fitting into a rectangle of size $4$ by $6.$ Thus $D$ has a probability of $1/735471$ of being chosen
at random as eight points from $24$ points.  
\bigbreak

\begin{figure}
     \begin{center}
     \begin{tabular}{c}
     \includegraphics[width=6cm]{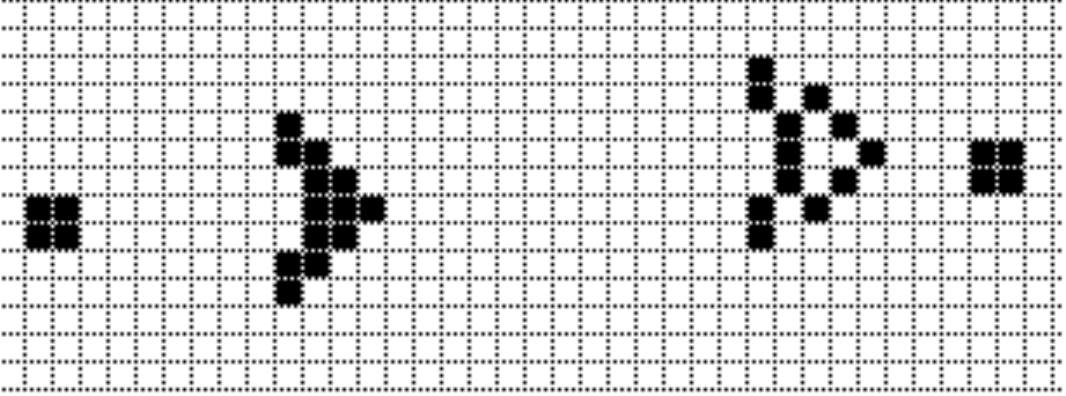}
     \end{tabular}
     \caption{\bf P(left) and Q(right) Compose the Glider Gun}
     \label{Figure 9}
\end{center}
\end{figure}

Should we regard self-replication as simply an instance of periodicity under iteration?
Perhaps, but the details are more interesting in a direct view.  The glider gun in Life is a
structure $GUN$ such that $L^{30}(GUN) = GUN + GLIDER.$ Further iterations move the disjoint glider away
from the gun so that it can continue to operate as an initial condition for $L$ in the same way.
A closer look shows that the glider gun is fundamentally composed of two parts $P$ and $Q$ such that 
$L^{10}(Q)$ is a version of $P$ and some residue and such that $L^{15}(P) = P^* + B$ where $B$ is a rectangular 
block, and $P^*$ is a mirror image of $P$, while $L^{15}(Q) = Q^* + B'$ where $B'$ is a small non-rectangular
residue. See Figure~\ref{Figure 9} for an illustration showing the parts $P$ and $Q$ (left and right) flanked by small blocks that
form the ends of the gun. One also finds that $L^{15}( B + Q^*) = GLIDER + Q + Residue.$ This is the internal 
mechanism by which the glider gun produces the glider. The extra blocks at either end of the glider
gun act to absorb the residues that are produced by the iterations. Thus the end blocks are catalysts
that promote the action of the gun. Schematically the glider production goes as follows:

$$P+Q \longrightarrow P^* + B + Q^*$$

$$B + Q^* \longrightarrow GLIDER + Q$$

\noindent whence

$$P+Q \longrightarrow P^* + B + Q^* \longrightarrow P + GLIDER + Q = P + Q + GLIDER.$$

\noindent The last equality symbolizes the fact that the glider is an autonomous entity no longer involved 
in the structure of $P$ and $Q.$ It is interesting that $Q$ is a spatially and time shifted version of 
$P.$ Thus $P$ and $Q$ are really ``copies" of each other in an analogy to the structural relationship of 
the Watson and Crick strands of the DNA. The remaining part of the analogy is the way the catalytic
rectangles at the ends of the glider gun act to keep the residue productions from interfering with 
the production process. This is analogous to the enzyme action of the topoisomerase in the DNA.
\bigbreak

The point about this symbolic or symbiological analysis is that it enables us to take an analytical 
look at the structure of different replication scenarios for comparison and for insight.
\bigbreak

There are a number of variants of Conway Life. We have earlier in this paper discussed HighLife and its self-replicator whose pattern is a direct relative to the self-replicator in the 1DRD.
Another variant of Conway Life is discussed in \cite{REF} by L. Kauffman and denoted by the name 7-Life in that paper.  The generative rule for 7-Life is $B37/S23,$ meaning that an empty square
gives birth to a marked square if it has either three neighbors or seven neighbors, and a marked square survives to the next generation if it has either two or three neighbors. Conway Life is defined by the 
distinction $B3/S23.$ In Conway Life one has gliders that occur naturally and we have discussed the glider gun that emerged from a design interaction with computer experiments using Conway Life.
7-Life behaves differently from Conway Life. There are still naturally occuring gliders, but relatively small initial configurations tend to behave dynamically, interacting via the gliders to produce self-sustaining, 
slowly growing configurations. These configurations can eventually give birth to more complex self-reproducing entities as described and illustrated in \cite{REF}. The entity that emerges, usually after thousands
of iterations, is more complex (a pair of mirror-imaged configurations) than the glider, but by our experience, not so improbable as to never emerge! This leads to the question of the possiblity and 
probability of the emergence of complex structures, analagous to biological structures, in the forward history of an RD automaton. We mention the cases of non-orthodox RD and experiments of this kind since
structurally, all these automata do operate recursively on the basis of distinctions made at each step. The variants of Conway Life and the Wolfram automata are all very simple instances of recursive distinctioning
where the basic language is binary and there is only one distinction made at each step.\\

\section{Laws of Form}
In this section we discuss a formalism due the G. Spencer-Brown in his book ``Laws of Form" \cite{GSB} that is often called the ``calculus of indications". This calculus is a study of mathematical foundations with a topological notation based on one symbol, the mark:
$$\M{ } \, .$$
This single symbol represents a distinction between its own inside and outside.
The mark is seen as making a distinction and the calculus of indications is a calculus of distinctions where the mark refers to the act of distinction. The mark is self-referential and refers to its own action and to the distinction that is made by the mark itself. Spencer-Brown is quite explicit about this identification of action and naming in the conception of the mark, and by the end of the book he reminds the reader that ``the mark and the observer are, in the form, identical".  We make this discussion here because it is important to trace the origins of the idea of distinction that is so central to the present paper.\\

The concept of distinction as used in Laws of Form is very close to that used implicitly in set theoretic mathematics. There the fundamental distinction is represented by set brackets (the act of collecting into a set) and the empty set $\{ \,\, \}$ is the ``first distinction".\\

\bigbreak

In the calculus of indications the mark can interact with itself in two possible ways. The resulting formalism becomes a version of Boolean arithmetic, but fundamentally simpler than the usual Boolean arithmetic of $0$ and $1$ with its 
two binary operations and one unary operation (negation).  \\

 Remarkably, the calculus of indications provides a context in which we can say exactly that a certain logical particle, the mark,  can act as negation {\it and} can interact with itself to produce itself.
\bigbreak

The mathematics in Laws of Form begins with two laws of transformation about these two basic expressions. Symbolically, these laws are:
\begin{enumerate}
\item Calling : $$\M{} \, \M{} \, =   \M{}$$  
\item Crossing: $$\M{ \M{ } }  =  \,\,\,\,.$$
\end{enumerate}
The equals sign denotes a replacement step that can be performed on instances of these patterns
(two empty marks that are adjacent or one mark surrounding an empty mark).
In the first of these equations two adjacent marks condense to a single mark, or a single mark expands to form two adjacent marks.  In the second equation  two marks, one inside the other, disappear to form the unmarked state indicated by nothing at all. 
That is, two nested marks can be replaced by an empty word in this formal system.  Alternatively, the unmarked state can be replaced by two nested marks. These equations give rise to a natural calculus, and the mathematics can begin.  For example,  {\it any expression can be reduced uniquely  to either the marked or the unmarked state.}  The following example illustrates the method:
$$     \M{\M{\M{\M{} \M{}} \M{}} \M{}} \M{}  =   \M{\M{\M{\M{}} \M{}} \M{}}\M{} =   \M{\M{ \M{}} \M{}}\M{} $$
$$ = \M{\M{}}\M{} = \M{} \,\,\,.$$
The general method for reduction is to locate marks that are at the deepest places in the expression
(depth is defined by counting the number of inward crossings of boundaries needed to reach the given mark). Such a deepest mark must be empty and it is either surrounded by another mark, or it is adjacent to an empty mark. In either case a reduction can be performed by either calling or crossing. 
\bigbreak 

Laws of Form begins with the following statement.
``We take as given the idea of a distinction and the idea of an indication, and that it is not possible to make an indication without drawing a distinction. We take therefore the form of distinction for the form."  
Then the author makes the following two statements (laws):
\begin{enumerate}
\item {\it The value of a call made again is the value of the call.}
\item {\it The value of a crossing made again is not the value of the crossing.}
\end{enumerate}
The two symbolic equations above correspond to these statements. First examine the law of calling. It says that the value of a repeated name is the value of the name. In the equation
$$\M{} \, \M{} \, = \M{}$$
one can view either mark as the name of the state indicated by the outside of the other mark.  
In the other equation
$$\M{ \M{ } } = \,\,\,\,.$$
the state indicated by the outside of a mark is the state obtained by crossing from the state indicated on the inside of the mark. Since the marked state is indicated on the inside, the outside must indicate the unmarked state.  The Law of Crossing indicates how opposite forms can fit into one another and vanish into nothing, or how nothing can  produce opposite and distinct forms that fit one another, hand in glove.  The same interpretation yields the equation
$$\M{} \, = \, \M{}$$
where the left-hand side is seen as an instruction to cross from the unmarked state, and the right hand side is seen as an indicator of the marked state. The mark has a double carry of meaning. It can be seen as an operator, transforming the state on its inside to a different state on its outside, and it can be seen as the name of the marked state. That combination of meanings is compatible in this interpretation.  
\bigbreak

From the calculus of indications, one moves to algebra.  Thus 
 $$\M{\M{A}}$$
stands for the two possibilities
  $$\M{\M{\M{}}} \, = \, \M{} \,  \longleftrightarrow  \, A = \M{}$$
$$\M{\M{}} \, = \, \, \, \,  \longleftrightarrow \, A \,  = $$
In all cases we have
$$\M{\M{A}} \, = \, A.$$
 
 By the time we articulate the algebra, the mark can take the role of a unary operator
 $$ A \longrightarrow \M{A}.$$ But it retains its role as an element in the algebra.
Thus begins algebra with respect to this non-numerical arithmetic of forms.  The primary algebra that emerges is a subtle precursor to Boolean algebra.  One can translate back and forth between elementary logic and primary algebra:
\begin{enumerate}
\item $\M{} \longleftrightarrow T$
\item $\M{\M{}} \longleftrightarrow F$
\item $\M{A} \longleftrightarrow \sim A$
\item $AB \longleftrightarrow A \vee B$
\item $\M{\M{A} \M{B}} \longleftrightarrow A \wedge B$
\item $\M{A}B \,\, \longleftrightarrow \,\,A \Rightarrow B$
\end{enumerate}
The calculus of indications and the primary algebra form an efficient system for working with basic symbolic logic.
\bigbreak

By reformulating basic symbolic logic in terms of the calculus of indications, we have a ground in which negation is represented by  the mark {\em and} the mark is also interpreted as a value (a truth value for logic) and these two intepretations are compatible with one another in the formalism. 
At this point the reader can appreciate what has been done if he returns to the usual form of symbolic logic. In that form we that $$\sim \sim X = X$$ for all logical objects (propositions or elements of the logical algebra) $X.$ We can summarize this by writing $$\sim \sim \,\,\,= \,\,\, $$ as a symbolic statement that is outside the logical formalism. Furthermore, one is committed to the interpretation of 
negation as an operator and not as an operand. The calculus of indications provides a formalism where
the mark (the analog of negation in that domain) is both a value and an object, and so can act on itself in more than one way.
\bigbreak

The mark as linguistic  particle is its own anti-particle. It is exactly at this point that physics meets logical epistemology. Negation as logical entity is its own anti-particle.  In our view,  the world and the formalism we use to represent the world are not separate.  The observer and the mark are (formally) identical. A path is opened between logic and physics.
\bigbreak

\begin{figure}
     \begin{center}
     \begin{tabular}{c}
     \includegraphics[height=7cm]{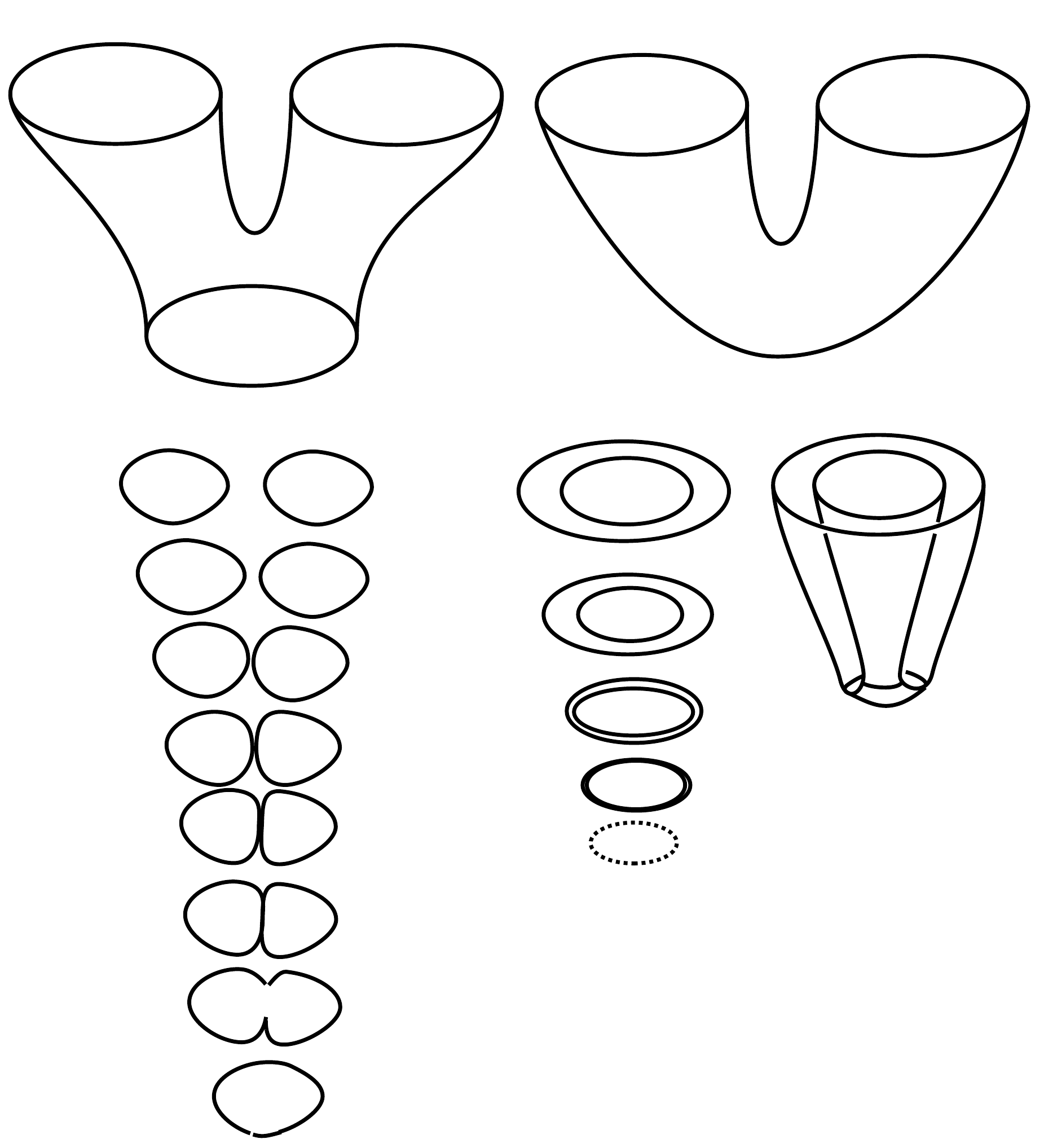}
     \end{tabular}
     \end{center}
     \caption{\bf Calling, Crossing and Cobordism}
     \label{callcross}
     \end{figure}

The visual iconics that create via the half-boxes of the calculus of indications a model for the mark as logical particle  can also be seen in terms of cobordisms of surfaces. View Figure~\ref{callcross}. There the boxes have become circles and the interactions of the circles have been displayed as evolutions in an extra dimension, tracing out surfaces in three dimensions. The condensation of two circles to one is a simple cobordism betweem two circles and a single circle. The cancellation of two circles that are concentric can be seen as the right-hand lower cobordism in this figure with a level having a continuum of critical points where the two circles cancel. A simpler cobordism is illustrated above on the right where the two circles are not concentric, but nevertheless are cobordant to the empty circle. Another way of putting this is that two topological closed strings can interact by cobordism to produce a single string or to cancel one another. Thus a simple circle can be a topological model for the mark, for the fundamental distinction.  

     \bigbreak

We are now in a position to discuss the relationship between logic and quantum mechanics. We go below boolean logic to the calculus of indications, to the ground of distinctions based in the phenomenology of distinction arising with the emergence of concept and percept together, in the emergence of a universe in an act of percreption. Here we find that the distinction itself is a logical particle that can interact with itself to produce itself, but can also interact with itself to annihilate itself. The fundamental state is a superposition of these two possibilities for distinction.
We are poised between affirmation of presence and the fall into an absence that we cannot know. This superposition is likely not yet linear in the sense of the simple model of quantum theory. Nevertheless, it is at this source, the place of arising and disappearing of awareness, that we come close to the quantum world in our own experience. As always, this experience is known to us in ways more intimate than the reports of laboratory experiments.  It is the uniqueness of every experience, of every distinction. There can be no other one. There is only  this and this and this yet again.\\

Nevertheless, one can go on and consider quantum states related to the aforementioned logical particle. Crossing this boundary into quantum theory proper one finds that topology and physics come together in this realm and there is a complex possibility of much new physics to come and a new basis for quantum computing. We refer the reader to \cite{Majorana} for more about this theme. It will take more thought and a sequel to this paper, to begin to sort out the relationships between quantum theory and recursive distinctioning at the level of this form of epistemology.\\

\noindent{\bf Remark.} In Laws of Form we can express $XOR(A,B) = A^{B} = B ^{A}$ by the formula $$A^{B} = B^{A} = \M{\M{A} B}\, \M{A \M{B}}.$$
Note that if $B$ is marked, then $$A^{\M{\,}} = \M{A}.$$  Thus the operation of $XOR$ is the action of the mark itself. We can regard diagrammatic circuits such as we used in Figure~\ref{xorcircuit} to be applications of the mark in the form of the $XOR$ operation above. In this way, the apparently awareness-dependent operations of Laws of Form shift to the automatic discrimination capabilities of computer circuits and the 
forms of RD can be seen as written in the language of the calculus of indications. These points of view inform each other circularly.\\

\section{Commentary}
Here is a collection of remarks and insights into recursive distinctioning that come from conversations between the authors of this paper over a number of years.

\begin{enumerate}
\item Joel: When distinction-making is applied to a pattern there is a new pattern that
is comprised of the variety of distinctions recorded.
Thus, a new pass of distinction-making can be applied to the pattern of distinctions,
and this kind of a process can repeat itself recursively, indefinitely.

\item Joel: I had made a discovery  (mathematical in nature) of
processes of recursive distinctioning (which is not patentable per se), and then
invented a physical embodiment that performs these processes. 

\item Joel:  
The sensing of gradients (chemical concentrations, nutrients, etc.) in bacteria is well
established and demonstrable.   These are elements of distinction-making at very
primitive levels.  It is much harder to demonstrate recursive distinction-making in bacteria,
because these are more abstract operations.  It can be done however with live neuron circuits,
and about 250K separate us now from results of such a demo.\\

Eshel Ben-Jacob proposed that recursive distinction-making may be easier to
demonstrate in genetic/immunological systems and it would also be much cheaper
 than the work planned with neurons.   I am awaiting for more details.
At any rate, Eshel's program is all interrelated, with recursive distinction-making
being a unifying theme.\\

\item Joel: I tend to think in terms of sensory-driven cognition
that is constructed bottom-up, beginning with stage 1 -
sensory distinctions; and proceeds to stage 2 -
indefinite recursion that starts out from stage 1
and builds up successive layers of distinctions-of-distinctions.  
It is unlikely that these two low-level stages involve awareness. 
A working hypothesis is that some sort of awareness emerges from
the primitive stages 1 and 2 towards a level that you identified as
type 1.  So, basically I tend to think of your type 1 as an epiphenomenon
that arises from stages 1 and 2.\\

[Type 1 for LK is a distinction that comes simultaneously with an awareness of that distinction.]\\
 
I believe (actually have shown) that stages 1 and 2 are mechanizable. 
A missing link, of course, is the transition from stages 1 and 2 to your type 1.
I am very sympathetic to constructivist dispositions and the place
of human beings in the order of things.   I agree that thought thinking
itself is all we got...  but I see no contradiction in proposing that
thought processes have their ultimate genesis in pre-cognitive and pre-aware
primitve processes of sensory-driven recursive distinctioning.\\
 
\item Joel: Spencer-Brown has been very seductive to a lot of people
and rightfully so.   For most of us, drawing a distinction is a cognitive act
that is performed by a full-blown human being.   Spencer-Brown, of course,
represents a distinction by some sketching of circles on a piece
of paper by a human.  I don't object to this! That's how much of mathematics
is done.   Scribbling of some symbols, sometimes in reference to some
drawings of geometric or topological configurations. \\

But doubts linger.   Is it possible to entertain a situation where distinctions
are drawn by acts that are short of being cognitive?
And if this is possible, where is the observer, the self?   And what constitutes
the other? What will happen to the expected dynamics of ``I and Thou".   Will there emerge
a ``becoming"   Becoming of what?
It seems utterly futile to concoct a scenario of distinction-making at a level that
is well below a cognizing person.   (And what's left of constructivism if
the cognizing person is `dissolved` to his sensory modalities?)\\

[LK: Note that Spencer-Brown never discusses how distinctions arise but always discusses distinctions that are accompanied by an awareness or an observer.]\\

Well, the thing is this.   Sensory modalities, all of them, must make local
distinctions in certain features (e.g., intensities) in signals that impinge on them. 
It has been studied in great detail in visual perception, beginning with the retina.
Photoreceptors in the retina make local distinctions of light intensities that
impinge on the retina.  (Absent this capacity for local distinctions amounts
to blindness.) This local distinction-making is accomplished by comparisons that ultimately
cause firing/non-firing.   These processes involve certain physiological/biochemical
processes, in conjunction with massive neural circuits.
The above type activity is clearly pre-cognitive, involuntary, and (with sufficient
abstraction) can be accomplished by computing machines as well.\\

 The essence of my patent document is RD (in one-dimension;  but it is
motivated directly by RD in 2-D, which operates on 2-D digital imagery. 
2-D RD is abstracted from local distinction-making at the retinal level,
as worked out by Weisel and Hubel in the early 1960s.)\\
 
I recently sketched for the history of my ideas (beginning in the
early 1960s) and how these are embedded in the patent document,
including the basis for ``fantomarks" and their streaks.\\
 
I think that the singular contribution of my particular RD processes 
is operationalizing the process of recursion on distinction-making.
For it gives precise and detailed trace of what it entails, including
an emergent dialectics, circularity, and so on.\\ 
 
To be sure, other people have talked about recursive distinction-on-distinction
(notably Maturana, in the context of his much higher-level ``languaging"),
but it should be clear that my RD is at a precognitive level, is mechanizable
and affords a thorough examination of its emergent properties.\\

\item Joel:  I noticed that thing -- the hypothetical distinction (or contingent distinction)
that hasn't actually been made.  It exemplifies the potency of distinction, even if not acted upon.
These are the wonders of distinction, actual, virtual, potential, contingent, and hidden,
to name only a few types.   Now, when these are compounded via recursion --
watch out!\\

\item Joel:   
I have no objection to make a (provisional) distinction between the kind of
distinction in RD automata and the Maturana and Varela kind of distinction.   In itself, this 
act of distinction between two distinctions is a good example of what
RD automata typically do.   I think that, in the end, we'll mutually discover
that the distinction between the two kinds of distinctions will gradually dissolve.\\
 
Here is a succinct description
of the roles of distinction in RD automata:\\

In RD automata, we have two basic elements that involve notions of distinction.\\
 
1. An element of distinction-making.   This element involves acts
of distinguishing (verb) and is a ``process".\\
 
2. The results of distinctioning are a collection of distinctions; where a distinction
is a product, object (noun).  \\
 
Usually these objects form a pattern of distinctions (the pattern as a whole is also an object)
that is subject to further acts of distinctioning.\\
 
Thus process and products alternate, recursively, where both process and products
involve notions that relate to distinction.\\
 
The process element involves distinction-making; and the product element is a
pattern of objects, referred to as distinctions. (Each such distinction is a local,
fragmentary boundary that records the result of prior acts of distinctioning.)\\
 
It is crucial to understand that the alternation between process/product is recursive
and indefinite in duration;  also, that such indefinite recursion is guaranteed to
drive the process into circularity. This, as a whole, represents the notion of
Recursive Distinctioning in RD automata.   (It is called BIP in the patent.)\\
 
The RD automata model is motivated by natural vision. The initial
stages are motivated by the retina, and the rest of the recursive process is
postulated to take place in the lateral geniculate nucleus (LGN) and the
visual cortex proper. \\
 
 In recent years, some researchers in advanced techniques in neural circuits
(not artificial neural nets, but rather actual, live neural tissue) entertain
the hypothesis that a certain version of RD automata takes place in normal
brain tissue activity.\\

\item Joel:
 This is to systematize RD by dimension.\\
 
*  1-D -- This is the case that is documented in the patent.  It was predated by the
2-D case. A neighborhood comprises 3 elements, where a central element has two neighbors.
There are exactly 4 combinations of relationships between an element and its
2 neighbors, representable by 4 ideoographs, as described in the patent.\\
 
* 2-D -- This is the case that relates to image processing;  goes back to 1964.
 A neighborhood (Moore neighborhood) is comprised of 9 elements,
where a central element has 8 neighbors.  There are exactly 256
combinations of relationships between an central element and its 8
neighbors.  These are representable by 256 ideographs.\\
 
The 2-D case can be decomposed into a network of 1-D's.
For comparison, John Conway's Game of Life is also run on a
Moore neighborhood but has only 2 states (as compared to
256 (!) states in the Game of RD).  The richness and complexity
of Game of Life is well known.  Imagine the complexity of this 2-D RD
game.\\
 
* 3-D -- A neighborhood is comprised of 27 elements, where a central
element has 26 near-neighbors.  There are exactly $2^{26}$
(i.e., 67,108,864) combinations of relationships between a
central element and its 26 near-neighbors.  Clearly, I didn't
investigate this case. Instead, I retreated to the 0-D case, see below.\\
 
* 0-D  -- This is the case where RD starts with a single speck against the 'void'.
It yields the scenario of the baryon octet, as described elsewhere. \\

\item Joel and Lou: 
Your comment is interesting.   There are RD processes that are uniquely in the
purview of human observers.  There are certain RD processes that can be
performed by automata; and there may also be RD processes in nature.
 The challenge is to integrate all three types into an encompassing framework
whose unifying theme is RD processing.   \\

As to experimenting with CA, there are obviously untold numbers of possible
CA, some of which have extremely interesting behaviors.  
 In RD we focus on a singular cellular automaton, the one CA whose rule
is recursive distinction-making.   Once we grasp that distinction-making
is a unique operation (in regards to perception and cognition) we realize
that we must focus on the particular class of RD automata, in preference 
to the other zillions of CA possibilities that are available for our consideration
and entertainment.   I submit that RD automata is the needle in the haystack
of CA.\\

\item Lou:   In programs that we design the initial automatic distinctions are
distinctions that are put in by design.
In the observation of such programs new distinctions arise for us, that
can be used for further designs.
But in nature it is not obvious how those structures that we are calling
distinction operators have arisen. We do not imagine that they occur by
design. We do not imagine that they were ideas in the mind of a designer.
I am very aware of this issue as I have experimented at other levels with
cellular automata and have seen how by varying rule structures one can
find extraordinary recursive structures that one would never have
imagined. Our relationship with our own constuctions and with nature is
complex.

\item Joel:
 Transdistinction operates on patterns of raw sensory data to produce a first
pass of local distinction-making in such patterns.  Further processing is relegated
to higher centers in the nervous system.   (For example, this is essentially what the retina
does (in part) in vision.)\\
 
This first pass is relatively easy to accomplish by computing devices.  Thus, impairment
in a sensory organ can be overcome by using such prosthetic devices.  The next
issue, of course, is how to connect the output of the prosthetic device to higher centers.
In vision, for example, a connection needs to be done to the optic nerve, or directly
to the lateral geniculate nucleus, from which the normal vision pathways would be followed 
to the visual cortex.\\
 
Assuming that such devices will become reality, would it modify our
notion of the observer?   Namely, a human observer so equipped would initiate
its observation by an automatic device that does distinction-making. So, there
you have it -- a hybrid of human/machine in a long sequence of distinction-making;
some automatic and some human-based.\\
 
 \item Joel: 
Yes, quids and quods seem to be generalized notions of containers/extainers.\\

[LK: Extainers have the formalism $E = > < $ while containers have the formalism $C = < >.$  Extainers are open to interaction from the outside.
Containers are closed forms not likely to iteract. But note that  $$EE = > < > < = > C <$$ and $$ CC = < > < > = < E >.$$ Thus an extainer interacts with an extainer to produce a container, and a container interacts with 
a container to produce an extainer. We can distinguish between containers and extainers by allowing containers to move freely (commute) with other elements. Then 
$$EE = > C < = C > < = CE$$ and we see that $C$ can be the catalyst for self-replication. And if we regard the extainer as the ``environment'', then the movement $$ < > \longrightarrow < E > $$ can be seen as our earlier abstraction of the emergence of Watson and Crick strands from the environment. We obtain the self-replication of DNA type:
$$< > \longrightarrow < E > \longrightarrow < > < >.$$
See \cite{LK10,LK11,LK12} for more about extainers and containers. ]

Inasmuch as quids and quods come about literally out of nowhere
(they are byproducts of RD that operates on arbitrary initial unspecified
things, including fantomarks), their natural algebra may be significant.\\

Quids and quods (discrete/continuous) are self-organized. They enter into
an elaborate dance that is not choreographed by external manipulation.
The dance has classical dialectical patterns.\\

Replication is part of the game. There are at least two types of replication:\\

1. For RD with fixed boundaries, there is guaranteed circularity. Thus
a whole bunch of strings are periodically replicated. These happen to be
4-letter strings with certain complementarity properties. Close enough
to DNA, but not quite the same.\\

2. For RD with shifting extainers (such as in the Baryon Octet scenario),
there is replication of patterns via self-similarity in the trace. In effect,
a basic pattern reappears periodically.\\

All in all, I propose to consider the algebra of quids/quods
(which extends your (Lou K.) notions of containers/extainers) somewhere
at the foundations of your marvelous edifice.

See this paper:  \cite{JI2}, Figure  2, page 11.
This is an RD that starts out with a first arbitrary distinction.
Focus on lines 1 thru 8.\\
 
'0' is like your Container that fuses $< >$ together.  It may contain at most
one thing.   There is a notion of Extended Container, written:  $<* \cdots *>$,
which may contain a bunch of things.  (It shows in Fig. 2 as $[==== \cdots =]$)
 ($C$ is an element of quids, and Extended Container is a quod, as defined in the patent document).
 Now, make the following substitutions in Fig. 2:\\
 
$0$  is $C$\\
 
$]$  is $>$\\
 
$[$ is  $<$\\
 
$=$ is $*$ \\
 
Lines $1$ thru $8$ will look look this:
                                                      $$ >C<$$
                                                     $$>CCC<$$
                                                   $$>C<*>C<$$
                                                 $$>CCCCCCC<$$ 
                                               $$>C<*****>C<$$
                                             $$>CCC<***>CCC<$$
                                           $$>C<*>C<*>C<*>C<$$
                                          $$>CCCCCCCCCCCCCCC<$$ 
and you can continue thru line 16 and beyond.
 Within that 16 line diagram you can identify 10 configurations that look like this:
                                  $$>C<$$
                                 $$CCC$$
                                 $$<*>$$
Those 10 configurations are self-organized similarly to the Pythagorean Tetractys.
 See Fig. 3 at page 12. Those configurations allow us to uncover the configuration of the baryon octet
that is embedded therein.   See  Figure 7 at page 16.\\

{\it Thus the physical interpretation of $>$ and $<$ are up and down quarks and $*$ is a strange quark.}\\
 
Let's recoup what we're doing.  We start out with a first distinction and
apply recursive distinctioning to it.  We develop the trace of a cellular
automaton that does RD.  Within that trace we discover the Pythagorean 
Tetractys;  within which we discover the 8 particles of the baryon octet
expressed in terms of their constituent quarks.
 Note:  There ought to be a link to$ SU(3)$, which still eludes me.\\
 
 \item LK: Clearly we have just begun this study. There is much more to come.
 \end{enumerate}

\smallbreak

\bigbreak


\begin{thebibliography}{131}

\bibitem{Bernd} Bernd Schmeikal,  Basic Intelligence Processing Space, (these proceedings).

\bibitem{JI} J. Isaacson, United States Patent No. 4286330. August 25, 1981.

\bibitem{IJK} J. Isaacson and L. H. Kauffman, letter-to-the-editor on recursive distinctioning,  Journal of Space Philosophy (JSP), Vol. 4, No. 1, Spring 2015. 

\bibitem{JI0} Private Communication with Eshel Ben-Jacob.

\bibitem{JI1}
Joel D Isaacson, ÒAutonomic String-Manipulation System,Ó U.S. Patent No. 4,286,330, Aug. 25, 1981, www.isss.org/2001meet/2001paper/4286330.pdf

\bibitem{JI2}
Joel D Isaacson, ÒSteganogramic Representation of the Baryon Octet in Cellular Automata.Ó Archived in 45th ISSS Annual Meeting and Conference: International Society for the System Sciences, Proceedings, 2001,
 www.isss.org/2001meet/2001paper/stegano.pdf

\bibitem{JI3}
Joel D Isaacson, ÒThe Intelligence Nexus in Space Exploration,Ó in Beyond Earth: The Future of Humans in Space, ed. Bob Krone (Toronto: Apogee Books, 2006), Chapter 24, 
www.thespaceshow.files.wordpress. com/2012/02/beyondearthch24-isaacson.pdf

\bibitem{JI4}
Joel D Isaacson, ÒNature's Cosmic Intelligence,Ó Journal of Space Philosophy 1, no. 1 (Fall 2012): 8-16

\bibitem{LK1}
Louis H Kauffman. ÒSign and Space,Ó in Religious Experience and Scientific Paradigms: Proceedings of the 1982 IASWR Conference (Stony Brook, NY: Institute of Advanced Study of World Religions, 1985), 118-64

\bibitem{LK2}
Louis H Kauffman, ÒSelf-reference and recursive forms,Ó Journal of Social and Biological Structures 10 (1987): 53-72

\bibitem{LK3}
Louis H Kauffman, ÒSpecial Relativity and a Calculus of Distinctions,Ó Proceedings of the 9th Annual International Meeting of ANPA (Cambridge: APNA West, 1987), 290-311

\bibitem{LK4}
Louis H Kauffman, ÒKnot Automata,Ó Proceedings of the Twenty-Fourth International Conference on Multiple Valued Logic Ð Boston (Los Alamitos, CA: IEEE Computer Society Press, 1994), 328-33

\bibitem{LK5}
Louis H Kauffman, ÒEigenform,Ó Kybernetes 34, no. 1/2 (2005): 129-50

\bibitem{LK6}
Louis H. Kauffman, ÒReflexivity and Eigenform Ð The Shape of Process,Ó Kybernetes 4, no. 3, (July 2009): 121-37

\bibitem{LK7}
Louis H Kauffman, ÒThe Russell Operator,Ó Constructivist Foundations 7, no. 2 (2012): 112-15

\bibitem{LK8}
Louis H Kauffman, ÒEigenforms, Discrete Processes and Quantum Processes,Ó Journal of Physics, Conference Series 361 (2012): 012034

\bibitem{LK9}
Marius Buliga and Louis H Kauffman, ÒChemlambda, Universality and Self-Multiplication,Ó in Artificial Life 14 Ð Proceedings of the Fourteenth International Conference on the Synthesis and Simulation of Living Systems, ed. Hiroki Sayama, John Rieffel, Sebastian Risi, RenŽ Doursat, and Hod Lipson (Cambridge, MA: MIT Press, 2014). 

\bibitem{LK10} Louis H Kauffman, Iterants, Fermions and Majorana Operators. In "Unified Field Mechanics - Natural Science Beyond the Veil of Spacetime", edited by
R. Amoroso, L. H. Kauffman, P. Rowlands, World Scientific Pub. Co. (2015).  pp. 1-32.

\bibitem{LK11} Louis H. Kauffman , Biologic. AMS Contemporary Mathematics Series,
Vol. 304, (2002), pp. 313 - 340.

\bibitem{LK12} Louis H. Kauffman, Self-Reference, Biologic and the Structure of Reproduction,
Progress in Biophysics and Molecular Biology, Special Issue edited by Plamen
Simeonov,Volume 119, Issue 3, December 2015, p. 382-409. 

\bibitem{BuligaKauff} Marius Buliga and Louis H Kauffman, Chemlambda, universality and self-multiplication, "Artificial Life 14 -Proceedings of the Fourteenth International Conference on the Synthesis and Simulation of Living Systems" Edited by Hiroki Sayama, John Rieffel, Sebastian Risi, RenŽ Doursat and Hod Lipson, MIT Press (2014). 8 pages.


\bibitem{REF} Kauffman,L. [2009] Reflexivity and Eigenform -- The Shape of Process.  - Kybernetes, Vol 4. No. 3, July 2009.



\bibitem{SRF}
Kauffman, L. [1987], Self-reference and recursive forms,  Journal of Social and Biological Structures 
(1987), 53-72.


\bibitem{BL2} L. H. Kauffman, Biologic II, in ``Woods Hole Mathematics" edited by Nils tongring and R.
C. Penner, World Scientific Series on Knots and Everything Vo.34 (2004), p. 94-132.

\bibitem{BL1} L. H. Kauffman , Biologic. AMS Contemporary Mathematics Series,
Vol. 304, (2002), pp. 313 - 340.

\bibitem{KL} L. H. Kauffman, Knot Logic,  In {\it Knots and Applications}  ed. by L. Kauffman, World Scientific Pub. Co.,
(1994), 1-110.

\bibitem {KP}
L.H. Kauffman, {\em Knots and Physics}, World Scientific Publishers (1991), 
Second Edition (1993), Third Edition (2002), Fourth Edition (2012).

\bibitem{Majorana} Louis H. Kauffman . Knot logic and topological quantum computing with majorana fermions. In ``Logic and algebraic structures in quantum computing and information", Lecture Notes in Logic, J. Chubb, J. Chubb, Ali Eskandarian, and V. Harizanov, editors,  124 pages Cambridge University Press (2016).

\bibitem{MUV} H. R. Maturana, R. Uribe and F. G. Varela, Autopoesis: The organization of living systems, its characterization and 
a model, {\em Biosystems}, Vol. 5, (1974), 7-13.


\bibitem{GSB} G. Spencer-Brown, {\it Laws of Form,} Julian Press, New York (1969).


\bibitem{FV}	F. J. Varela,  {\it Principles of Biological Autonomy,} North Holland Press (1979).


\bibitem{Wolfram} ``A New Kind of Science", (2002) Stephen Wolfram Publisher.



\end{thebibliography}
\end{document}